\documentclass[fleqn,usenatbib]{mnras} 
\usepackage[T1]{fontenc}
\usepackage{ae,aecompl}
\usepackage{aas_macros}
\usepackage{multirow}
\usepackage{layouts}
\usepackage{url}
\usepackage{graphicx}
\usepackage{float}
\usepackage[normalem]{ulem}
\usepackage{xargs}
\usepackage{comment}
\usepackage{verbatim, amsmath, amsfonts, amssymb}
\usepackage{breqn}
\usepackage[binary-units=true]{siunitx}
\usepackage[pdftex,dvipsnames]{xcolor}  
\usepackage[colorinlistoftodos,prependcaption,textsize=tiny]{todonotes}
\newcommandx{\unsure}[2][1=]{\todo[linecolor=red,backgroundcolor=red!25,bordercolor=red,#1]{#2}}
\newcommandx{\change}[2][1=]{\todo[linecolor=blue,backgroundcolor=blue!25,bordercolor=blue,#1]{#2}}
\newcommandx{\info}[2][1=]{\todo[linecolor=OliveGreen,backgroundcolor=OliveGreen!25,bordercolor=OliveGreen,#1]{#2}}
\newcommandx{\improvement}[2][1=]{\todo[linecolor=Plum,backgroundcolor=Plum!25,bordercolor=Plum,#1]{#2}}
\newcommandx{\thiswillnotshow}[2][1=]{\todo[disable,#1]{#2}}
\usepackage{color, soul}     

\newcommand{\metal}{\abund{Fe}{H}}
\newcommand{\teff}{\ensuremath{T_\mathrm{eff}}}
\newcommand{\logg}{\ensuremath{\log\,g}}
\newcommand{\abund}[2]{\ensuremath{[\mathrm{#1}/\mathrm{#2}]}}

\usepackage{acronym}
\usepackage[flushleft]{threeparttable} 
\acrodef{CMDs}{colour-magnitude diagrams}
\acrodef{CVs}{Cataclysmic variables}
\acrodef{FoV}{field-of-view}
\acrodef{IFU}{integral field unit}
\acrodef{SED}{spectral energy distribution}
\acrodef{SFR}{star-formation rate}
\title[The Southern Photometric Local Universe Survey]{The Southern Photometric Local Universe Survey (S-PLUS): improved SEDs, morphologies and redshifts with 12 optical filters}

 \author[C. Mendes de Oliveira et al.]
{C. Mendes de Oliveira,$^{1}$\thanks{E-mail: claudia.oliveira@iag.usp.br}
T.~Ribeiro$^{2,3}$, 
W.~Schoenell$^{4}$, 
A.~Kanaan$^{5}$,
R.~A.~Overzier$^{6,1}$, 
\newauthor
{A.~Molino$^{1}$,  L.~Sampedro$^{1}$, P.~Coelho$^{1}$,    C.E.~Barbosa$^{1}$
% ...
}
\\
\\
$^{1}$Departamento de Astronomia, Instituto de Astronomia, Geof\'isica e Ci\^encias Atmosf\'ericas da USP, Cidade \\ Universit\'aria, 05508-900, S\~ao Paulo, SP, Brazil\\
$^{2}$NOAO, P.O. Box 26732, Tucson, AZ 85726\\
$^{3}$Departamento de F\'isica, Universidade Federal de Sergipe, Av. Marechal Rondon, S/N, 49000-000, S\~ao Crist\'ov\~ao, SE, Brazil\\
$^{4}$Departamento de Astronomia, Instituto de F\'isica, Universidade Federal do Rio Grande do Sul (UFRGS), Av. Bento Gon\c{c}alves 9500, \\
Porto Alegre, RS, Brazil\\
$^{5}$Departamento de F\'isica, Universidade Federal de Santa Catarina, Florian\'{o}polis, SC, 88040-900, Brazil\\
$^{6}$ Observat\'orio Nacional, Minist\'erio da Ci\^encia, Tecnologia, Inova\c c\~ao e Comunica\c c\~oes, \\
Rua General Jos\'e Cristino, 77, S\~ao Crist\'ov\~ao, 20921-400, Rio de Janeiro, RJ, Brazil\\
\\
The remaining institutions are at the end of the paper.
\\
}

\newpage

\hypersetup{draft} 
\begin{document}

\date{Accepted . Received ; in original form }
\pagerange{\pageref{firstpage}-\pageref{lastpage}} \pubyear{2015}
\maketitle
\label{firstpage}
\begin{abstract}

The Southern Photometric Local Universe Survey (S-PLUS) is imaging $\sim$\SI{9300}{\deg^2} of the celestial sphere in twelve optical bands using a dedicated 0.8 m robotic telescope, the T80-South, at the Cerro Tololo Inter-american Observatory, Chile. The telescope is equipped with a 9.2k$\times$9.2k e2v detector with 10 $\micron$ pixels, resulting in a field-of-view of 2 deg$^2$ with a plate scale of 0.55\arcsec\ pixel$^{-1}$. The survey consists of four main subfields, which include two non-contiguous fields at high Galactic latitudes (|$b$| $>$ 30\degr, \SI{8000}{\deg\squared}) and two areas of the Galactic plane and bulge (for an additional \SI{1300}{\deg\squared}). S-PLUS uses the Javalambre 12-band magnitude system, which includes the 5 $ugriz$ broad-band filters and 7 narrow-band filters centered on prominent stellar spectral features: the Balmer jump/[OII], Ca H+K, H$\delta$, G-band, Mg b triplet, H$\alpha$, and the Ca triplet. S-PLUS delivers accurate photometric redshifts ($\delta_{z}/(1+z)=0.02$ or better) for galaxies with $r<20$ AB mag and $z<0.5$, thus producing a 3D map of the local Universe over a volume of more than $1 (Gpc/h$)$^3$. The final S-PLUS catalogue will also enable the study of star formation and stellar populations in and around the Milky Way and nearby galaxies, as well as searches for quasars, variable sources, and low-metallicity stars. In this paper we introduce the main characteristics of the survey, illustrated with science verification data highlighting the unique capabilities of S-PLUS. We also present the first public data release of $\sim$\SI{336}{\deg^2}  
of the Stripe 82 area, which is available at \url{datalab.noao.edu/splus}. 
\end{abstract}
 

\color{black}

\section{Introduction}

In the past decade, astronomy has firmly shifted towards the collaborative exploration of large observational surveys that provide homogeneous multi-wavelength data. In this sense, the Sloan Digital Sky Survey \citep[SDSS,][]{2000AJ....120.1579Y} opened up a new era of astronomy by covering a large area of the sky at Northern Galactic latitudes with photometry in 5 broad-band filters, supplemented by an efficient spectroscopic campaign with high completeness for Galactic stars, bright galaxies, and quasars. This has inspired numerous new survey projects in both hemispheres that are extending the SDSS legacy by covering larger areas, observing to greater depths or in other wavelengths. 

\begin{figure}
\includegraphics[width=\columnwidth]{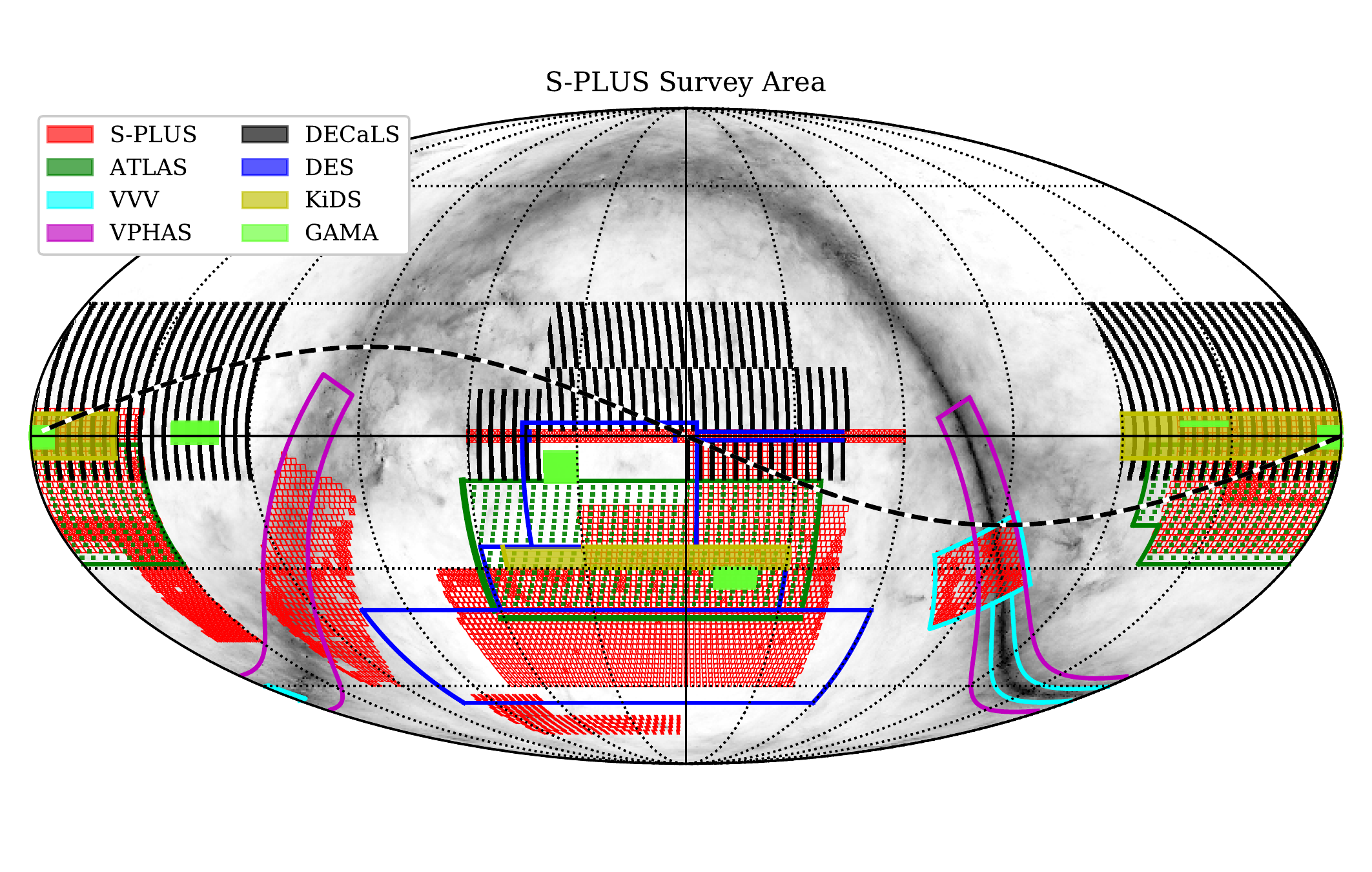}
\caption[ESO-SPLUS synergies]{Diagram in equatorial coordinates showing some of the main optical and near-infrared surveys in the Southern Hemisphere (we omit the surveys SkyMapper, Gaia, and LSST that cover the entire hemisphere or sky). For the optical surveys: ATLAS \citep{atlas} is shown in hatched green, VPHAS+ is the pink rectangular contour over the bulge and disk of the Galaxy, DECaLS is in hatched black, DES \citep{DES} is shown in blue contours,  KiDS \citep{kids} in filled-yellow, and GAMA in filled-green areas. The only near-infrared survey displayed is VISTA-VVV, in light blue contours, mainly over the Galactic bulge, overlapping with S-PLUS.   The area covered by S-PLUS is shown in red. The dashed-black line represents the ecliptic. The background image is the extinction map of \citet{1998ApJ...500..525S}
}
\label{T80Ssinergies}
\end{figure}

\begin{figure}
\includegraphics[width=8.5cm]{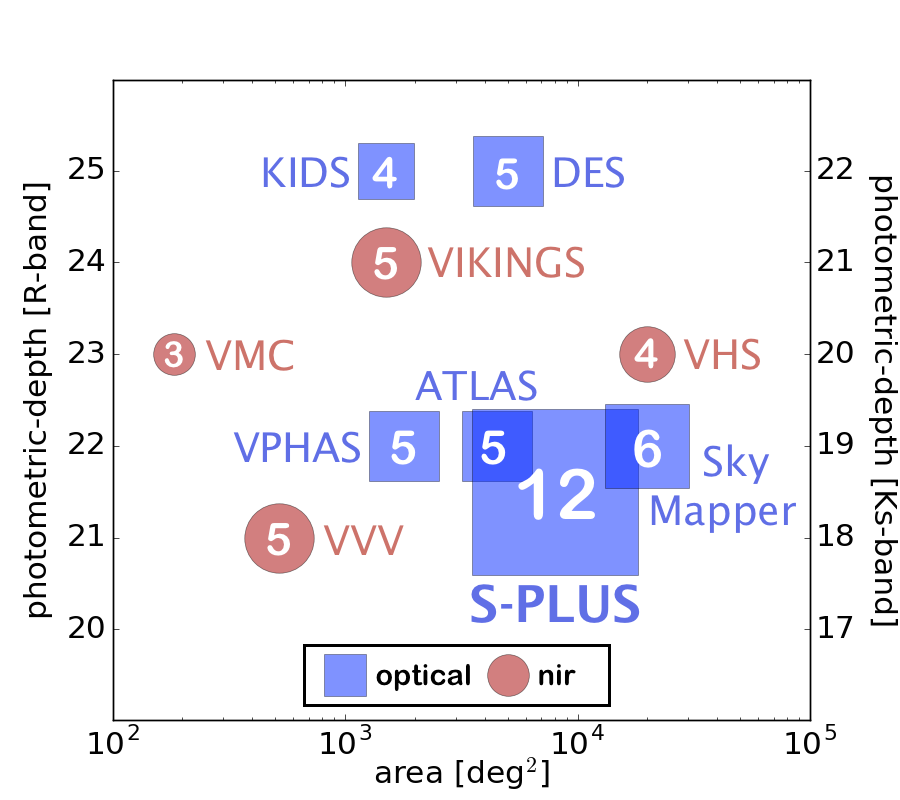}
\caption[ESO-SPLUS comparison]{Comparison of several Southern Hemisphere optical and near-infrared imaging surveys. The scales on the left and right side of the figure show the approximate depths for the optical surveys  (blue boxes) and near-infrared surveys (red circles), respectively, both in AB magnitudes. The number in each box indicates the number of filters in the survey; the box size is proportional to the number of filters.}
\label{T80Scomparison}
\end{figure}

The Southern Photometric Local Universe Survey (S-PLUS\footnote{\url{www.splus.iag.usp.br}}) is an imaging survey that will cover $\sim$9300 deg$^2$ in twelve filters, using a robotic \SI{0.8}{\meter}-aperture telescope at the Cerro Tololo Interamerican Observatory (CTIO), Chile. Besides the standard optical bands $u$, $g$, $r$, $i$, and $z$, filters centred on the following features of stars and nearby galaxies are used: [O{\sc{ii}}], Ca H$+$K, G-band, H$\delta$, Mg$b$, H$\alpha$, and CaT. As has been shown in Cenarro et al. (2019), this 12-band system is ideally suited for stellar classification, especially for very ([Fe/H] $< -2.0$) and extremely ([Fe/H] $< -3.0$) metal-poor stars, and carbon-enhanced metal-poor (CEMP) stars, as well as for a  significantly improved photometric redshift estimation of galaxies in the nearby universe.  
Although there are many current and future large-area imaging surveys in the Southern Hemisphere, S-PLUS provides
a unique sampling of the optical spectrum thanks to its seven narrow-band filters. Figs. \ref{T80Ssinergies} and \ref{T80Scomparison} show comparisons of different optical and near-infrared surveys conducted with telescopes located in the Southern Hemisphere, with respect to their area coverage, photometric depth, and number of filters. 

S-PLUS will also offer synergies with the Gaia mission \citep{Perryman2001, 2018A&A...616A...1G} that ultimately will deliver (planned for second half of 2021) low-resolution blue and red spectrophotometry for compact sources obtained through prisms, over a similar wavelength range as probed by S-PLUS. Especially in the case of resolved galaxies, the S-PLUS images will be useful for identifying which areas  contributed to the Gaia spectra, and what information is being missed.  In addition, as pointed out by Cenarro et al. (2019), the Javalambre $u$-band, in combination with the Gaia data, may be useful for improving the Gaia sensitivity at these wavelengths. When the Large Synoptic Survey Telescope \citep[LSST,][]{2008arXiv0805.2366I} comes online, it will provide deep observations of the sky observable from CTIO with temporal information, but still using only five broad-band filters. Therefore, it is foreseen that multi-band narrow-band surveys using even modest telescopes like S-PLUS can still play a useful role by providing important spectral information that is needed for a wide range of astrophysical applications. Stellar typing and photometric redshifts from multi-band surveys such as S-PLUS will provide a valuable resource for cross-checking the calibration of LSST and other surveys. 

It is important to note that J-PLUS\footnote{\url{www.j-plus.es/}}, performed with the T80/JAST telescope in Spain, has been generating data for the last several  years.  T80-South and its large-format camera, including the filters, are a duplicate of that system installed at Cerro Javalambre.  Besides doing excellent science (e.g., \citealt{2019A&A...622A.176C}), J-PLUS is also important for calibrating J-PAS, the Javalambre Physics of the Accelerating Universe Survey\footnote{\url{www.j-pas.org}}, which will take the narrow-band filter strategy to the extreme, by using 54 equally spaced narrow-band filters (\SI{145}{\angstrom}-wide) and 5 broad-band filters covering the entire optical spectrum. J-PAS will be performed with a dedicated 2.5 m telescope and a wide \ac{FoV} camera at the Javalambre Astrophysical Observatory in Spain  \citep{jpasredbook}.  However, as of yet, no such survey has been planned for the Southern Hemisphere.

This paper describes S-PLUS, highlights its
various niches, based on the results from our science
verification data obtained during the second semester of 2016 and the second semester of 2018, and presents the first 
public S-PLUS data release (DR1) in the Stripe 82 region\footnote{The Stripe 82 region covers the rectangular area within the coordinates 4$^{h}$$<RA<$20$^{h}$ and -1.26$^{o}$$<Dec<$1.26$^{o}$, \citet{2015ApJS..219...12A}.}. S-PLUS DR1 is available at \url{datalab.noao.edu/splus}, and it is characterised in Section 4 and in the NOAO data lab site,
as well as in Molino et al. (in prep.).
Section 2 describes
the technical aspects of the survey - the telescope, optics,
control system, camera, filter system - and survey strategy, including a description of the five sub-surveys of S-PLUS. 
Section 3 presents the key science areas of each sub-survey.  In Section 4, a brief
description of the data reduction pipeline is given. In addition,
this section specifies the production of catalogues and data calibration strategies,
tests of the PSF stability over the images,
photometric and photometric redshift depths, and
our plans for future data releases.  In Section 5, we present a table with the characteristics of S-PLUS DR1 and  describe some preliminary results from the analysis of the first S-PLUS dataset. Finally, Section 6 summarizes the paper.
 
\section{The S-PLUS Project} 

S-PLUS is carried out with the T80-South (hereafter, T80S), a new \SI{0.826}{\meter} telescope optimised for robotic operation; T80S is equipped with a wide \ac{FoV}  camera (\SI{2}{\deg^2}). The telescope, camera, and filter set are identical to those of the 
Javalambre Auxiliary Survey Telescope (T80/JAST), installed at the Observatorio Astrof\'{\i}sico de Javalambre. {T80/JAST} is currently performing the Javalambre Photometric Local Universe Survey (J-PLUS), a 12-band survey of a complementary area in the northern hemisphere \citep[see][for details]{2019A&A...622A.176C}.

\subsection{The S-PLUS Consortium}

The S-PLUS project, including the T80S robotic telescope and the S-PLUS scientific survey, was founded as a partnership between the S\~ao Paulo Research Foundation (FAPESP),  the Observat\'orio Nacional (ON), 
the Federal University of Sergipe (UFS), and the Federal University of Santa Catarina (UFSC), with important financial and practical contributions from other collaborating institutes in Brazil, Chile (Universidad de La Serena), and Spain (Centro de Estudios de F\'{\i}sica del Cosmos de Arag\'on, CEFCA). The consortium is open to all scientists from the participating institutes, as well as any other scientist through a vigorous external collaborator program.

\subsection{Site}

\begin{figure}
\includegraphics[width=8.3cm]{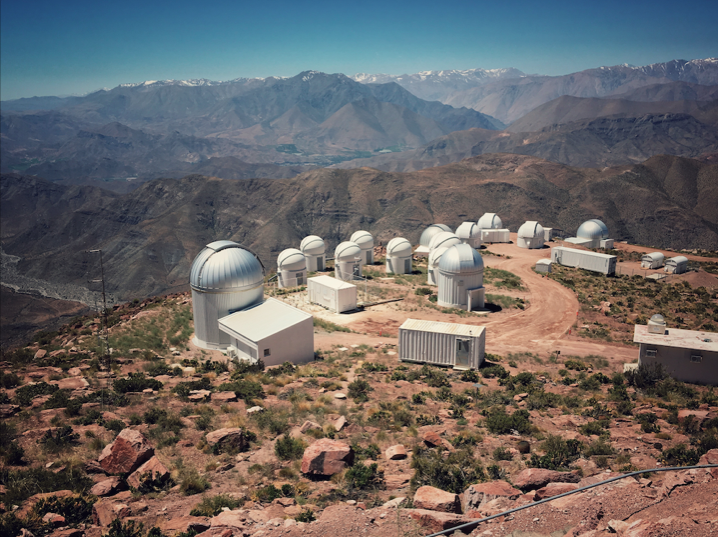}
\caption[site_figure]{T80S is located on Cerro Tololo, beside the PROMPT telescopes. In this photo, taken in October 2017, T80S is the largest dome on the left.}
\label{figure_site}
\end{figure}

The T80S is located near the summit of Cerro Tololo in central Chile, approximately two hundred meters Northeast of the \SI{4.0}{\meter} Blanco telescope. Fig. \ref{figure_site} shows a picture of the telescope and its neighbourhood. T80S sits at an altitude of \SI{2178}{\meter} above sea level, at geodetic position (World Geodetic System 84, South latitude and West longitude are negative) -30:10:04.31, -70:48:20.48 \citep{2012arXiv1210.1616M}.
CTIO has highly stable weather conditions, with 82.3\% of time 
used for wide-field survey observations over the period 2013-2016 (S. Heathcote, private communication - note that the last two years included an El Ni{\~n}o cycle). The median total seeing is \SI{0.95}{\arcsec} (FWHM), and the best 10-percentile is \SI{0.64}{\arcsec} \citep{Tokovinin.Baumont.Vasquez.2003a}.

\subsection{Telescope, Optics and Control system}

The T80S has a German equatorial mount (model NTM-1000), manufactured by the company ASTELCO\footnote{\url{www.astelco.com}}, under a contract with the company AMOS\footnote{\url{www.amos.be}}. The optical and telescope designs were done in a close collaboration between CEFCA and AMOS/ASTELCO. The same NTM-1000 universal mount, in EQ configuration, used in T80S, has since then been used in six other telescopes produced by ASTELCO, for the SPECULOOS\footnote{\url{www.speculoos.uliege.be}} and the SAINT-EX\footnote{\url{www.saintex.unibe.ch}} projects. 

\begin{figure}
\centering
\includegraphics[width=0.98\linewidth]{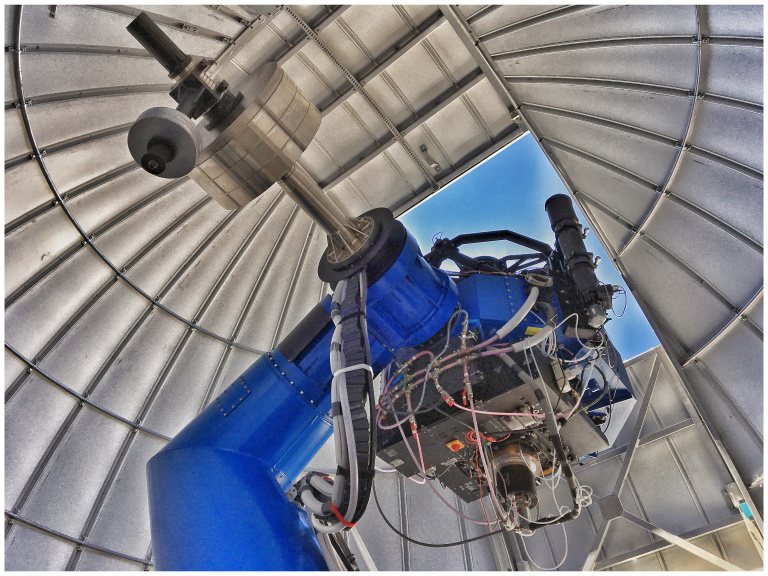}
\caption[t80scam]{T80S and its wide-field camera.}
\label{t80scam}
\end{figure}

The optical system of T80S consists of a f/4.31 Ritchey-Chretien with one axial Cassegrain focal plane and a clear aperture of \SI{860}{mm}. This provides a plate-scale of 55.56\arcsec mm$^{-1}$, a total \ac{FoV} of \SI{130}{\milli\meter} (translating to a \SI{2}{\deg} diameter on the sky), and an optimal FoV of \SI{110}{\milli\meter} (\SI{1.7}{\deg} diameter on the sky). The field corrector lens built by AMOS ensures an aberration degradation less than $1\%$. A picture of the telescope and its camera is shown in Fig.~\ref{t80scam}. T80S is housed in an \SI{8}{m} Ash dome. The telescope can slew between two opposite sky positions in less than \SI{1.5}{\minute}, the limiting factor being the time it takes for the dome to move between the two positions. T80S is robotically operated by the \verb/chimera/\footnote{\url{github.com/astroufsc/chimera/}} observatory control system. Developed in Python, \verb/chimera/ uses the Pyro3 library to convert the  observatory sub-systems into Python objects that are accessible over the local network in a distributed way. On top of this framework, a supervisor algorithm takes care of checking the weather conditions, and executes the observations according 
to constraints imposed by the astronomical conditions.

\subsection{Camera}

T80S is equipped with an optical imager, T80Cam-S, consisting of a 12-filter system distributed in two filter wheels (see Section 2.5), shutter, entrance window, cryostat, detector, and the corresponding electronics and control system.
The camera T80Cam-S is a duplicate of T80Cam \citep{2012SPIE.8446E..6HM}; both cameras were produced by the company Spectral Instruments\footnote{\url{www.specinst.com/}}.  
T80Cam-S is operated through the Observatory Control System \verb/chimera/. 

The detector used is a 9232$\times$9216 \SI{10}{\micron}-pixel array manufactured by the company e2v\footnote{\url{www.e2v.com}}. The telescope plate scale at the detector is 0.55\arcsec pixel$^{-1}$, and the \ac{FoV} of the camera is $1.4\times1.4$ \si{deg^2}.  The CCD is read out with 16 amplifiers organized in an $8\times 2$ array. During readout of  the amplifiers, the camera controller adds 27 pre- and post-scan pixels along the serial direction, and 54 post-scan pixels in the parallel direction for the overscan correction. The detector can be operated at four different readout speeds, and two different gains, with either the  1x1 unbinned option or binned 2x2. By default, we only use the regular 1x1 unbinned option through our control system.   See Table~\ref{tab:ccd_read_out} for the available readout modes, where the values over all 16 amplifiers have been averaged, for each mode, binning option, and gain. The last column shows the time needed for reading out an entire frame. We regularly use mode 5 for scientific observations since December 2017, which provides the best compromise between readout speed and readout noise. 
	
\begin{table}
\caption{\small Available T80Cam-S readout speed and gain modes.}
\begin{center}
\label{tab:ccd_read_out}
\begin{threeparttable}
\begin{tabular}{c|c|c|c|c|c}
\hline
\hline
Mode & Read rate &  Bin & Gain & RON & time \\
     &    (KHz)  &  &($e^{-}$/ADU)& ($e^{-}$)& (s) \\
\hline
0       & 1010& 1x1 &   2.03&   6.60&   10.83 \\
1        & 1010& 1x1 &  0.91&   5.27&   10.54 \\
2        & 1010& 2x2 &  1.93&   6.28&   6.77 \\ 
3        & 1010& 2x2 &  0.89&   5.15&   6.78 \\ 
4        & 500& 1x1  &   2.12 &  4.47&   15.97 \\
5\tnote{*}        & 500& 1x1  &   0.95 &  3.43&   16.57 \\
6        & 500& 2x2  &   2.02 &  4.25&   8.14 \\ 
7        & 500& 2x2  &   0.93&   3.34&   8.13 \\ 
8&      250&    1x1   &  2.15&   3.49&   26.60 \\
9&      250&    1x1   &  0.96&   2.74&   26.60 \\
10&     250&    2x2  &   2.04&   3.33&   10.80  \\
11&     250&    2x2  &   0.94&   2.69&   10.81 \\
12&     100&    1x1  &   2.15&   2.79&   57.69 \\
13&     100&    1x1  &   0.96&   2.34&   57.69 \\
14&     100&    2x2   &  2.05&   2.67&   18.58 \\
15&     100&    2x2   &  0.94&   2.32&   18.58 \\
\hline
\hline
\end{tabular}
\begin{tablenotes}
    \item[*] S-PLUS observing mode since December 2017.
\end{tablenotes}
\end{threeparttable}
\end{center}
\end{table}
					
\begin{figure*}
\includegraphics[width=17cm]{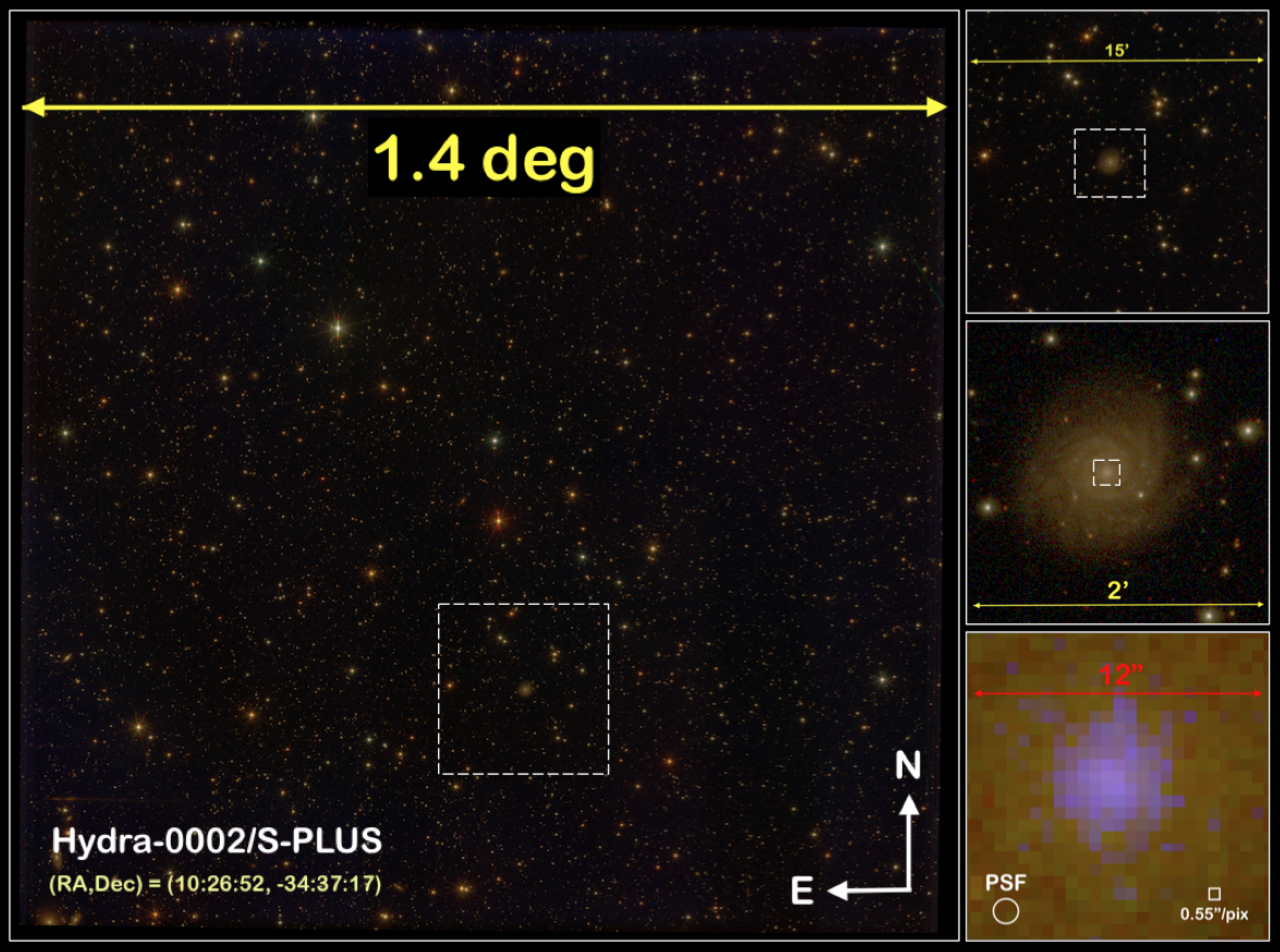}
\caption[Example of a S-PLUS field.]{Example of a S-PLUS field, illustrating the potential of combining a very wide FoV telescope  with a 9232$\times$9216 \SI{10}{\micron}-pixel array CCD detector.  The large image on the left shows the full S-PLUS field of view. The right-hand panels show consecutive zoom-in images of the centre of the Hydra cluster (\SI{15}{\arcmin} on a side, top panel), of one galaxy (\SI{2}{\arcmin} on a side, middle panel), and of a galaxy bulge (\SI{12}{\arcsec} on a side, bottom panel).} 
\label{T80SFoV}
\end{figure*}

Fig.~\ref{T80SFoV} illustrates the potential of S-PLUS in probing different astronomical scales. The left panel shows the whole field of a single image, with dimension 1.4 x 1.4 \si{\deg^2}, while the right-side panels display successive zoom-ins of the same image, including a \SI{15}{\arcmin}$\times$\SI{15}{\arcmin} field which corresponds to the scale of a nearby group or cluster, a \SI{2}{\arcmin}$\times$\SI{2}{\arcmin} field representing the scale of a nearby galaxy, and a \SI{12}{\arcsec}$\times$\SI{12}{\arcsec} field indicating the scale of the bulge of a nearby galaxy.

\subsection{The S-PLUS Filter System}

\begin{figure*}
\includegraphics[width=\linewidth, trim = 80 5 120 50, clip ]{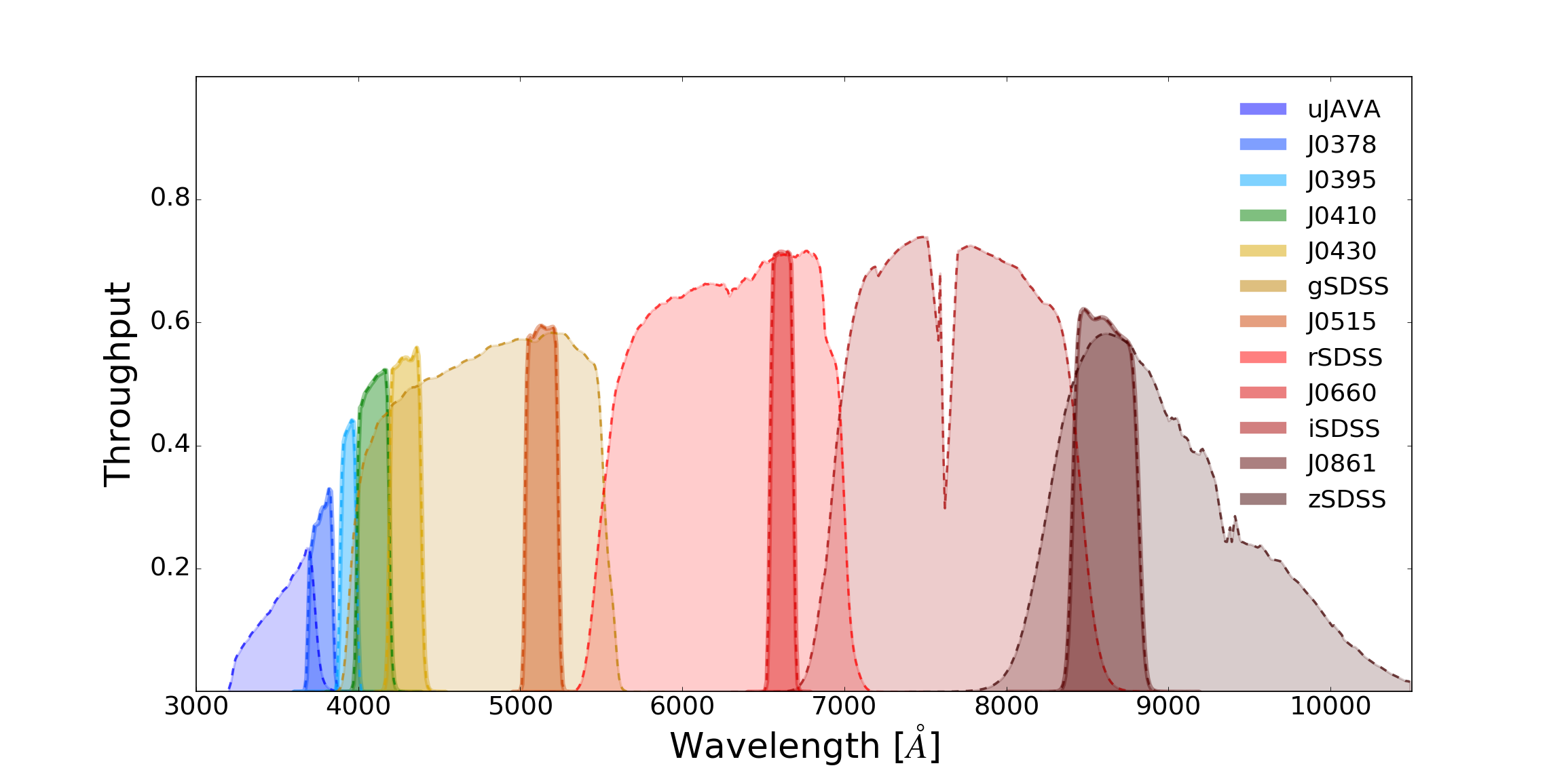}
\caption[the S-PLUS filter system]{The Javalambre 12-filter system. The y-axis shows the total efficiency of the S-PLUS filters, obtained through the multiplication of the average filter transmission curves, the atmospheric transmission, the CCD efficiency, and the primary mirror reflectivity curves. Different filters are coloured according to the labels shown in the legend at the right.}
\label{filtersystem}
\end{figure*}

S-PLUS uses the 12-filter photometric system devised for the J-PLUS project. Through a combination of broad-  and narrow-band filters that serve to identify the main stellar spectral features (absorption lines and continuum), this photometric system was designed for the optimal classification of stars \citep{Gruel12,MarinFranch12}. As illustrated in Fig.~\ref{filtersystem}, the filter system is composed of 7 narrow-band filters (J0378, J0395, J0410, J0430, J0515, J0660, J0861) that coincide with, respectively, the [OII], Ca H+K, H$\delta$, G-band, Mgb triplet, H$\alpha$, and Ca triplet features.   The system also includes the $u$, $g$, $r$, $i$, and $z$ broad-band filters which serve to constrain the spectral continuum of sources.  The $g$, $r$, $i$, and $z$ bands are similar to those from SDSS \citep{1996AJ....111.1748F},  with some small zero-point differences, listed in Table A1.
The $u$-band filter is the Javalambre $u$-band filter, which has a slightly more efficient transmission compared to the SDSS $u$-band, as described in \citet{2019A&A...622A.176C}.  

Fig.~\ref{filtersystem} presents the total transmission curves of the S-PLUS photometric system. It includes contributions from the filter transmission themselves (measured in CEFCA, in 2015 - available in the project website\footnote{\url{github.com/splus-survey/filter_curves}}), the atmospheric transmission (see below),  the efficiency of the CCD (as measured by e2v) and the primary mirror reflectivity curve (as measured in CTIO, in 2016 - the curve had no measurements beyond 880 nm; an extrapolation guided by the aluminum reflection curve was applied).   The 12 filters are  distributed between two filter wheels, which are installed inside T80Cam-S. The 2-D filter transmission maps were obtained by performing laboratory measurements over a 10 $\times$ 10 evenly spaced grid across the filter surface. The atmospheric transmission for Cerro Tololo was computed using fig. 3 of \citet{2018AJ....155...41B}. Note that curves for the secondary mirror and the corrector were not included in the computation of the total transmission curves shown in Fig.~\ref{filtersystem}.  The central wavelengths and FWHM of the filters+atmosphere+CCD+M1 transmission curves are listed in Table \ref{tablefilter}. 

\begin{figure}
\includegraphics[width=9.3cm]{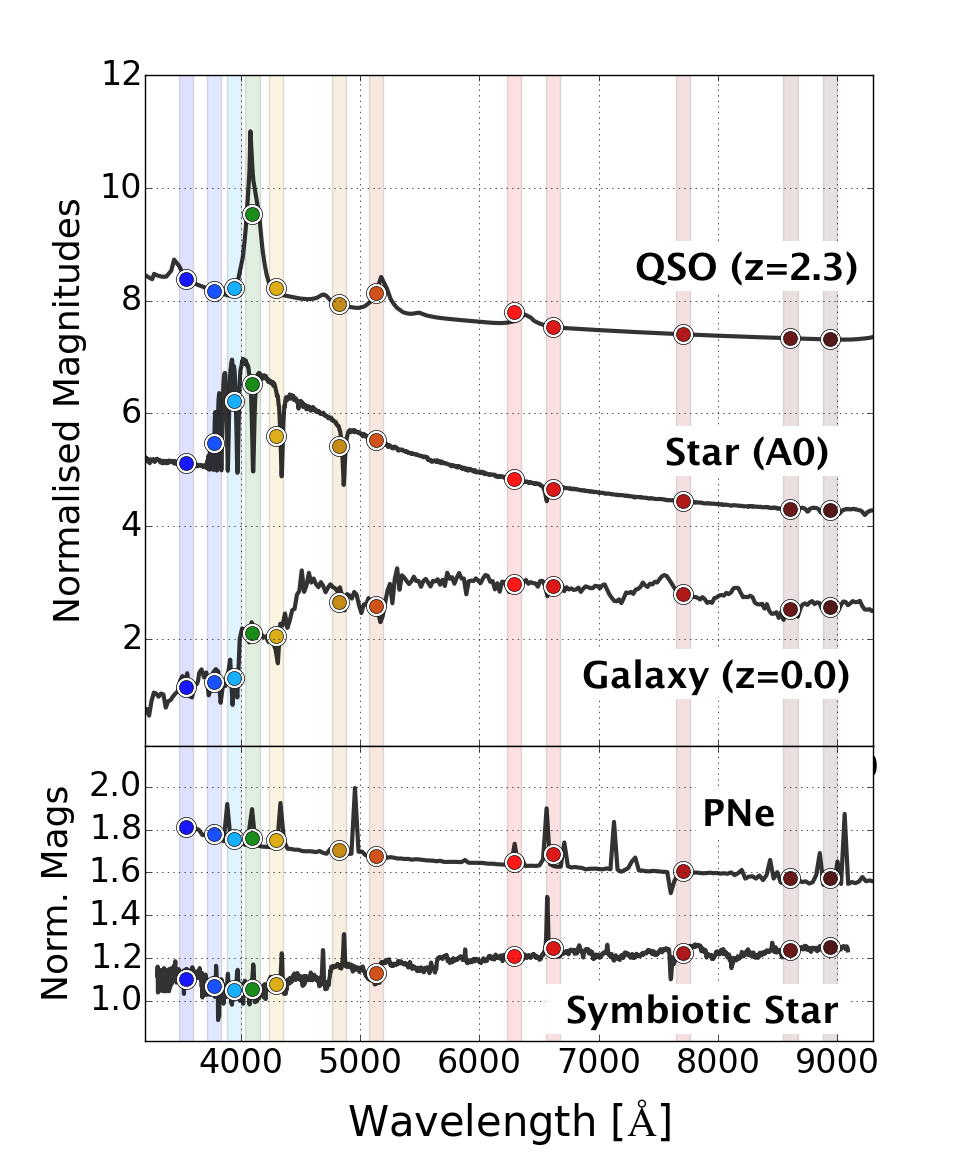}
\caption[SPLUS_photo_spectra]{Examples of different spectra (solid black lines) and their convolution with the S-PLUS 12-filter photometric system (coloured dots). From top to bottom: a quasar, a main-sequence star, an early-type galaxy, a planetary nebula, and a symbiotic star. The vertical bands correspond to the effective wavelengths of the S-PLUS filters. The coloured dots indicate the expected magnitudes after convolving the spectra with the S-PLUS filter transmission curves. }
\label{T80Sstellspectrum}
\end{figure}

Fig.~\ref{T80Sstellspectrum} shows examples of spectra of different objects (a quasar, a galaxy, an A0 star, a planetary nebula and a symbiotic system) convolved with the filters, indicating that the photometric system naturally captures the spectral information in greater detail than the 5-band SDSS or the broad-band $UBVRI$ photometric systems. 
\color{black}

\begin{table}
\caption{{\small Summary of S-PLUS filters.}}
\begin{center}
\label{tablefilter}
\begin{tabular}{lccc}
\hline
\hline
Filter	&	$\lambda_{\rm eff} $	&	$\Delta \lambda$	&		Comment	\\	
name	&	[\AA]	&	[\AA]	&		\\	\hline
uJAVA & 3574 & 330 & Javalambre $u$ \\ 
J0378 & 3771 & 151 & $[\mathrm{O}\,\textsc{ii}]$ \\ 
J0395 & 3941 & 103 & Ca H+K \\ 
J0410 & 4094 & 201 & H$\delta$ \\ 
J0430 & 4292 & 200 & G-band \\ 
gSDSS & 4756 & 1536 & SDSS-like $g$ \\ 
J0515 & 5133 & 207 & Mgb Triplet \\ 
rSDSS & 6260 & 1462 & SDSS-like $r$ \\ 
J0660 & 6614 & 147 & H$\alpha$ \\ 
iSDSS & 7692 & 1504 & SDSS-like $i$ \\ 
J0861 & 8611 & 408 & Ca Triplet \\ 
zSDSS & 8783 & 1072 & SDSS-like $z$ \\ 
\hline
\hline
\end{tabular}
\end{center}
\end{table}   

\subsection{Overview of the S-PLUS scheduling strategies}
  
S-PLUS is composed of five sub-surveys, described in detail in the next section.
The robotic operation of the telescope allows autonomous management of the observations of these sub-surveys. The observatory control system (\verb/chimera/) contains a built-in queue execution module capable of conducting different modes of observations. In standard configuration mode, a set of observations is planned and fed into the queue before the night starts. A separate module automatically selects suitable target fields belonging to the different sub-surveys, given a set of sky conditions and assigned priorities, and feeds them into the queue execution module.  
 
During day-time operations, the module pre-selects suitable target fields and simulates the observing night for different sky conditions. Remote operators check the results of the simulation and, if required, apply corrections to the scheduling parameters. During night-time operations, the module is fed with telemetry on sky and system data, and is able to make scheduling adjustments depending on the conditions. 

\section{Overview of the S-PLUS}
\label{sub_surveys}

\begin{figure*}
\centering
\includegraphics[width=\linewidth, trim=40 60 40 45, clip=true]{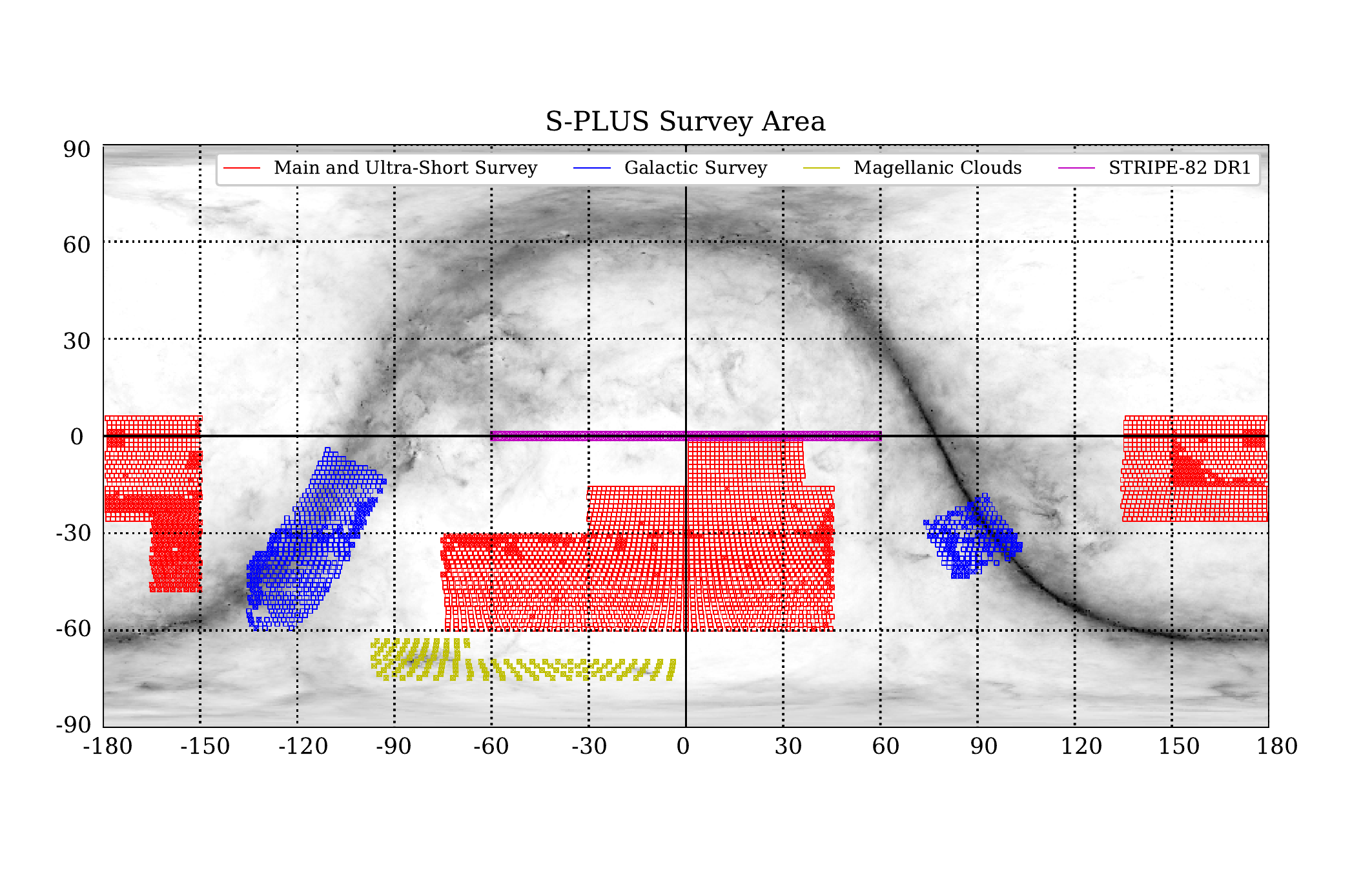}
\caption{Footprint of three of the five S-PLUS sub-surveys, over-plotted onto the extinction map of \citet{1998ApJ...500..525S} 
in Cartesian projection.  Red squares show the Main and Ultra-Short Surveys, which share the same area. 
Blue squares show the Galactic fields. Yellow squares highlight the area of the Magellanic Clouds, which are included in the Main Survey. Filled areas have already been observed at the time of this writing, in March 2019. Magenta is the area of the Stripe 82 contained in DR1 - this is part of the Main Survey but we highlight it with a different colour for clarity.}
\label{fig:surveys}
\end{figure*}

In order to optimize the usefulness of S-PLUS data for the different science topics of interest to the collaboration, the S-PLUS is divided into five sub-surveys, which are detailed in \S\ref{MAIN} to \S\ref{MARBLE} below. Additional information on the sub-survey areas, exposure times, filters, and cadences are summarised in Table~\ref{tab:surv-strategies}; their sky coverage is shown in Table \ref{table:coordinates} and Fig. \ref{fig:surveys}. 

\begin{table*}
\caption{Overview of the S-PLUS sub-survey strategies.}
\begin{tabular}{cccccccc}
\hline
Sub-survey &  Area & Visits & Filters & $T_{exp}$ & Sky & FWHM & Moon  \\
\hline
\hline
MS         &  8000 deg$^2$    &   1 &  all &  Table  \ref{tab:MSexptimes} &  phot & $<$ 2.0\arcsec & grey/dark  \\
Footprint:           &         &     &     &            &       \\
 see Fig.~\ref{fig:surveys}           &  & & & &  \\
\hline
USS  &     8000 deg$^2$          & 1 &  all & 1/12 of MS &  non-phot & any & any                 \\
\hline
VFS &  TBD                &   TBD     &       TBD     &  TBD          
& non-phot & any & any  \\
\hline
GS   &  1300 deg$^2$                    &    1  &   all     &  Table \ref{tab:MSexptimes}   &              phot & any & any  \\
bulge (400 deg$^2$):     &              &       &          &    & &                 \\
 $-10^o<l<10^o$,   &    &       &          &    & &                 \\
\& $-15^o<b<5^o$    &    &       &          &   & &                  \\
disk (1020 deg$^2$):     &              &  &          &    & &                 \\
$220^o<l<278^o$,       &    &   &  &        & &   \\
\& $-15^o<b<5^o$       &    &   &  &       & &    \\
&  &  1 & $r'$,~$i'$,~H$\alpha$ & 1/12 of MS & non-phot & any & any  \\
For selected GS fields  &  &  $>25$ & $r'$,~$i'$,~H$\alpha$ & Table \ref{tab:MSexptimes}  & non-phot & any & any   \\
\hline
MFS       &{\it Dorado group},~{\it M83}&           &    all        & Table \ref{tab:MSexptimes}     &      phot &  $>$ 2.0\arcsec & grey/dark\\
See Table 6   &  {\sc SMC/47Tuc},~{\it Hydra Cluster}    &           &                      &     & &                                     \\
\hline
\hline
\end{tabular}
\label{tab:surv-strategies}
\end{table*}

\begin{table*}
\caption{Survey Coordinates}
\resizebox{\textwidth}{!}
{\begin{tabular}{|c|c|c|c|}
\hline
                                                                                      &                            & \multicolumn{2}{c|}{(RA, Dec)}                                                                         \\ \hline
\multirow{2}{*}{Galactic Survey}                                                      & Disk (polygon with vertices)  & \multicolumn{2}{c|}{$(136^o, -40^o); (133^o, -60^o); (110^o, -4^o); (92^o, -14^o)$}                                                                                                                                                \\ \cline{2-4} 
                                                                                      & Bulge (polygon with vertices) & \multicolumn{2}{c|}{$(287^o, -26^o); (276^o, -44^o); (268^o, -17^o); (256^o, -34^o)$}                                                                                                                                              \\ \hline
\multirow{10}{*}{Main and Short Surveys}                                                          & Stripe 82                      & $0^o < $RA$ < 60^o$ and $300^o < $RA$ < 360^o$ & $-1.4^o < $DEC$ < +1.4^o$   \\ \cline{2-4} 
                                                                                      & Hydra Cluster                         & $150^o < $RA$ < 165^o$                                                                                                             & $-48^o < $DEC$ < -23.5^o$                                       \\ \cline{2-4}
                    & \multirow{2}{*}{Magellanic Clouds}             & $65.5^o < $RA$ < 98^o$                                                                                                              & $-69^o < $DEC$ < -62.5^o$                                       \\ \cline{3-4} 
                                                                                      &                               & $2^o < $RA$ < 98^o$                                                                                                                & $-75.5^o < $DEC$ < -69^o$                                       \\ \cline{2-4} 
                 & \multirow{6}{*}{{\begin{tabular}[c]{@{}c@{}}Remaining\\ S-PLUS \\ fields\end{tabular}}}             & $323.5^o < $RA$ < 359.5^o$                                                                                                         & $-15.5^o < $DEC$ < -1.4^o$                                      \\ \cline{3-4} 
                                                                                      &                               & $0^o < $RA$ < 30^o$ and $315^o < $RA$ < 360^o$                                                                           & $-30^o < $DEC$ < -15.5^o$                                       \\ \cline{3-4} 
                                                                                      &                               & $0^o < $RA$ < 75^o$ and $315^o < $RA$ < 360^o$                                                                           & $-60^o < $DEC$ < -30^o$                                         \\ \cline{3-4} 
                                                                                      &                               & $150^o < $RA$ < 165^o$                                                                         & $-23^o < $DEC$ < +5^o$   \\ \cline{3-4} 
                                                                                      &                               & $165^o < $RA$ < 225^o$                                                                       & $-26.5^o < $DEC$ < +5^o$ \\ \hline
\end{tabular}}
\label{table:coordinates}
\end{table*}

\subsection{The Main Survey}
\label{MAIN}

The {\it Main Survey} (MS) covers an area of $\sim$8,000 deg$^2$ with a single epoch observation of each field, per filter, under photometric conditions and seeing from \SI{0.8}{\arcsec} to \SI{2.0}{\arcsec}. Three consecutive dithered exposures are taken in each filter, for a total exposure time of approximately one hour and 30 minutes per field. Each of the three individual exposures of the MS (taken with the exposure times shown in Table~\ref{tab:MSexptimes}) are taken at slightly different positions in order to minimize the contribution from bad pixels and to facilitate cosmic ray cleaning. The dither offsets amounts to 10 arc sec along the RA direction ($\sim$18 pixels).  In order to mitigate differences in S/N in the edges of the images due to the dithering strategy, we ensure an overlap between images of at least 30 arcsec. This procedure is also useful to produce a homogeneous photometric calibration across the fields.

\begin{table}
\caption{Main Survey exposure times.}
\begin{center}
\label{tab:MSexptimes}
\begin{tabular}{lc}
\hline
\hline
Filter  &  T$_{\rm exp}$\\
name    &       (s)             \\
\hline
$u$   &       3$\times$227            \\
J0378   &       3$\times$220            \\
J0395   &       3$\times$118            \\
J0410   &       3$\times$59             \\
J0430   &       3$\times$57             \\
$g$   &       3$\times$33             \\
J0515   &   3$\times$61         \\
$r$   &       3$\times$40             \\
J0660   &       3$\times$290            \\
$i$   &       3$\times$46             \\
J0861   &       3$\times$80             \\
$z$   &   3$\times$56         \\
\hline
\hline
\end{tabular}
\end{center}
\end{table}

Our MS observing strategy is a modification of the J-PLUS strategy, and it is expected that the datasets from both S-PLUS and J-PLUS can be combined in the future for scientific projects where a large area ($\sim$ 16000 square degrees) is desirable. The S-PLUS MS strategy is mainly motivated by the requirements set by the extragalactic science. The original goal was to match the photometric depth of SDSS in the broad-band filters; however, S-PLUS images are, on average, shallower than SDSS (see \S\ref{photodepth} 
and Table 8). The MS has significant overlap with Pan-STARRS \citep{2012ApJ...756..158S}, DES \citep{DES}, KiDS \citep{kids}, and ATLAS \citep{atlas}, and can thus provide improved photometric redshifts for objects in these fields down to $r_{AB}\sim20$ (see \S\ref{photozdepth}).  

The determinations of photo-z, environment indicators, and star-galaxy separation (described in Section 5) using DR1,  will form the basis for a number of important extragalactic studies. For example, we expect to detect several million  galaxies in the MS -  from these data we plan to build a  new multi-wavelength galaxy catalogue, with uniform environment criteria, choosing from isolated galaxies to groups/clusters. This will extend previous Southern Hemisphere catalogues to a complete, volume-limited sample, mitigating projection effects by using the more precise S-PLUS photometric redshift information ($\delta_{z}/(1+z)=0.02$  or better, see Section 4.7). 

Exploring the 12-band filter information, we will be able to recover galaxy morphologies and stellar populations, in order to perform a pixel-by-pixel or region-by-region spectral energy distribution (SED) analysis, in an integral-field-unit approach (IFU-like science). The narrow-band filters used in S-PLUS are tailored to study absorption and emission lines at z=0. In particular, the filter J0660 is suitable to study H$\alpha$ ($\lambda=$\SI{6563}{\angstrom}) up to redshifts $z\lesssim 0.015$, providing an important tool to measure the \ac{SFR} of galaxies in the local Universe. 

S-PLUS will also be of fundamental importance for studies in our Galaxy.  It will allow searches for streams and substructures not yet known in the Galactic halo. In this respect,  blue horizontal-branch (BHB) stars and blue stragglers may be excellent indicators of structure.  Based on an extrapolation of the SDSS survey \citep{2000AJ....120.1579Y}, we should be able to detect over 50,000 BHB stars and 100,000  blue stragglers in the MS footprint. Both types of stellar objects are interesting to evaluate the stellar density of the Galactic halo profiles, and their colours may provide valuable information about the age gradient across the halo system of the Milky Way \citep{Santucci15b, 2016NatPh..12.1170C}.

\begin{figure}
\includegraphics[trim= 12 17 5 5 , clip, width=\columnwidth]{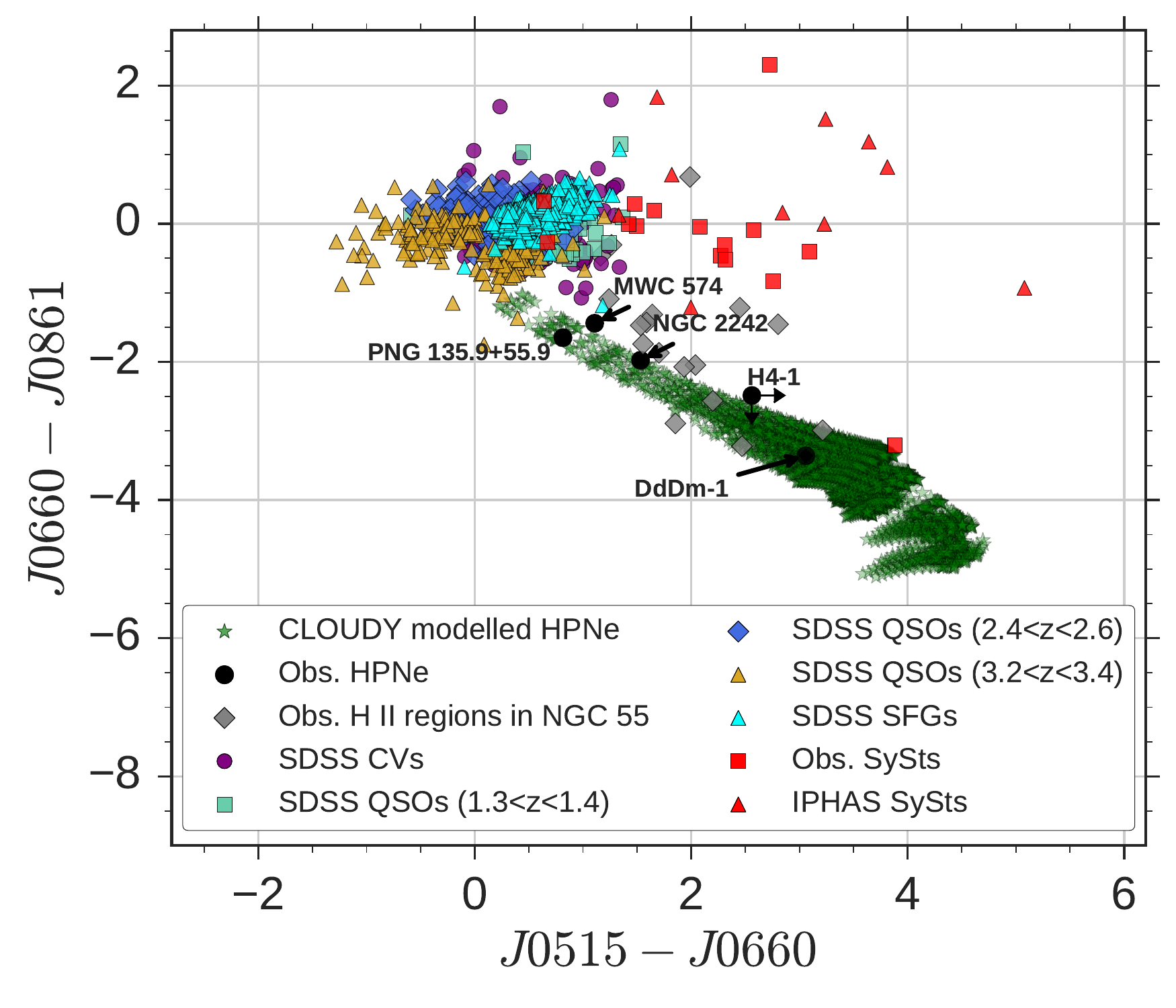}
\caption[]{The colour-colour diagram J0515-J0660 vs. J0660-J0861, used here to separate halo planetary nebulae (HPNe) and symbiotic stars (SySts). Symbols correspond to different emission line objects: modelled HPNe (dark green stars - seen from the middle to the right of the diagram); observed HPNe (black circles); SDSS quasars with redshift in the range from 1.3 to 1.4 (light-green boxes), 2.4 to 2.6 (blue diamonds), and  3.2 to 3.4 (orange triangles); SDSS cataclysmic variables (\ac{CVs}, violet circles); SDSS star-forming galaxies (SFGs, cyan triangles); symbiotic stars from \citet{2002A&A...383..188M} (red boxes, see also the new catalogue of SySts, Akras et al. 2019); symbiotic stars from IPHAS (red triangles) and extragalactic H~{\sc ii} regions (grey diamonds). Note that the halo planetary nebulae (dark green stars and black circles) and symbiotic stars (red boxes and triangles) comprise a fairly well-defined locus (and mostly away from other objects) in this colour-colour diagram, not occupied by any other emission-line objects except for the extragalactic H~{\sc ii} regions (grey diamonds).}
\label{Fig9}
\end{figure}

Other important and complementary tracers of the structure of our Galaxy are planetary nebulae and globular clusters. Statistical tools, such as principal component analysis, and classification tree analysis, among others, will help evaluating which combinations of magnitudes and colours work best to identify and study different classes of objects.  As an example, colour-colour plots using filters J0515, J0660, and J0861 are a useful selection tool for identifying halo planetary nebulae and symbiotic stars, given their characteristic spectra (see Fig. ~\ref{Fig9}). Furthermore, the 12-band filter system is sensitive to changes in stellar atmospheric parameters, including effective temperature (\teff), surface gravity (\logg),  metallicity (\metal), and abundance ratios such as [C/Fe] and [$\alpha$/Fe], and appear superior in the determination of stellar parameters compared to the 5-band SDSS system (Whitten et al. 2019). 

Finally, as each MS pointing consists of observations in 12 filters, each having 3 exposures, we obtain 36 time-steps that could also be used to detect (bright) objects that move or vary in brightness. By alternating observations in blue and red filters, we increase the temporal window in which an object is observed in two or more adjacent narrow bands.  This will allow building light curves on time-scales shorter than about 30 min, for many tens of thousands of variable stars. Thus, it is clear that the MS data can be used for a wide range of scientific topics, from Solar System to Cosmology.

\subsection{The Ultra-Short Survey}

The Ultra-Short Survey (USS) has the same footprint as the MS, with exposure times that are 1/12th of the values shown in Table \ref{tab:MSexptimes}. Therefore, the saturation limit is brighter in all 12 filters (typically 8 mag, instead of the typical 12 mag for the MS). This allows covering an important scientific niche, the search for bright low-metallicity stars.\

The most metal-poor stars in the Galactic halo carry important
information about the formation and early evolution of the chemistry in
the early Universe, as well as in the assembly of the Milky Way. Two
sub-classes are of great interest: 

\begin{enumerate}
\item { 
The ultra metal-poor (UMP; [Fe/H] 
$<-4.0$,
e.g., \citealt{beers2005,frebel2015}) stars, which are believed to be formed by
gas clouds polluted by the chemical yields of the very first (Population
III) stars \citep{iwamoto2005}. More than 80\% of the observed UMP stars in the
Galaxy present enhancements in carbon \citep[e.g.,][]{lee2013,
placco2014c}, the so-called carbon-enhanced metal-poor (CEMP) stars, and }

\item {
The highly $r$-process-element enhanced stars ($r$-II; with [Fe/H]$ <
-2.0$ and [Eu/Fe]$ > +1.0$, \citealt{beers2005}), which provide crucial information about the astrophysical site(s) of the rapid neutron-capture process. The production of $r$-process elements has remained elusive since the seminal work of \citet{b2fh}, but recent observations of the electromagnetic counterpart of the first neutron star merger detected by LIGO can possibly provide the final piece of this cosmic chemical puzzle \citep{abbott2017,shappee2017}.}
\end{enumerate}

\begin{figure}
\includegraphics[width=\columnwidth]{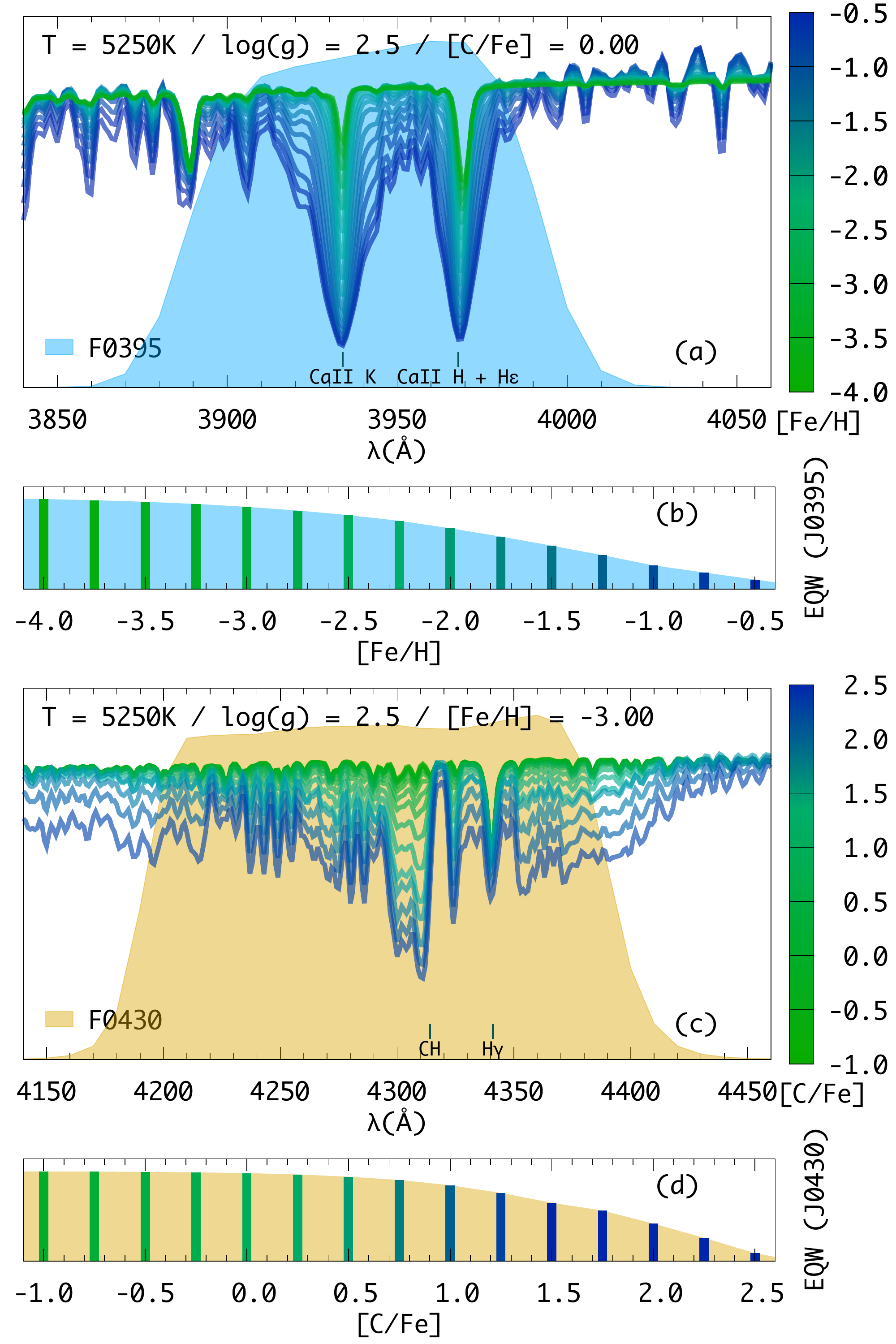}
\caption{(a) J0395 filter sensitivity curve, compared with synthetic spectra of different metallicities. (b) Behaviour of the integrated flux in the J0395 area for the synthetic spectra shown in (a). (c) J0430 filter sensitivity curve, compared with synthetic spectra of different carbon abundances. (d) Behaviour of the integrated flux in the J0430 area for the synthetic spectra shown in (c).}
\label{metal_poor}
\end{figure}

UMP stars are intrinsically rare (Placco et al. 2015a, 2016; Yoon et al. 2016), and can only be properly classified spectroscopically. Most UMP stars found to date are faint, which limits the amount of spectroscopic information that can be obtained within reasonable exposures times, even with 8-10 meter class telescopes. Previous photometric searches for such stars, using SDSS and the SkyMapper Survey (Wolf et al. 2018) data, were limited by the use of broad-band photometry. In this context, the narrow-band filters from S-PLUS show a clear improvement in the success rate of identifying low-metallicity stars (Whitten et al. 2019), in addition to reaching a saturation limit similar to SkyMapper, which is considerably brighter than SDSS. Fig. \ref{metal_poor} shows the effect of changes in metallicity and carbon abundances, compared with the sensitivity curves of J0395 (panel a) and J0430 (panel c), for selected synthetic spectra of stars with fixed temperatures and surface gravities (Whitten et al. 2019). Panels (b) and (d) show the behaviour of the integrated fluxes along the filter areas. In both cases the narrow-band filters used are capable of successfully capturing the changes in [Fe/H], down to $\sim$--3.0, and changes in [C/Fe], starting at $\sim$+0.5.

The 12-band filter system is far more efficient for the identification of these stars. S-PLUS will deliver a catalogue of likely metal-poor stars, suitable for
the immediate study of their spatial distributions, which constrains the assembly
history of the Milky Way. 
In this context, given that the candidates from the MS will be fainter than $r=12$ mag, due to saturation effects, the S-PLUS USS was devised to find 
bright low-metallicity star candidates suitable for high-resolution spectroscopic follow-up 
and studies in the near ultra-violet using the
Hubble Space Telescope. 
Follow-up studies have already been done for a limited number of bright
low-metallicity stars \citep[e.g.,][]{placco2014b,placco2015b}, and additional
work is clearly needed to support theoretical studies
\citep{meynet2010,nomoto2013}.
Of central importance, S-PLUS USS will then provide
targets for subsequent high-resolution spectroscopic studies needed
to separate the UMP, CEMP, and $r$-II sub-classes.

\subsection{The Variability Fields}

The \textit{Variability Fields Survey} (VFS) will perform repeated observations with a cadence set by the frequency of non-photometric nights, covering a number of fields already observed by the MS. At least 30\% of the total time of the survey will be dedicated to the VFS. 

Throughout the duration of S-PLUS, the VFS target fields and observing strategies will be set based on calls for proposals for the use of non-photometric nights. This will result in improved detection of each given class of objects, and for the follow-up of targets of opportunity, including cataclysmic variables, eclipsing binaries, variable low-mass stars, asteroids, SNe, AGNs (specially blazars), GRB afterglows, \textit{Fermi} LAT sources \citep{3fgl}, and gravitational wave events. We may also identify other transient events, such as the fast radio bursts and tidal disruption events  \citep{Burrows2011}.  

The VFS data will be inspected for new asteroids and other moving objects. Some SNe may also be identified, although this is not a primary goal of VFS. In addition, the follow-up of \textit{Fermi} LAT triggers is interesting due to the matching of the typical error box of these triggers (of about 1 degree diameter) to the field of view of the camera. 
About one third of the sources in the latest Fermi/LAT Source Catalogue (3FGL) are of unknown  type \citep{3fgl}, and 
their identification may result in a large number of new blazars. Finally, identification and follow-up of the  electromagnetic counterparts of gravitational wave events \citep{abbott2017} are areas in which VFS may bring important contributions. 

At the time of this writing, there is one long-term program that was awarded VFS observing time in 2018B and continuing through 2019, aiming to detect cataclysmic variable stars. 

\subsection{The Galactic Survey}

\begin{figure*}
\includegraphics[width=17.5cm]{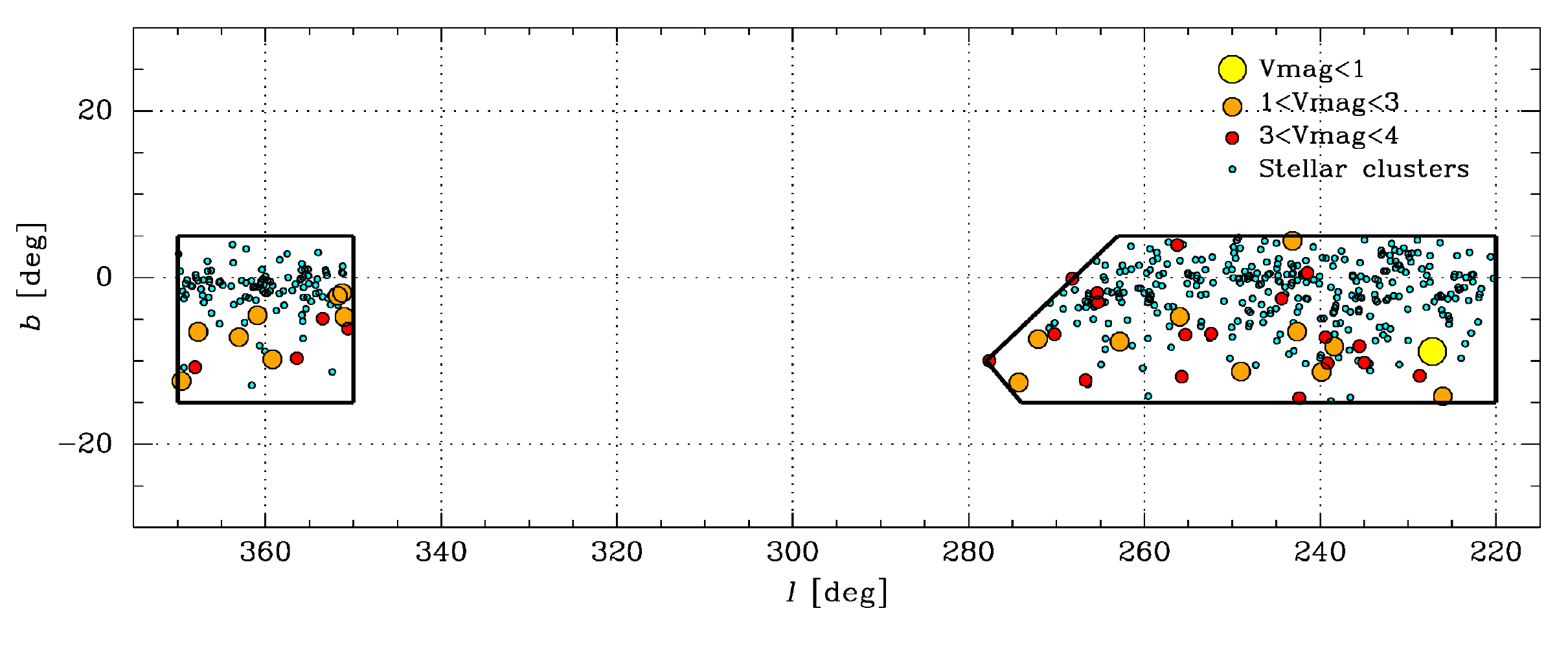}
\caption{Distribution of stellar open clusters within the S-PLUS Galactic Survey area, totalling 444 objects from \citet{2002A&A...389..871D}. Stars brighter than $V=4$~mag are also marked. The Galactic area is divided in ``bulge" (large square) and ``disk" (large pentagon), with a total surveyed area of \SI{1420}{\deg\squared}. The region around each $V<4$\,mag stars is excluded from the observations, reducing the total GS area to \SI{1300}{\deg\squared}.}
\label{fig:OpenClustersFootPrint}
\end{figure*}

The {\it Galactic Survey} (GS) covers an area of about 1420 deg$^2$ in the Milky Way plane in all 12 filters, including regions of the bulge ($-10^o<l<10^o$ and $-15^o<b<+5^o$, for a total of $\sim$400~deg$^2$) and the disk ($220^o<l<278^o$ and $-15^o<b<+5^o$, for a total of $\sim$1,020~deg$^2$, see Fig.~\ref{fig:OpenClustersFootPrint}). The bulge area, as well as the disk area within $-5^o<b<+5^o$, overlap with VPHAS+ in the optical \citep{2014MNRAS.440.2036D} and VVV/VVVX in the near-IR \citep{2010NewA...15..433M}. 

The tiling pattern was designed in equatorial coordinates, thus when seen in Galactic projection the tiles are not aligned. The outline of the GS area has a ``saw-tooth'' profile similar to other Galactic surveys (e.g., VPHAS+). The GS area contains 41 stars brighter than $V=4$\,mag, the brightest of which is Sirius ($\alpha$ CMa, $V=-1.46$\,mag). Because of saturation problems related to these stars, a total of 62 tiles are excluded from the GS area (reducing the effective area to \SI{1,300}{\deg\squared}). 

The first epoch of the GS will have the MS exposure times, followed by two sets of shallower observations (taken with exposure times of duration 1/12th of the MS), only through the $r$, $i$, and J0660 filters. Finally, the GS will obtain, for selected fields, at least 25 more epochs in the $r$, $i$, and J0660 bands at random cadence over several years, at the same depth as the first-epoch observations (same exposure times as MS). The range of exposure times will probe a wide interval of magnitudes, allowing the sampling of different stellar populations, while  observations at different epochs will suit the detection of variable sources, including pulsating RR Lyrae and Cepheids. 

In the regions where the extinction is high, the narrow-band colours will break the degeneracy between reddening and spectral type for a large number of stars.  
Two main studies that are planned with these data are:

\begin{itemize}
\item \textbf{Variable stars}: The cadence  and number of observations in the GS is suitable for the detection of variable sources, including pulsating RR Lyrae and Cepheids, \ac{CVs} and eclipsing binaries, as well as transient sources such as  microlensing events. Since the ecliptic crosses the GS bulge area, asteroids will also  be detected  in the  variability data. Moreover, the narrow-band observations will provide more stringent constraints on  the colours  of stars  undergoing microlensing  events and stars harbouring planet candidates, as  well as classification of variable sources such as RR Lyrae  and \ac{CVs}.  The variability data will be complementary to those obtained by  LSST, given that S-PLUS will discover variable stars as bright as g = 9 mag, well below the saturation limit of LSST. 

\item \textbf{Stellar Open Clusters}: 
 
 A cross-match between the unprecedented high-precision measurements from the Gaia mission \citep{Perryman2001, 2018A&A...616A...1G} and the multi-band photometry of the S-PLUS survey will allow a systematic study of open clusters down to a magnitude deeper than current analyses. Gaia/DR2 \citep{2018A&A...616A...1G} will allow a clean determination of cluster membership by applying tools specially designed for this goal \citep[see][]{Samp1, Samp2}. Taking advantage of the S-PLUS filters will allow us to carry out reliable spectral-type classification for all cluster members, and thus explore the general physical properties of open clusters, such as radius, ages, metallicities, and masses, down to fainter magnitudes.  
  
\end{itemize}

\subsection {Marble Field Survey}
\label{MARBLE}

The \textit{Marble Field Survey} (MFS) is composed of a set of specific fields that will be revisited as often as possible under dark or grey nights and photometric conditions, when the seeing is too poor for MS observations, i.e. $>$\SI{2}{\arcsec}.  Objects selected for the MFS at the time of this writing are the M83 galaxy,
the SMC, the Dorado Group, and the Hydra cluster (see Table 6).
The repeated observations of the MFS will increase the depth of the MS images, and is suitable for the study of nearby galaxies, galaxy groups and clusters, and their surroundings, i.e., galaxy halos, intragroup and intra-cluster light. The MFS may also be used for identification and characterisation of variable sources.

\begin{table}
\centering
\begin{minipage}{90mm}
\scriptsize
\caption{Marble Field Survey.}   
\begin{tabular}{p{1.7cm} l l l l}    
\hline\hline   
Name & RA & DEC &  Obs. & Notes \\
     &    &     &                & (filter, airmass) \\
     
\hline
M\,83 & 13 37 01 & -29 51 57 & Feb-Jun & all, 1.1\\
SMC  & 00  17  47 &  -72 13 10 & Jul-Dec  & all, 1.4-1.6 \\
 (+47\,Tuc) & 00  35  33 & -72  13  10 & &  \\
 & 00  53  20 & -72 13  10 & & \\
 & 01  11 07 & -72  13  10 & & \\
 & 00  18  57 & -73  27  42 & & \\
 & 00  37 54 & -73  27  42 & & \\
 & 00 56  51 & -73  27  42 & & \\
 & 01  15  47 & -73  27 42 & & \\
Dorado group& 04 17 35 & -55 12 10 & Sep-Jan & all, 1.1-1.4\\
              & 04 17 35 & -55 30 00 &  & \\
Hydra cluster & 10 37 54 & -26 41 23 & Jan-May & all, $<$1.3\\
              & 10 37 10 & -28 04 38 & & \\
\hline                  
\end{tabular}
\end{minipage}
\label{table:Marble_Fields}
\end{table}   

\section{Data Flow, from raw data to scheduled data releases}
\label{Section4}

This paper presents the first S-PLUS data release, DR1, on Stripe 82. This section characterises these data. 
Further characterisation of DR1 is reported in Molino et al. (in prep.) and SamPedro et al. (in prep.).

The raw imaging data of S-PLUS are processed daily and data catalogues are generated at the data centre, located in the T80S technical room on Cerro Tololo. 
Full backups of the raw data are made with LTO6 tapes, for any eventual re-processing, if needed. The processed data are transferred through fibre connection to IAG/USP, in S\~ao Paulo.
An overview of the data reduction process is given in \S\ref{Reduction}.

Multi-band photometric catalogues are generated by running the \texttt{SExtractor} software \citep{1996A&AS..117..393B, BertinSExt} on a combined reduced image, which is the weighted-sum of the reddest ($griz$) broad-band images. This process is described in \S\ref{Catalogues}.

Photometric calibration of the images is performed with a novel technique 
using stellar models, as described in detail by Sampedro et al. (in prep.) and in Section~\S\ref{Calibration} below. Zero points are also obtained through standard techniques, by observing typically two spectrophometric standard stars each night, at three different air masses. These are also described in the same section.

The astrometric accuracy of the S-PLUS observations and the variation of the FWHM across the fields are investigated in \S\ref{astrometry} and \S\ref{sec:psf}. The typical photometric depths and photo-z depths of the Main Survey images are derived in  \S\ref{photodepth} and \S\ref{photozdepth}.  Information on the data products that will be offered to the community and scheduled data releases is provided in \S\ref{datareleases}.

\subsection{Overview of the Data Reduction Process} 
\label{Reduction}

The S-PLUS raw data are reduced using an early version (number 0.9.9) of the data processing pipeline \textsc{jype} (developed by CEFCA's Unit for Processing and Data Archiving, UPAD) designed to reduce data for the J-PLUS and the J-PAS surveys \citep{2014SPIE.9152E..0OC}. This, in turn, is based on the photometric pipeline originally developed for the ALHAMBRA survey  \citep[see][]{cristobal09, jpasredbook, Molino14}. 

The basic reduction strategy consists of four steps: {\sc i}) Generating a master bias; {\sc ii}) Creating a master flat; {\sc iii}) Reducing the individual frames; and {\sc iv}) Combining the individual frames into the final astrometrically-aligned images. Bias frames are obtained every night, and twilight flats are obtained, whenever the sky is clear, at dawn and at dusk.  Twilight flats work well for our purposes. Bias and twilight flat fields are stable over a period of about a month, and therefore these are obtained for such a period, encompassing the observations of the object. Master flats are obtained for each filter. Only flat fields with counts between 8000 and 45000 are used. Overscan subtraction, trimming and bias subtraction is applied to each individual flat field. Master flats are then created by obtaining, for each pixel, the median value, with 3-sigma clipping, of all usable flats of a given filter, after scaling each image by its mode. This is performed using the task \textsc{imcombine} of Image Reduction and Analysis Facility IRAF\footnote{\url{iraf.noao.edu}} with options \textsc{median}, \textsc{sigclip}, \textsc{scale=mode}, and \textsc{zero=none}. Finally, the master flats are normalized to have a mean of unity.

The reduction of individual images consists of applying the overscan
subtraction, trimming, bias subtraction, and master flat division. Then, cosmetic corrections (removing satellite
tracks and cosmic rays) and fringing subtraction are performed. Satellite track and cosmic ray subtraction is performed using either \textsc{SatDetect}, in the first case, and \textsc{LACosmic} \citep{2001PASP..113.1420V} or retina filter in the second case. Fringing frames are obtained by combining the final individual frames that suffer from fringing, usually only in the $z$ filter. The fringing patterns are stable over several months, so a single fringing frame is made by combining all images over such a period that do not have any bright objects. The last step is the combination of the individual images, which is done by obtaining the median, with 3 sigma clipping, pixel by pixel, for typically three images of each field and filter. This is performed using the task \textsc{imcombine} of IRAF with options \textsc{median}, \textsc{sigclip}, \textsc{scale=none}, and \textsc{zero=mode}. 

After the final images are produced,  data catalogues are generated, as described in the next subsection. The data also need to be calibrated, as described in Section \S\ref{Calibration}. After calibration is accomplished, the instrumental magnitudes are replaced by calibrated magnitudes in the final catalogues.  

\subsection{Deriving Multi-band Photometric Catalogues}
\label{Catalogues}

\vspace{0.2cm}

Deriving accurate multi-band photometric catalogues suitable for all of the scientific cases described throughout \S\ref{sub_surveys} is challenging.  It requires an optimised photometric tool, capable of identifying and correcting the specific observational effects that make images inhomogeneous, in particular, the smearing of objects due to variations in the point spread function (PSF) across bands. This is an effect that, if not taken into account, can cause the photometric apertures to integrate light from different regions of an object.  

We have written an additional pipeline code, based on the \texttt{SExtractor} software, that analyses the images that come out of the \textsc{jype} pipeline. Photometric catalogues are constructed both in single image mode for individual filters, and in double image mode  when performing multi-band aperture-matched photometry. The use of a deep detection image is desirable in order to enhance the detectability of faint (or low surface-brightness) sources, and to better define the photometric apertures when computing multi-band photometry. We automatically generate a detection image for each pointing as a weighted combination of the reddest ($griz$) broad-band images. This combination makes use of the automatically generated weight-maps \citep[produced by the \texttt{SWARP} software,][]{BertinSwarp} to account for potential inhomogeneities in the exposure times (i.e., effective depths) across each field, and FWHM differences between bands.

The next steps are the following:

\begin{itemize}

\item The PSF-corrected photometry is obtained. Initially, the software 
defines several photometric apertures based on the detection image. Then, for each filter, it estimates how much flux has been missed within that aperture, as a result of the different sizes of the PSF for a single-filter image compared to the detection-image. A corresponding correction is then applied, yielding PSF-corrected magnitudes. The full procedure is explained in detail in \cite{Molino14}, in their section 3.2.

\item The aperture-matched photometry based on the detection images is obtained.  This produces accurate colour determinations for SED-fitting analysis and photometric redshift determinations. 

\item An empirical estimation of the photometric noise in the images is performed, taking into account artificial correlations among pixels (i.e., smoothing) induced during the image-reduction process. The degree of  correlation, along with other pieces of information directly related to the sources (such as aperture sizes or integrated fluxes), are used to recompute the noise estimate provided by \texttt{SExtractor}. A correction of the photometric uncertainties estimated by SExtractor is then applied.

\item Derivations of photometric upper-limits are obtained for sources detected on the detection-images and not detected on individual bands. Although there exist several approaches to estimate these photometric upper-limits, in S-PLUS we choose to simply convert the integrated enclosed signal within the photometric aperture into a  magnitude. 
These upper limits are of considerable importance for the computation of photometric redshifts.    

\item Weight-maps and rms-maps are created to minimize the detectability of spurious sources on the detection images.

\end{itemize}

More details on each of these procedures are given in Section 3 of \cite{Molino14}.

\subsection{Data Calibration and Final Catalogues}
\label{Calibration}

A new photometric calibration technique is employed here, specifically developed for wide-field multi-band photometric surveys such as S-PLUS. A similar version of such technique is planned to be used for calibrating J-PAS (Gruel et al. 2012). The calibration takes advantage of other surveys such as SDSS \citep{2007AJ....134..973I, 2008ApJ...674.1217P}, Pan-STARRS \citep{2012ApJ...756..158S}, DES \citep{2018ApJS..235...33D, 2018AJ....155...41B} or KiDS \citep{kids}, which derived photometric calibrations for millions of stars, typically in 4-5 bands, in areas overlapping with S-PLUS. In addition, instead of using complex (and sometimes inaccurate) transformation equations between filter systems, our calibration strategy relies on libraries of stellar models as if they were spectrophotometric standard stars.
 
As a first step, we select typically one thousand stars in an S-PLUS tile that have known magnitudes from one of the surveys cited above. For each star, a template fitting algorithm is used to find the most likely model that fits the literature photometric information. The stellar templates used are from the Next Generation Spectral Library (NGSL, \citet{NGSL_lib}) and the Pickles library \citep{Pickles}. The best model is then used to compute a preliminary model stellar magnitude, in each of the 12 bands. The initial zero-points of the S-PLUS filter system are determined through convolution of the filters with the best model, and comparison between the resulting magnitudes and the instrumental magnitudes obtained for each star in the S-PLUS image (obtained with \texttt{SExtractor} as described in \S\ref{Catalogues}).
  
Once the initial zero-point values have been derived for the S-PLUS filter system, the process is iterated by fitting again the stellar models, but now to the newly derived 12-band photometry for each object. After a few iterations, in which the model and instrumental magnitudes are compared, the methodology converges to a final solution for the zero-points in every filter, with typically a few percent uncertainties. Note that the success of the technique comes from the fact that we are deriving a single number (the zero-point) from the fit to close to one thousand stellar spectra. All zero-points are then absolute-calibrated to match Gaia's photometry \citep{2017A&A...599A..50A}.
 
\begin{figure*}
\centering
\includegraphics[width=\linewidth]{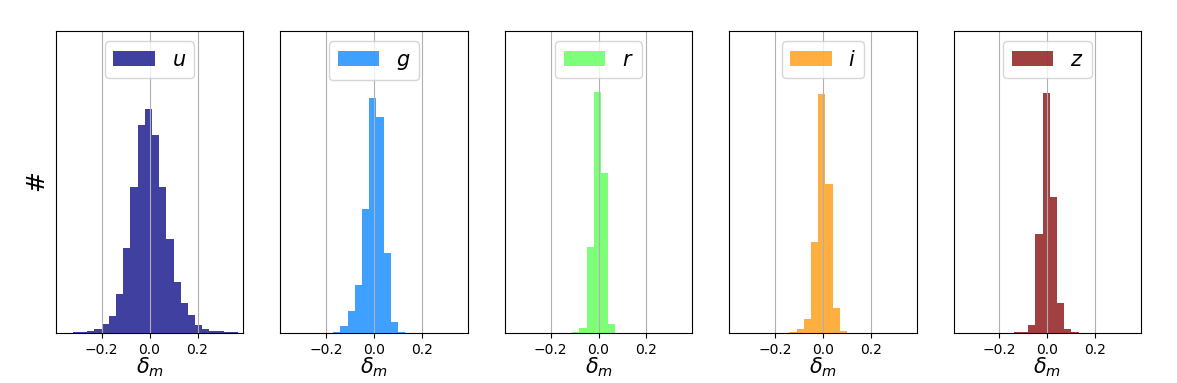}
\caption[]{Comparison of S-PLUS and SDSS photometry (mag${_{S-PLUS}}$ - mag_${_{SDSS}}$) for objects in DR1 with magnitudes below 20. The rms of the distributions for the five filters, ugriz, are 0.06, 0.05, 0.03, 0.05 and 0.03 mag, respectively, proving the good consistency between the two data sets. The mean differences between the SDSS and the S-PLUS filter systems give an offset in the x-axes of 0.06, -0.02, -0.03, -0.01 and 0.03 mag for the five bands respectively. This is due to small differences in the filter systems described in Table A1.}
\label{Difference_phot_SPLUS_SDSS}
\end{figure*}

As the calibration strategy is based on the use of stellar libraries, it does not require large campaigns with multiple observations of standard fields. Comparisons were made to the photometry obtained by S-PLUS and SDSS, for the five bands in common (ugriz), with good agreement, as can be seen in Fig. \ref{Difference_phot_SPLUS_SDSS}. The rms of the distributions for the five filters, ugriz, are 0.06, 0.05, 0.03, 0.05 and 0.03 mag, respectively. Nevertheless, two spectrophotometric stars are observed in three different airmasses every clear night to check the zero-points. 
Extinction coefficients for the site were obtained using the standard fields observed over 200 nights, for 10 bands ($u$ and $z$ excluded). Average values for the mean atmospheric extinction coefficient obtained for each band are listed in Table \ref{tab:meanzps}. Details on the comparisons between the two types of calibrations (standard calibration and using stellar libraries) will be presented in Sampedro et al. (in prep.).

\begin{table}
\centering
\caption{Mean atmospheric extinction coefficients obtained from the analysis of standard stars.}
\begin{tabular}{ccc}
\hline
\hline
FILTER & extinction coefficient \\
\hline
J0378 &  $0.414 \pm 0.025$ \\
J0395 &  $0.356 \pm 0.011$ \\
J0410 &  $0.306 \pm 0.008$ \\
J0430 &  $0.268 \pm 0.014$ \\
gSDSS &  $0.188 \pm 0.015$ \\
J0515 &  $0.141 \pm 0.013$ \\
rSDSS &  $0.099 \pm 0.005$ \\
J0660 &  $0.078 \pm 0.008$ \\
iSDSS &  $0.067 \pm 0.009$ \\
J0861 &  $0.035 \pm 0.011$ \\
\hline
\hline
\end{tabular}
\label{tab:meanzps}
\end{table}
 
Once the zero-points are obtained, the final catalogues with calibrated magnitudes are derived. The final data catalogues include the basic astrometric (coordinates), photometric (e.g., fluxes and magnitudes), and morphological (e.g., ellipticity, position angles, major and minor axis ratio, and stellarity) information for all sources detected in the images. Releases of specific Value-Added-Catalogues (VACs) will be made available 
as part of S-PLUS collaboration science projects. VACs may include photometric redshift measurements, the results of SED fitting analysis, star/galaxy classification, or other higher-order information derived from the S-PLUS images.
 
\subsection{Astrometric Accuracy}
\label{astrometry}

In this section we describe the level of accuracy reached by our image reduction pipeline. We note that the coordinates  computed by the reduction pipeline for DR1, following the ICRS (International Celestial Reference System) and taking the 2MASS catalog \citep{2003yCat} as a reference, are not meant to be used in astrometric investigations per se, but they are useful for locating the great majority of the objects.  We have compared the astrometric position of the S-PLUS DR1 sources with those from the SDSS DR12 data on Stripe 82 \citep{2015ApJS..219...12A} for $\sim$1M stars in common. To avoid saturated or poorly detected sources, we considered a magnitude interval of 14$<$r$<$21. 

\begin{figure}
\includegraphics[width=8.4cm]{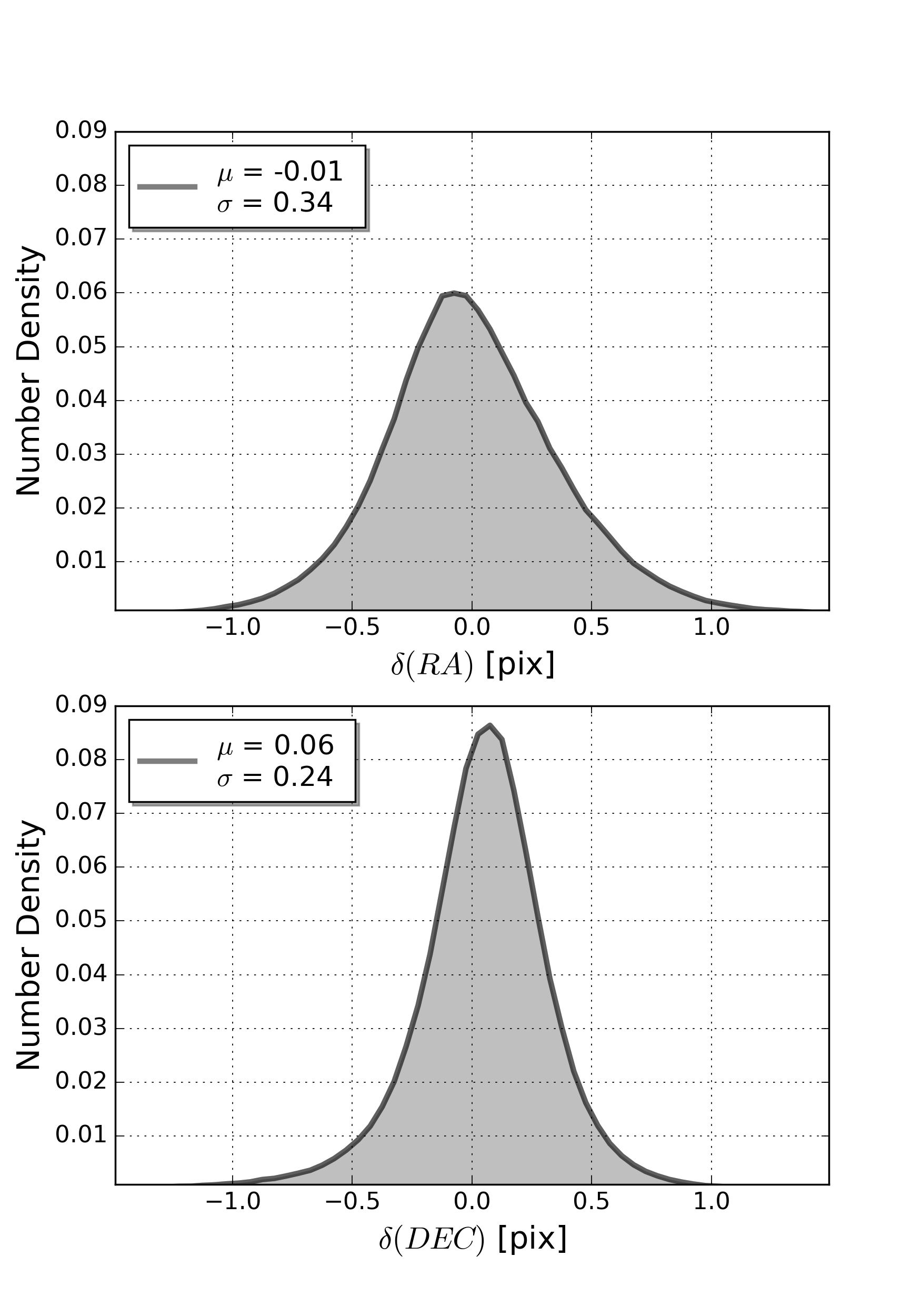} 
\caption[Astrometric Precision]{Astrometric accuracy of S-PLUS sources. The two panels show comparisons between SDSS/DR12 and S-PLUS for a common sample of $\sim$1M stars.  A very small mean difference for both RA and DEC is observed, with a scatter of 0.34 and 0.24 of a pixel  respectively (i.e., 0.19 and 0.13 arcsec respectively).}
\label{accastrom}
\end{figure}

As illustrated in Fig.  \ref{accastrom}, where the differences between coordinates are represented separately for RA and DEC, we find an average astrometric accuracy of the order of -0.01 pix and 0.06 pix respectively, with an average rms scatter of 0.34 and 0.24 pixels (0.19 and 0.13 arcsec) respectively.  Thus, we assert that our images have been properly corrected, and the coordinates given in our catalogues are robust. 

\subsection{Determination of the Stellar FWHM across the Field}
\label{sec:psf}

\begin{figure}
\includegraphics[width=8.7cm]{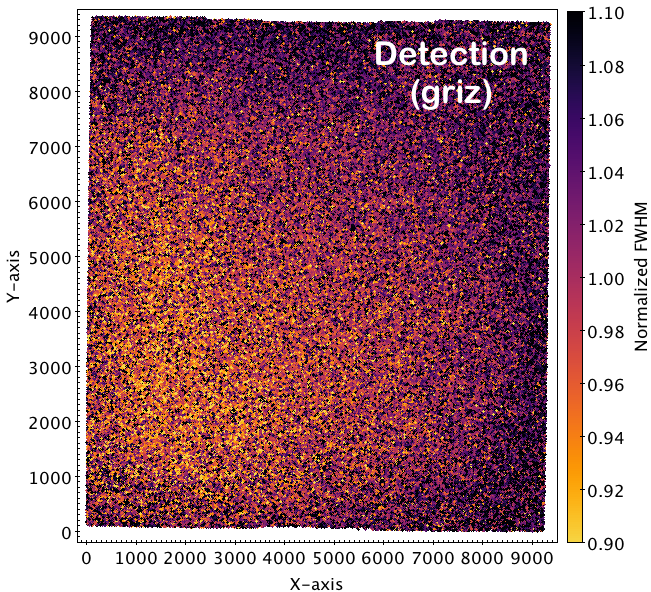}
\caption[PSF-stability1]{FWHM average variation across images. The figure shows the result obtained using a stack of all images of DR1 in four bands, $griz$ (see \S\ref{sec:psf} for details). The normalisation of the FWHM values for a given field was performed using the average FWHM value derived from a sample of $\sim$ 500 bright non-saturated point sources across that field. Note that the variation of the FWHM from the centre of the field to the outskirts is on average 10\%. }
\label{psfvar1}
\end{figure}

The S-PLUS DR1 Stripe 82 data were used for checking the average variation of the FWHM of stellar objects across the field. Detection images (i.e., a combination of $griz$ bands) were used for this exercise. The differences in the FWHM measurements for a given star, in the four bands, $g$, $r$, $i$, $z$, was never more than half a pixel, therefore a simple combination of the four images was appropriate (using only the $r$-band yields very similar results). The FWHM values of typically 500 bright non-saturated stars across each field were measured (using \texttt{SExtractor}) and they were normalised to the average FWHM of the bright, isolated, and non-saturated stars in each image. The result is shown in Fig.~\ref{psfvar1}. Note that the average FWHM corresponds to unity, on the scale shown in the right-hand side of the figure, and the variation from the centre to the border is 10\%.

\subsection{Photometric Depths}
\label{photodepth}

\begin{figure}
\includegraphics[width=9.0cm]{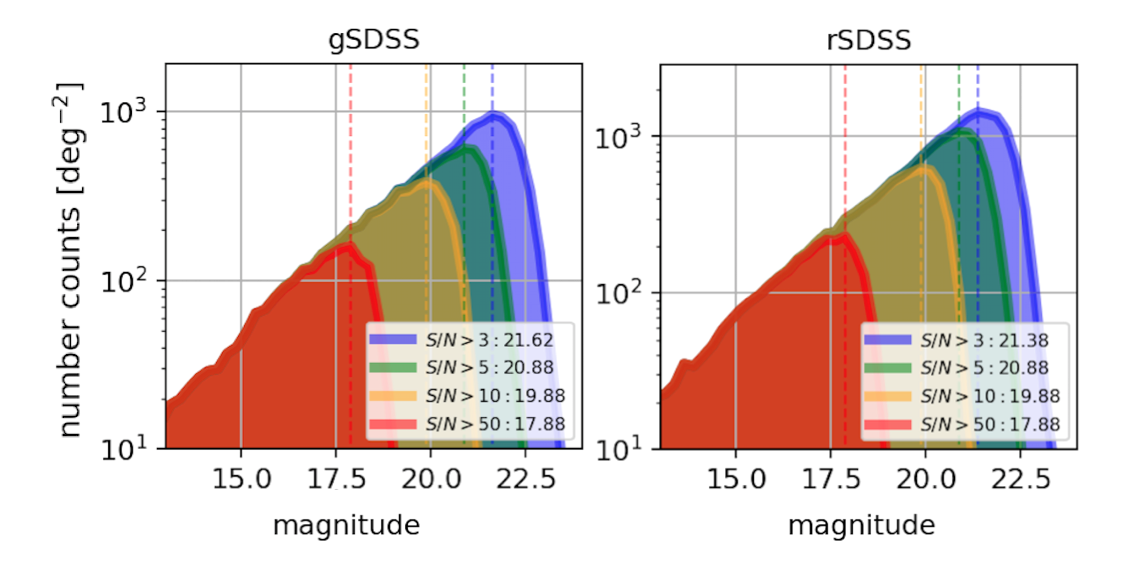}
\caption[Photometric-depth]{Photometric depths of the S-PLUS $g$ and $r$-band images at different signal-to-noise levels, as derived from the S-PLUS DR1 Stripe 82. The dashed lines show the magnitudes for which the samples are considered complete (where the derivatives are zero). As an example, sources with a signal-to-noise S/N$\sim$$3$, are expected to be complete down to a magnitude $g$$<$21.62 and $r$$<$21.38 respectively. For sources with other signal-to-noise ratios, the magnitudes of completeness are shown in the legend at the top left.}
\label{rdepth}
\end{figure}

The S-PLUS DR1 Stripe 82 data were used to estimate the average photometric depth of the S-PLUS images. As summarised in Table 8,
the photometric depths were calculated using five different definitions for sources detected in a given filter with a signal-to-noise ratio $\geq$3. Here, \textit{$m_{peak}$} corresponds to the \textit{Petrosian} magnitude at which detections start declining rapidly (i.e., the derivative is zero); \textit{$m_{50\%}$}, \textit{$m_{80\%}$}, and \textit{$m_{95\%}$} correspond to the magnitudes at which it includes 50\%, 80\%, and 95\% of the total detected sources and \textit{$m_{3arcs}$} corresponds to the integrated magnitude within circular apertures of 3 arc-second diameter. As can be seen in Fig. \ref{rdepth}, where the estimated photometric depths of $r$ and $g$-band images at different signal-to-noise ratios are shown, the S-PLUS images are expected to be complete down to a magnitude $g$ $<$ 21.62 and $r$ $<$ 21.38 for all sources (point and extended) with a S/N $>$3.   

\vspace{0.2cm}

\begin{table}
\centering
\begin{minipage}{80mm}
\scriptsize
\caption{Photometric depth of images. The table shows the estimated photometric depth of the S-PLUS images using five different definitions, and selecting only sources detected with a minimum signal-to-noise of S/N$\geq$3 on individual filters: \textit{$m_{peak}$} corresponds to the \textit{Petrosian} (i.e., total) magnitude at which detections start declining rapidly (i.e., the derivative is zero); \textit{$m_{50\%}$}, \textit{$m_{80\%}$}, and \textit{$m_{95\%}$} correspond to the magnitudes at which it includes 50\%, 80\%, and 95\% of the total detected sources; \textit{$m_{3arcs}$} corresponds to the magnitude integrated within circular apertures of 3 arc-second diameter.}
\begin{tabular}{cccccc}     
\hline\hline   
FILTER &  $m_{peak}$  & $m_{50\%}$  & $m_{80\%}$  & $m_{95\%}$  & $m_{3arcs}$  \\
\hline
$u$    & 21.07  &  22.10  &  23.11  & 24.12  & 22.56 \\
J0378  & 20.64  &  21.83  &  22.86  & 23.88  & 22.27 \\ 
J0395  & 20.11  &  21.47  &  22.52  & 23.65  & 21.87 \\ 
J0410  & 20.30  &  21.53  &  22.57  & 23.67  & 21.94 \\ 
J0430  & 20.38  &  21.54  &  22.59  & 23.67  & 21.94 \\ 
$g$    & 21.79  &  21.88  &  22.85  & 23.88  & 22.16 \\ 
J0515  & 20.61  &  21.33  &  22.42  & 23.53  & 21.64 \\ 
$r$    & 21.63  &  21.12  &  22.07  & 22.88  & 21.32 \\ 
J0660  & 21.36  &  21.02  &  21.98  & 22.93  & 21.12 \\ 
$i$    & 21.22  &  20.54  &  21.41  & 22.07  & 20.72 \\ 
J0861  & 20.32  &  20.23  &  21.29  & 22.36  & 20.39 \\ 
$z$    & 20.64  &  20.27  &  21.05  & 21.77  & 20.37 \\ 
\hline                  
\end{tabular}
\end{minipage}
\label{DepthTable}
\end{table}   

\subsection{Photometric Redshift Depth}
\label{photozdepth}

S-PLUS DR1 Stripe 82 data were used to characterise the performance of the photo-z estimates for different magnitude and redshift ranges.  This dataset is ideal  because of the availability of a high number of spectroscopic redshifts for galaxies and quasars. 
For the present exercise, we compiled a sample of galaxies in  S-PLUS DR1 of Stripe 82 with magnitudes $r<$21 and redshifts $z<1.0$. Our photometric redshift determinations were tested against a sample of galaxies with spectroscopic information taken from the literature.  The following datasets were used for constructing our reference sample: SDSS  \citep{2018ApJS..235...42A}, 2SLAQ \citep{2005MNRAS.360..839R}, 2dF \citep{2001MNRAS.328.1039C}, 6dF \citep{2004MNRAS.355..747J}, DEEP2 \citep{2013ApJS..208....5N}, VVDS \citep{2005AAS...207.6334L}, and PRIMUS \citep{2011ApJ...741....8C}, as well as surveys such as the SDSS- III Baryon Oscillation Spectroscopic Survey, BOSS  \citep{2013AJ....145...10D}, SDSS-IV/eBOSS \citep{2017ApJS..233...25A} and WiggleZ \citep{2010MNRAS.401.1429D}. The distribution of blue and red galaxies in this combined sample peak at magnitudes $r=19$ and $r=19.6$ respectively.  The procedure adopted for computing photometric redshift depths of S-PLUS is similar to that explained in \cite{Molino14} for the ALHAMBRA survey.
 
\begin{figure*}
\includegraphics[width=17.5cm]{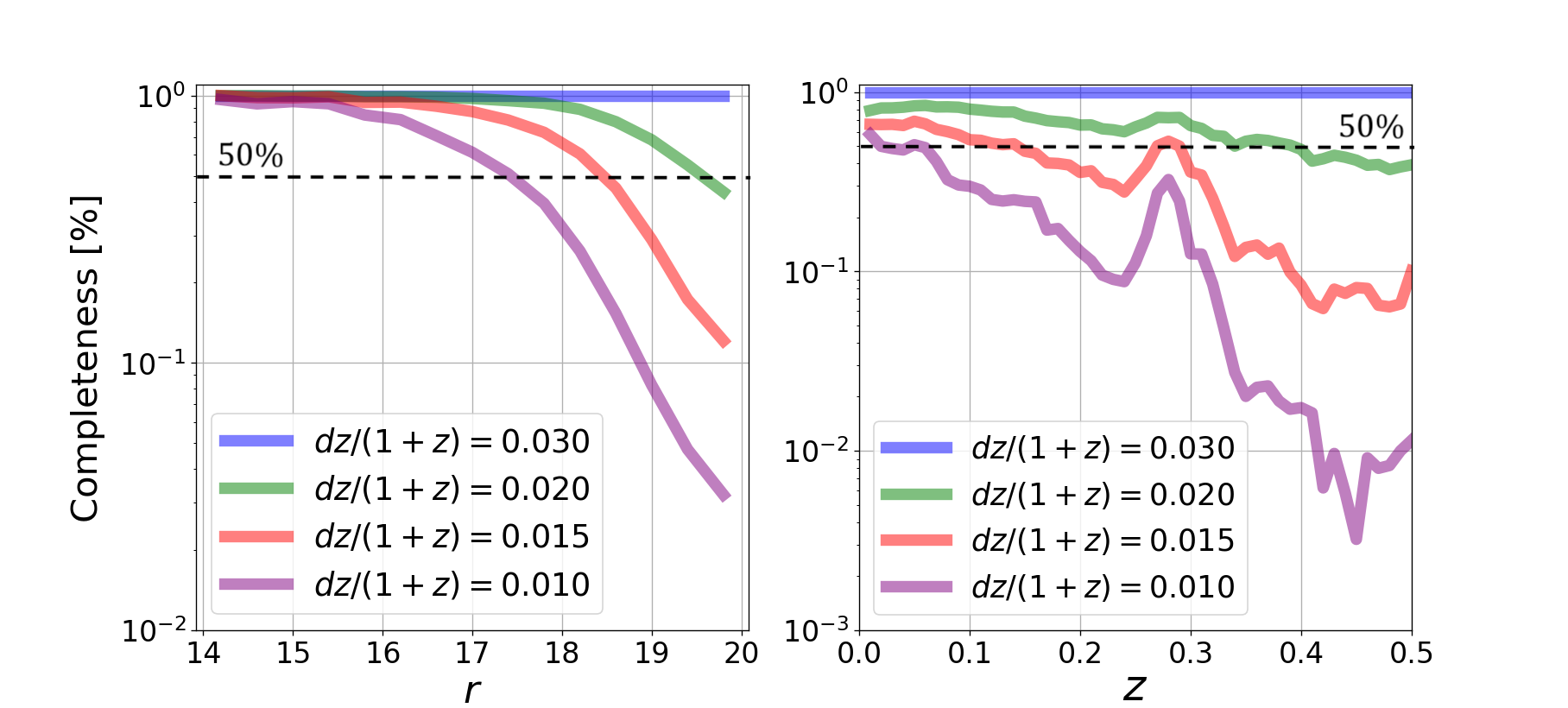}
\caption{Photometric redshift completeness. The panels represent the expected fraction of galaxies per magnitude $r$ (left panel) or redshift $z$ (right panel) bin with a maximum photometric redshift error. Solid lines correspond to the results obtained from the sample of galaxies in S-PLUS DR1 with spectroscopic redshift information (see text for details). A photo-z precision of $\delta_{z}/(1+z)=0.02$ or better is expected for 50\% of galaxies with a magnitude $r\sim19.7$ or a redshift $z<0.40$. Likewise, a precision of $\delta_{z}/(1+z)=0.01$ or better is expected for 10\% of galaxies with a magnitude $r<18.8$ or a redshift $z<0.32$. The magnitude range of the galaxies on the right panel is the same as for the left panel.}
\label{photozaccuracy}
\end{figure*}

Fig. \ref{photozaccuracy} shows the expected fraction of galaxies per magnitude $r$ (left panel) or redshift $z$ bin (right panel) with a maximum photometric redshift error. These values are estimated using the \texttt{Odds} parameter from the \texttt{BPZ} code, which allows retrieving samples with a maximum photo-z error. As drawn from the figures, we expect a photo-z precision of $\delta_{z}/(1+z)=0.02$  or better for 50\% of galaxies with a magnitude $r\sim19.7$, or a redshift $z<0.40$. Likewise, a precision of $\delta_{z}/(1+z)=0.01$ or better is expected for 10\% of galaxies with a magnitude $r<18.8$, or a redshift $z<0.32$. About 100\% completeness is expected for galaxies with a $\delta_{z}/(1+z)=0.03$ or better, down to a magnitude $r<20$, or a redshift $z<0.5$. Similarly, but now in global terms, the same analysis shows that after its completion (i.e., after observing  
\SI{8000}{\deg\squared}), the S-PLUS survey will provide photometric redshift estimates for $\sim$ 2 million galaxies with a precision of $\delta_{z}/(1+z)\leq 0.01$, for $\sim$16 million galaxies with $\delta_{z}/(1+z) = 0.02$, and for $\sim$32 million galaxies with $\delta_{z}/(1+z) = 0.025$, down to a magnitude $r$ = 21.
 
In terms of photo-z precision, the benefit of extending classical 5-filter broad-band surveys  \citep[such as SDSS;][]{2000AJ....120.1579Y} can be assessed directly using the S-PLUS data. Molino et al. (2019, submitted) uses the S-PLUS DR1 Stripe 82 data and compare photo-zs obtained with 5 bands with those obtained with 12 bands, using the SED-fitting code \texttt{BPZ} \citep{2000ApJ...536..571B}.   As shown in their paper, the 12-band system leads to an improvement in photo-z over the 5-band system of a factor of 4, for galaxies with magnitudes $r<15$, a factor of 2.5 for magnitudes $15<r<17$, and/or a factor of 1.7 for magnitudes $17<r<19$. As a function of redshift, the 12-band system leads to a factor of 2 improvement for galaxies with $z<0.1$ and of 1.5 for $0.1<z<0.4$. SDSS-like surveys cannot surpass a certain precision in the photo-z estimates irrespective of the signal-to-noise of the images. This limitation is imposed by the poorer wavelength resolution provided by the broad-band filters, causing a degeneracy in the colour-redshift space (this actually applies to every survey independent of the filter set).  

Note that besides the overall improvement in the photo-z estimates at all redshifts, the S-PLUS filter system provides a special redshift window at which the photo-z estimates undergo a significant improvement (see the right panel of Fig.~\ref{photozaccuracy}). 
At the redshift interval $z\sim0.26-0.32$, the [OIII] line ($\lambda=5007$ \si{\angstrom}) enters the J0660 filter and the H${\alpha}$ line ($\lambda=6600$ \si{\angstrom}) enters J0861, improving the photo-z precision.

\subsection{\label{datareleases}Data Releases}

The public data releases (DR) will be primarily  
hosted by NOAO data lab\footnote{\url{datalab.noao.edu/}} and the Brazilian Virtual Observatory (BRAVO) server at Laborat\'orio de Astroinform\'atica Data Centre\footnote{\url{lai.iag.usp.br}}.  
The DRs include multi-band images, single-mode and dual-mode photometric catalogues, and value added catalogues produced by the consortium. 
Raw images or intermediate-step reduction products (e.g., weight maps or segmentation images) may be made available upon request. The data will also be accessible through the S-PLUS data portal\footnote{\url{www.splus.iag.usp.br}} and through queries using the International Virtual Observatory Alliance (IVOA\footnote{\url{www.ivoa.net}}) interoperability standards Cone Search, SIA, TAP, and SSAP \citep{conesearch,sia,tap,ssap}. 

The baseline survey plan foresees five years to complete the survey. We intend to have six yearly data releases (DR), starting $\sim$~26 months after the start of operations (in August 2017). These are then scheduled for approximately the month of October in six consecutive years starting in 2019. The release of S-PLUS data on fields
coinciding with the SDSS Stripe 82 region, DR1, accompanies this paper\footnote{\url{datalab.noao.edu/}}.

\section{Results from the Science Verification Data}

This section first summarizes some key features of the S-PLUS DR1 in Table 9.
DR1 is composed of 170 contiguous pointings, adding up to $\sim$\SI{336}{\deg\squared} of the Stripe 82 area, observed in 12 filters.
The main characteristics of DR1 including a description of the reduction and calibration methods used, an analysis of the spatial distribution of the PSF along the images, as well as the photometric and photo-z depths attained in the DR1 dataset, have been described in \S\ref{Section4}. 

As described in \S\ref{MAIN}, files from the MS are generally dithered by 10 arc sec along the RA direction. However, in the case of the fields of Stripe 82 in the DR1, only those fields observed in 2018 were dithered, the ones in 2016 were not. DR1 fields had no overlapping area. These were decisions made early in the project that were then changed (to include dithering and overlapping areas for the remaining of the survey). 
 
S-PLUS DR1 contains about 3M sources, 2$/$3 are point-like and 1$/$3 extended sources.
From the sources classified as galaxies, nearly 35\% are classified as early/quiescent galaxies and 65\% as late/star-forming galaxies. In absolute numbers, S-PLUS DR1 includes $\sim$350k early and $\sim$650k late-type galaxies. S-PLUS DR1 catalogues released with this paper have a magnitude cut of $r$ = 21 mag.

\label{sec:results}
\begin{table}
\begin{center}
\label{DR1_}
\caption{Summary of DR1 characteristics - Stripe 82 area.} 
\end{center}
\begin{tabular}{|c|c|}
\hline
\multicolumn{2}{|c|}{S-PLUS DR1 data}                                                                                                                                                   \\ \hline
Area Covered               & $\sim$ \SI{336}{\deg\squared}                                                                                                                                       \\ \hline
Bands                      & \begin{tabular}[c]{@{}c@{}}Broad: u, g, r, i, z\\ --------------------------------------------\\ Narrow: J0378, J0395, J0410, \\  J0430, J0515, J0660, J0861\end{tabular} \\ \hline
Number of Sources          & $\sim$ 3 M                                                                                                                                              \\ \hline
Number of Tiles            & 170                                                                                                                                                         \\ \hline
Astrometric Accuracy       & 0.25 pix (0.14 '')                                                                                                                                         \\ \hline
Depth (S$/$N$>$3, r-band) & 21.38 mag                                                                                                                                                     \\ \hline
Zero point accuracy    & 1\% - 2\%                                                                                                                                                  \\ \hline
Seeing                     & $\sim$ 1.5 ''                                                                                                                                           \\ \hline
\end{tabular}
\end{table}

This section also presents preliminary results obtained using DR1, which will be detailed in future papers. MS data on Stripe 82 are used to exemplify the usefulness of S-PLUS at improving star/galaxy classification (\S\ref{stargalaxy}), at the determination of galaxy cluster/group membership (\S\ref{clustermembership}), in deriving environment density indicators (\S\ref{galaxyenvironment}), in quasar searches (\S\ref{quasarsearch}), in morphological studies (\S\ref{morphologicalparameters}), and for IFU-like science projects (\S\ref{IFUscience}). 

\subsection{Star/galaxy Separation Applied to the Stripe 82 Field
}
\label{stargalaxy}

The separation between stars and galaxies is a crucial step for every photometric survey. In the last decades there have been many solutions proposed to deal with this classification issue. Here, a new approach, specifically for multi-colour surveys, is presented.  
\vspace{0.2cm}

\begin{figure*}
\includegraphics[scale = 0.3]{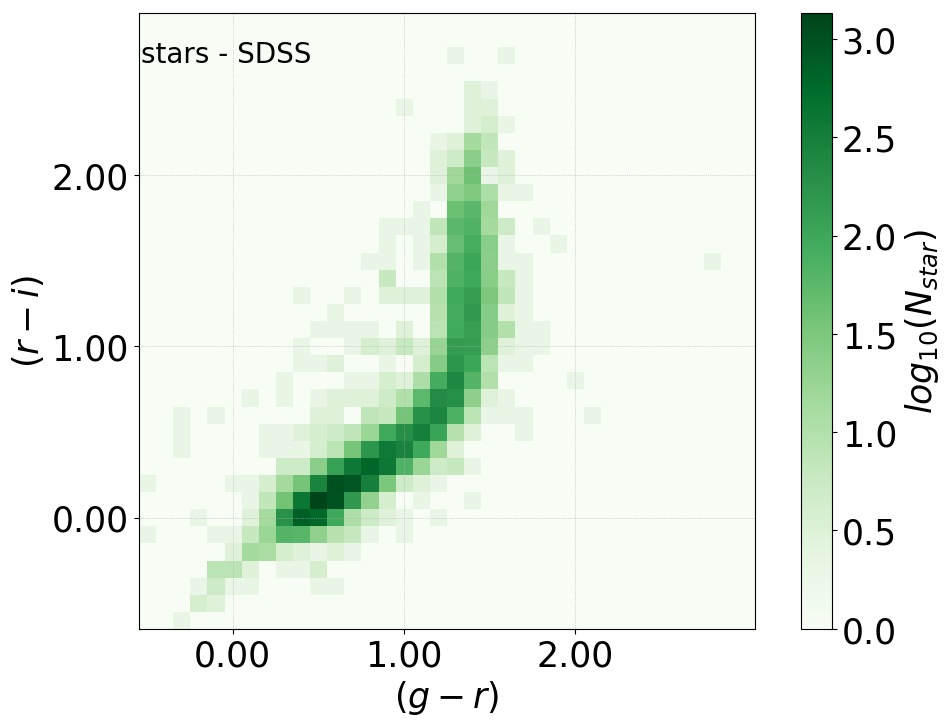}
\includegraphics[scale = 0.3]{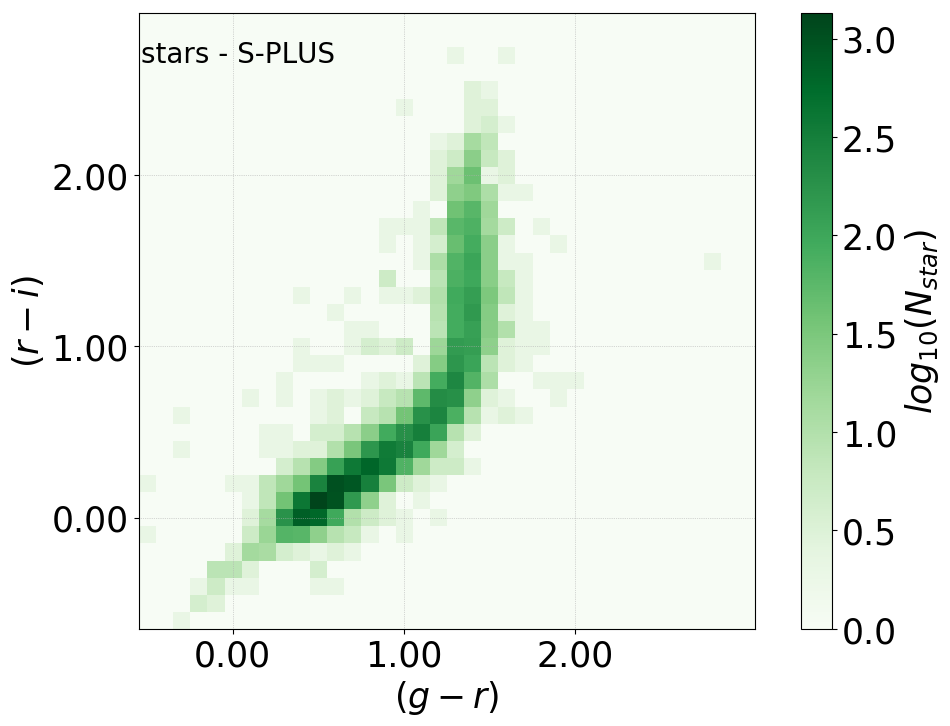}

\includegraphics[scale = 0.3]{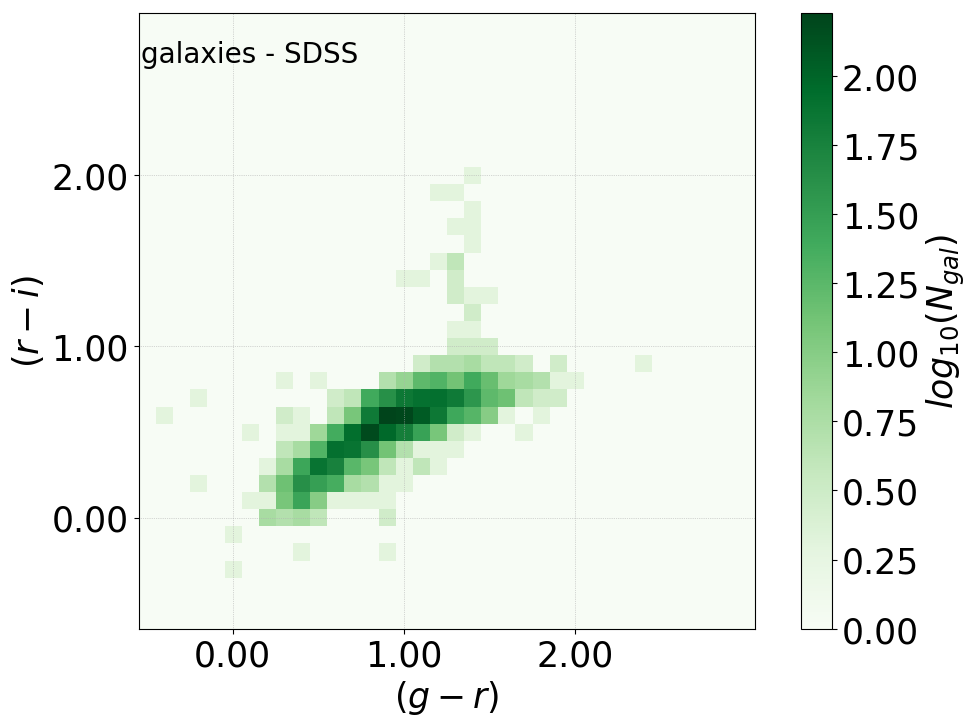}
\includegraphics[scale = 0.3]{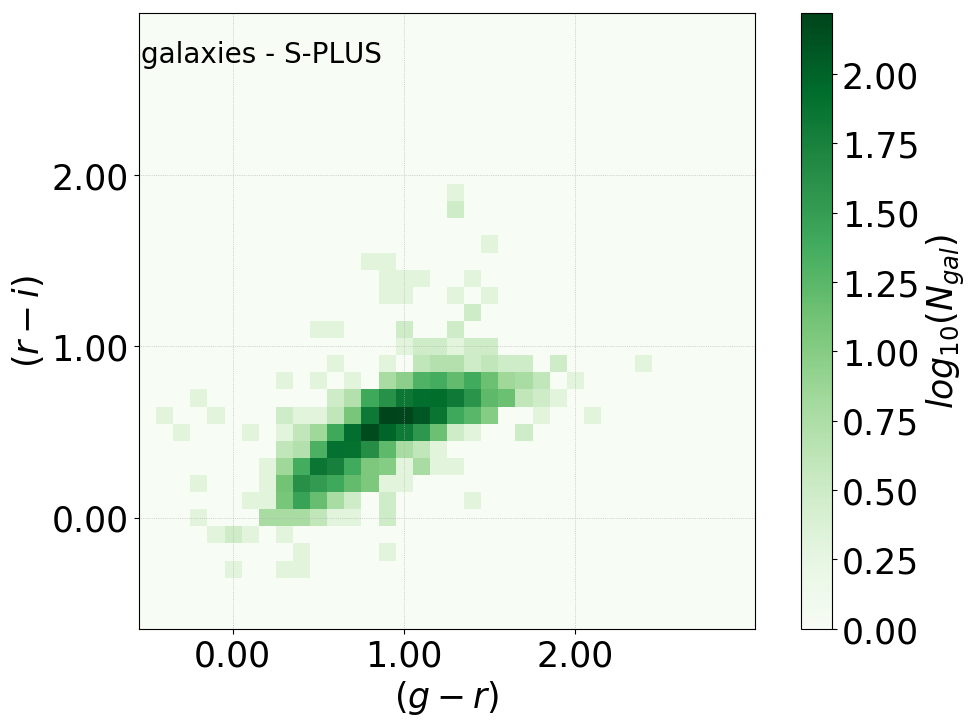}

\caption{The colour-colour diagram $(g-r)$ versus $(r-i)$ using S-PLUS magnitudes down to $r$=19. The overall performance of the code indicates that 97.9\% of the objects are correctly classified down to this magnitude limit (see \S\ref{stargalaxy}). Upper panels: The stellar \textit{locus} of objects classified as stars based on SDSS (left) and S-PLUS (right) data. Lower panel: The galaxy \textit{locus} in the same diagram using the SDSS and S-PLUS classifications.}
\label{fig.sg_separation}
\end{figure*}

In the experiment described here, the Random Forest technique \citep{Breiman2001} was used, combining the S-PLUS photometric  and morphological information (ellipticity, concentration, and FWHM) to classify objects into stars or galaxies.
A matched sample between S-PLUS DR1 Stripe 82 data and the photometric SDSS/Stripe 82 catalogue (Jiang et al. 2014 - the latter is complete to $\sim$ $r=24.6$) provided reliable classifications for $\sim$200k objects. This matched sample was used to properly train the Random Forest algorithm. The inclusion of the morphological parameters in the input set of features was crucial for improving the performance of the S-PLUS star/galaxy classifier. The overall performance of the code indicated that 95.7\% of the objects are correctly classified down to $r$=21. This assumes SDSS photometric classification as a truth table down to $r$ = 21, which is reasonable, given that the magnitude limit of the SDSS sample is
$\sim$ $r=24.6$, more than three magnitudes deeper than S-PLUS. Fig.~\ref{fig.sg_separation} shows colour-colour diagrams, $(g-r)$ versus $(r-i)$, for galaxies and stars, as classified by SDSS and S-PLUS, to $r$=19.  
Down to this magnitude limit, S-PLUS gets the correct classification for the sources in 97.9\% of the cases. We conclude that the star/galaxy classification employed in S-PLUS is able to classify objects correctly, and recovers the stellar and galactic \textit{loci} in the colour-colour diagrams expected based on the SDSS classification. 

\subsection{Determining Group and Cluster Membership with Accurate Photometric Redshifts}
\label{clustermembership}
 
Photometric redshifts (photo-z) have become an essential tool in astronomy, since they represent a quick and  inexpensive (in terms of observing time) way of retrieving redshift estimates for a large number of galaxies. Photo-zs are among the primary deliverables of S-PLUS, given that the 12-band photometric system allows higher photo-z precision compared to those derived with, for example, SDSS data \citep[see fig. 8 of][]{2018arXiv180403640M}. The high quality of the S-PLUS photometric redshifts will enable detailed studies of large-scale structure and galaxy evolution over the entire $\sim$\SI{8000}{\deg\squared}  area of the Main Survey.

In contrast with areas of the Northern Hemisphere covered by SDSS, where all galaxies with $r$ < 17.7 have an observed spectrum, the areas covered by S-PLUS typically do not have an abundance of easy-to-access fully-reduced SDSS-like spectra, even if partial areas have been surveyed spectroscopically with other Southern Hemisphere telescopes. The new generation of redshift surveys utilising multi-filter  photometric systems can play an important role in mitigating this North/South imbalance. Classical 3-4\% photometric redshift errors, computed from standard 4-5 broad-band filter systems, can be dramatically diminished to the 1-2\% level by simply including narrow-band filters (see Fig.~\ref{photozaccuracy}). Improved photometric redshift estimates also lead to narrower (i.e., less uncertain) probability distribution function (PDFs) needed for robust statistical analysis. In particular, accurate PDFs can play a key role in the identification of groups and galaxy clusters from photometric data (e.g. \citealt{2018arXiv180403640M}). In this regard, the S-PLUS will be used to construct the most accurate photo-z nearby-galaxy catalog yet produced over a large area of the Southern sky. 

\begin{figure*}
\includegraphics[width=18.0cm]{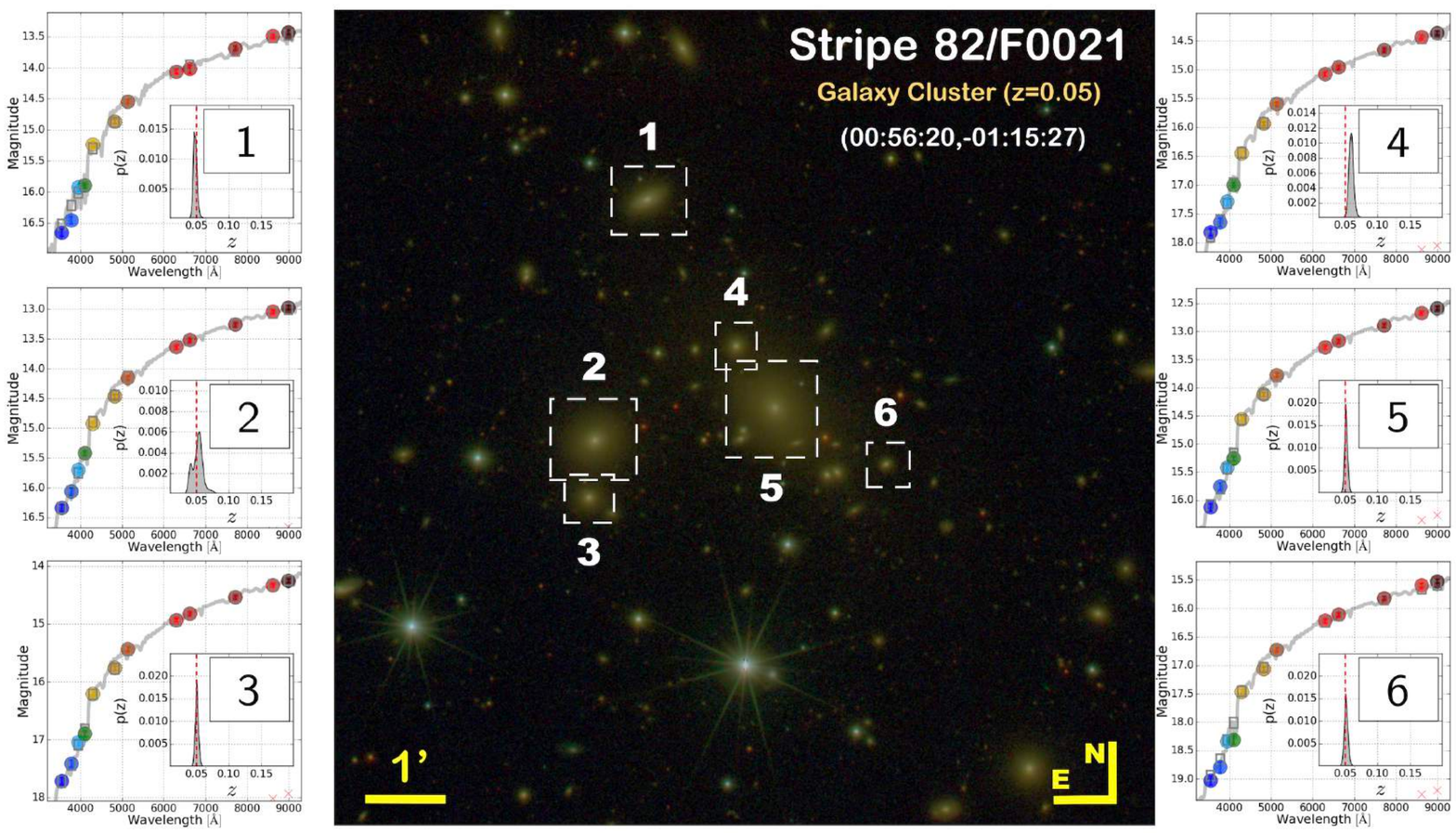}
\caption[SED-fitting]{Membership analysis in a cluster of galaxies. The figure shows an example of the typical SED-fitting and photo-z analyses for a sample of six early-type galaxies within a galaxy cluster at redshift $z$=0.05. Coloured points in the six stamps around the main figure correspond to the observed S-PLUS magnitudes, and the solid-grey line to the most likely galaxy template. The inner-panels in the stamps show the corresponding redshift probability distribution function (PDF) computed by  \texttt{BPZ} (grey curve) compared to the cluster redshift (dashed-red vertical line). Note the sharp PDFs, indicating precise photo-z determinations, for the brightest cluster members, allowing the identification of galaxies at the same redshift.}
\label{photoz1}
\end{figure*}

In order to illustrate this statement, we selected a galaxy cluster within the Stripe 82 at a redshift $z=0.05$, and picked six random early-type galaxies with different apparent magnitudes. Based on the S-PLUS photometry, we computed their photometric redshifts, estimating the most likely redshift and spectral-type as well as their PDFs. In Fig. \ref{photoz1} we present a zoom-in of the cluster core in the central region surrounded by six stamps, where in each stamp different coloured points correspond to the observed S-PLUS magnitudes and the solid grey lines correspond to the most likely galaxy templates from \texttt{BPZ}. The inner panels in these stamps show the corresponding redshift PDFs computed by the \texttt{BPZ} code for each galaxy compared to the cluster redshift (dashed-red vertical line), proving the capability of the S-PLUS data in detecting galaxies with similar redshifts; i.e., groups and galaxy clusters.  

The high precision of S-PLUS photometric redshifts (see Fig. \ref{photozaccuracy}) of $\sim$ 1.5\% (2\%) for a significant number of galaxies with r < 18.5 (19.7) will allow membership analysis in existing clusters and groups of galaxies down to intermediate magnitudes and redshifts, complementing already existing spectroscopic samples in the Southern Hemisphere. At least parts of some important nearby superclusters are in the MS footprint, such as Hydra-Centaurus, Pisces-Cetus, Phoenix, and Horologium. On the other hand, searches in the MS for new structures using techniques that can take advantage of the photometric redshift probability distributions will deliver new catalogues of clusters and groups of galaxies. This will then produce a 3D map of the local Universe over a volume of more than $1 h^{-3}$ Gpc$^3$. 
While the photometric redshift accuracy will not be sufficient for estimating dynamical masses of such systems, masses can be derived for systems in common with those selected in X-ray surveys or in surveys done using the Sunyaev-Zel'dovich effect,  or by establishing relations between mass and optical richness or luminosity.

\subsection{Galaxy Environment and Large-Scale Structure}
\label{galaxyenvironment}

The environment of a galaxy plays an important role in the current galaxy evolution scenario \citep{2004ApJ...615L.101B, 2009ARAA..47..159B, 2010ApJ...721..193P}. Several processes are proposed as being responsible for galaxy quenching, such as ram-pressure stripping \citep{GunnGott1972}, galaxy mergers \citep{MihosHernquist1994}, and galaxy harassment \citep{Mooreetal1996}. However, all these physical mechanisms act on different scales and in different environments, and their exact relative contributions to the general galaxy evolution scenario have been difficult to establish. The MS will provide accurate photometric redshifts and sufficiently large sky areas, suitable for characterizing galaxy environments in the local Universe. This will allow us to probe the connections between structure formation and galaxy formation, and thus constrain popular approaches such as the halo model \citep{CooraySheth}, halo occupation distribution \citep{Berlind2002,Zehavi2005}, and halo abundance matching \citep{Trujillo2011}. 

In the following, we show how MS data will be able to constrain the local density contrast of galaxies using a promising tool,
the k-NN (k-Nearest Neighbour) technique, adapted to take into account the photo-z uncertainties in our calculations \citep[as also done for the KiDS survey,][]{kids}. As shown by \cite{CostaDuarteetal2017}, the relation between galaxy luminosities, density contrasts and galaxy colours is recovered when applying this technique to photometric redshift data (see their fig. 5). 

\begin{figure} 
\includegraphics[width=\columnwidth]{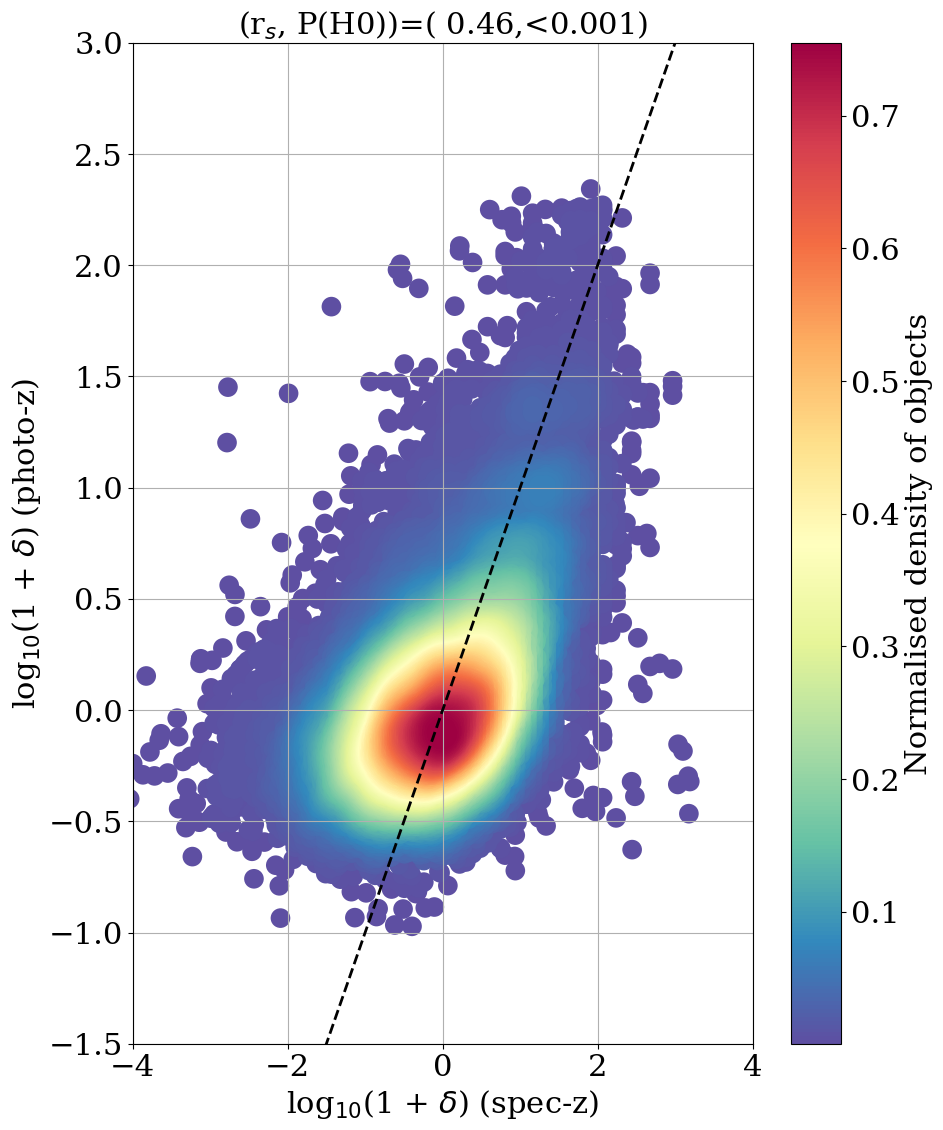}
\caption{The local density contrasts of galaxies in spectroscopic and photometric redshift spaces. These have been calculated using an S-PLUS mock volume-limited sample ($z<0.25$ and $M_r<-19.5 + 5 \log h$). The Spearman correlation coefficient (shown at the top) shows a significant correlation between both density contrasts, indicating that we can reliably recover the galaxy environment with S-PLUS photo-zs.}
\label{Figure_Marcus}
\end{figure}

In order to show the potential of the S-PLUS filter system in retrieving parameters indicative of galaxy environment, we constructed a mock catalogue of S-PLUS, which mimicked the predicted S-PLUS photometric depth and redshift uncertainties. A mock volume-limited sample was generated, including galaxies up to $z=0.25$ and $M_r < - 19.5 + 5 \log h$. 
The local density of galaxies was then calculated using the approach of \cite{CostaDuarteetal2017}. The comparison between the density contrasts ($1 + \delta = \frac{\rho}{\bar{\rho}}$ and k = 5) in spectroscopic and photometric redshift spaces is shown in Fig. \ref{Figure_Marcus}, presenting a Spearman correlation coefficient of $r_s=0.46$, and a probability of the null hyphotesis $p(H_0)<10^{-3}$. This exercise confirms that this technique is able to recover the galaxy environment, as measured by local densities, in photometric surveys, in particular using S-PLUS.

\subsection{Searches for Quasars}
\label{quasarsearch}

Searches using the SDSS and WISE have provided the largest and most reliable quasar catalogues yet compiled. \cite{2012AJ....144...49W} first presented the criterion z $-$ W1 > 0.66(g$_{SDSS}-z_{SDSS}$) + 2.01 to separate stars and quasars using SDSS and WISE bands, recovering 98.6\% of 3089 quasars with redshifts less than 4. For quasars with redshifts lower than 3.2, they suggested a criterion that only depended on WISE bands: W1 $-$ W2 > 0.57.  \cite{2018A&A...613A..51P} made use of W1 and W2 WISE bands along with SDSS bands to identify quasar candidates, resulting in the most recent SDSS catalog containing 526\,356 quasars.  Several other authors have also used WISE bands to separate quasars from stars, in particular to increase the numbers of quasars at the bright end of the luminosity function (e.g., \citealt{2017ApJ...851...13S}; \citealt{2017AJ....154..269Y}; \citealt{2018A&A...618A.144G}).  Earlier works based on the COMBO-17 survey had already discussed direct detection of emission lines in quasars through narrow-band optical SED (e.g., \citealt{2001A&A...377..442W}; \citealt{2003A&A...408..499W}; \citealt{2004A&A...421..913W}). The work described briefly in this section (and further presented in subsequent papers) will complement these previous works.

\begin{figure} 
\includegraphics[width=8.9cm]{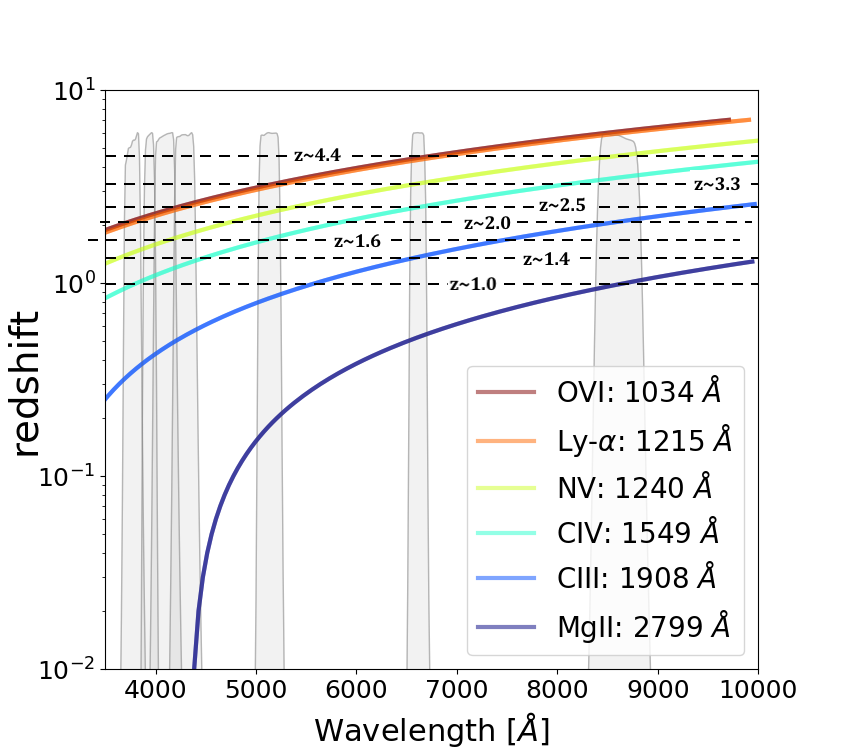} 
\caption[The best windows for quasars]{The redshift of typical quasar emission lines as a function of the rest-frame wavelength. This figure summarizes the potential usefulness of S-PLUS magnitudes in the detection and redshift determination of quasars. The transmission curves of the 12-filter system used by S-PLUS are indicated. Several quasar lines are detected by different filters, depending on the quasar redshifts. For an example, the Ly-$\alpha$ line becomes detectable in the first blue filter of S-PLUS at $z\sim 2.0$ and at the J0515 filter at  $z\sim 3.3$. The S-PLUS narrow-band filters will allow the simultaneous detection of two emission-lines from QSOs in at least 7 redshift windows up to a redshift z<5.  This sensitivity may increase the QSO detectability and redshift computation. }
\label{quasar}
\end{figure}

At specific redshifts, the broad emission lines of quasars
can be resolved spectrally by several narrow-band filters of S-PLUS. The best lines to be used for z $>$ 1 quasar detection can be clearly seen in Fig. \ref{quasar}. The CIII line ($\lambda=$ \SI{1908}{\angstrom})  passes through the H$\alpha$ filter at $z\sim 1.4$. The CIV and Ly-$\alpha$ lines become detectable in the bluest narrow-band filter,  at $z \sim 1.0$ and $z\sim 2.0$, respectively.  
Therefore, S-PLUS will be able to identify quasars, not only through standard UV dropout selection and colour cuts \citep[e.g.,][]{XDQSO}, but also through the direct detection of emission lines \citep[e.g.][]{2012MNRAS.423.3251A,ELDAR}.
The combination of both broad- and narrow-band filters alone, as well as in combination with WISE bands, will thus allow us to construct a large sample of quasars in the Southern Hemisphere, many of which will have accurate photometric redshifts and spectral information from the S-PLUS data alone. Moreover, once the photometric redshifts are known, one can make an estimate of the equivalent widths of the lines that lie within the narrow-band filters.

In this section, we present some preliminary results on quasar searches using S-PLUS DR1 combined with WISE photometry and using S-PLUS DR1 alone. 

\begin{figure}
\includegraphics[width=0.48\textwidth]{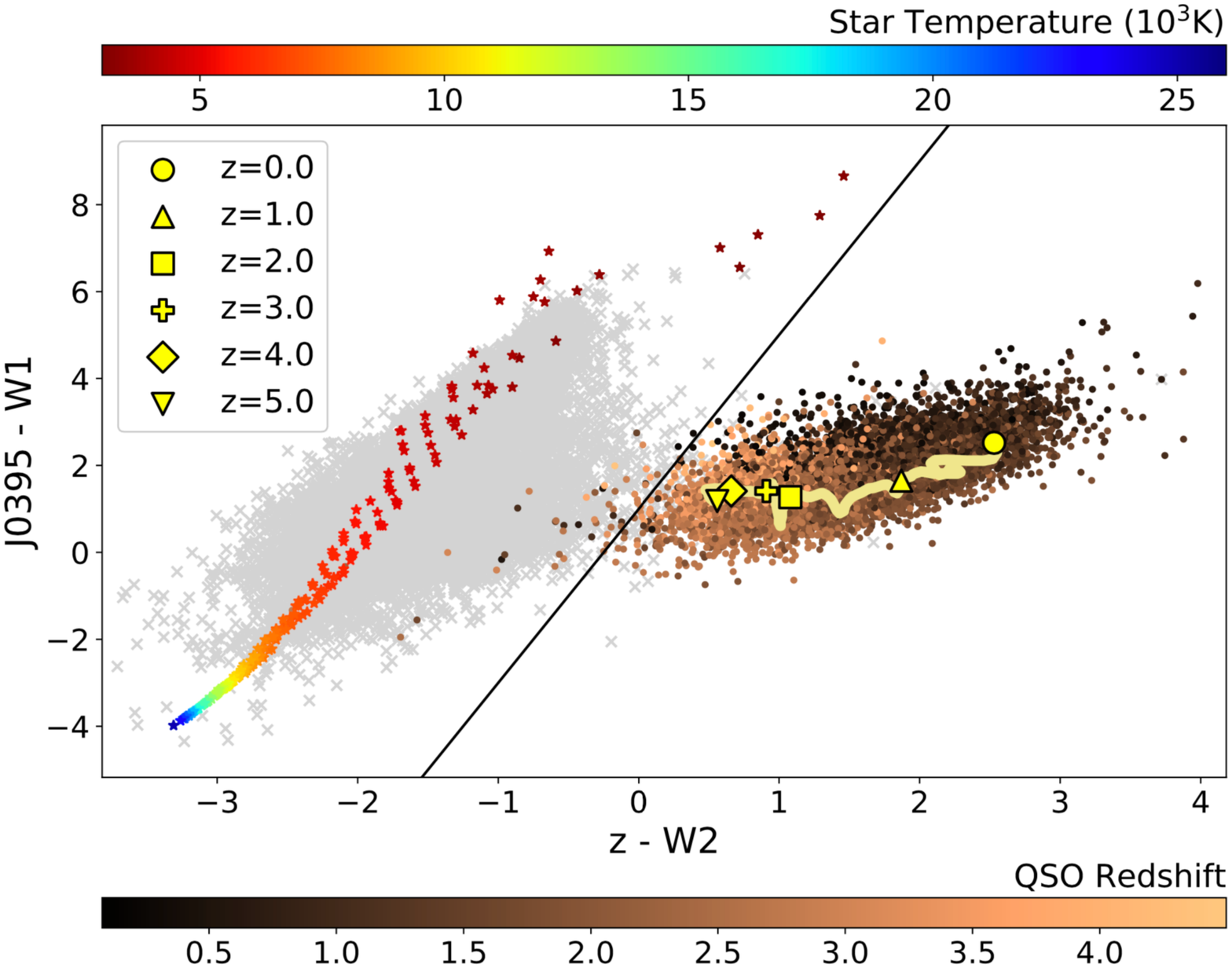}
\caption{S-PLUS/WISE colour-colour diagram of stars and quasars in the Stripe 82 area from Nakazono et al. (in prep.). Stars with reliable classification in S-PLUS are shown as gray Xs, whereas confirmed quasars (with known spectroscopy) are indicated by circles whose colours depend on the redshift, indicated on the bottom colour bar. Stellar models for different effective temperatures are indicated by star symbols, coloured according to the scale on the top. The yellow curve represents the evolution of the simulated colours for a QSO template with the redshift. On top of the curve, the yellow symbols mark the integer values of redshift. The dashed-black line represents an empirical relation (Eq. \ref{eq:colorcolor}) to separate stars from quasars. A total of 99.5\% of the known quasars in Stripe 82 occupies the expected region in the figure (to the right of the dashed black line), confirming the efficiency of the method. }
\label{fig:data_color}
\end{figure}

The star-galaxy classification used to select point-like sources for this work is described in \S\ref{stargalaxy} above, with the caveat that only objects with \texttt{SExtractor} photometric flags set to zero were selected (indicating isolated objects with good photometry).  Considering only objects with at least S/N $>$ 3 in S-PLUS DR1, to the limiting magnitude of the DR1 catalogue of $r$ = 21, Fig. \ref{fig:data_color} shows a colour-colour diagram combining S-PLUS and WISE data in the Stripe 82 field, indicating a good separation between stars and quasars. The empirical relation:
\begin{equation}
J0395 - W1 $<$ 4 \times (z - W2) + 1
\label{eq:colorcolor}
\end{equation}
was established in order to define a locus with the highest chance to find quasars. We found 1027 quasar candidates without spectroscopic classification in SDSS, with $r$<19, in an area of 336 deg$^2$, considering the relation above. This doubles the number of known quasars in the area, given that there are 914 known quasars identified spectroscopically in SDSS, in the area S-PLUS Stripe 82, with r < 19. Only three of the known quasars (0.33\%) fall outside the quasar locus defined by the above empirical relation. Note the limiting magnitude here is due to significant galaxy contamination at fainter magnitudes. Down to r < 19, only  0.41\% of the 9756 galaxies in Stripe 82, i.e. 40, galaxies are classified as point sources and fall over the quasar locus. Likewise, only 0.04\% of the stars, out of 25873, i.e.10, are found in the quasar locus. These numbers are for spectroscopically confirmed quasars from the catalogue of \cite{2018A&A...613A..51P} and the stars and galaxies are from SDSS DR15. Thus, the S-PLUS quasar catalogue on the Stripe 82 area matches the \cite{2018A&A...613A..51P} sample at 99.7\% completeness with 99.5\% purity. Such a high recovery rate of 99.7\% for the previously known quasars combined with a very low contamination rate illustrates the enormous potential for a 12-band survey like S-PLUS to find additional quasars by exploring the full colour space to our avail. Therefore, we expect to find improved results with a more robust and less strict analysis based on machine learning and these will be fully discussed in Nakazono et al. (in prep.). In the analysis described above we found about one previously unidentified quasar candidate for each known quasar selected using the SDSS dataset in the Stripe 82 field. Follow-up spectroscopy of these new candidates (r < 19 mag) will allow us to further test and calibrate our selection methods. Future tests with S-PLUS data in the GAMA fields, for which the spectroscopic samples are complete down to the limiting magnitude of our study, will furthermore allow us to assess quasar selection completeness and the contamination rate.  

In Queiroz et al. (in prep.) we perform the object classification without any near-infrared data, by employing a machine learning technique which provides the probabilities that any given point-like source detected with S-PLUS is a quasar, a star, or a galaxy. The method implements a Random Forest algorithm using a training set for each type of object, containing synthetic fluxes with the same level of noise as in S-PLUS, constructed from SDSS-DR12 spectra. Our training sets contain main sequence stars and white dwarfs, quasars in the redshift range $0.0<z<4.0$, as well as red and blue galaxies. 

\begin{figure}
\centering
\includegraphics[scale=0.8]{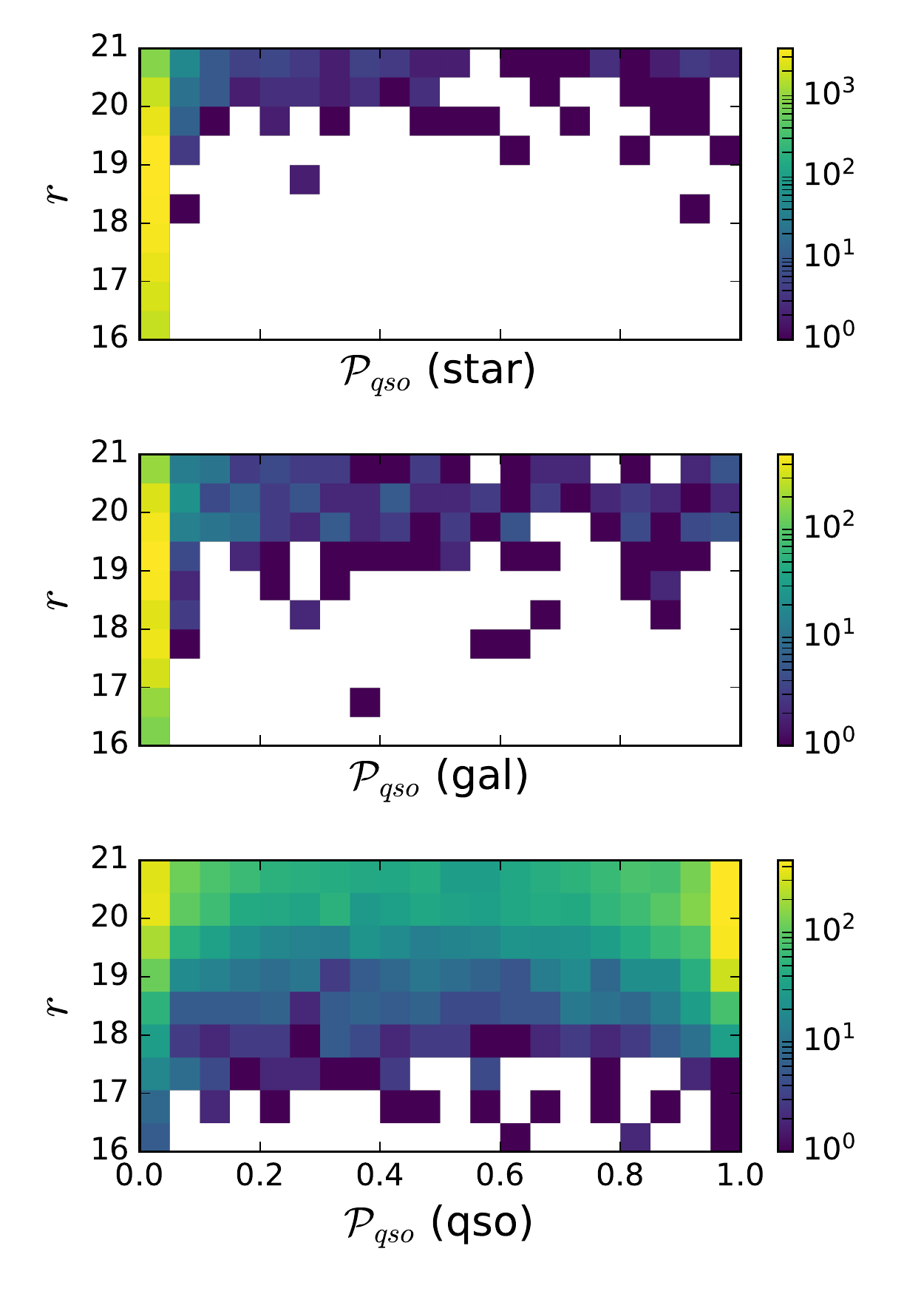}
\caption{Number of S-PLUS quasars detected in the Stripe 82 that are classified as stars (upper panel), galaxies (middle panel) and quasars (bottom).}
\label{fig:qso_completeness_purity}
\end{figure}

The performance of this technique was tested in a sample of about 40k point-like sources detected in S-PLUS DR1 of Stripe 82, again using the star/galaxy classification described in \S\ref{stargalaxy} above. By applying probability cuts on a magnitude-limited sample ($r<20.5$), we reach a completeness of $76\%$ and a purity of $\sim 94\%$ for the quasars (of which 1.6\% are stars, and 4.2\% are galaxies). The purity of the sample with different magnitude cuts is shown in Fig. \ref{fig:qso_completeness_purity} -- note that for $r<18$ no stars are classified as quasars. Extrapolating these results to the MS we forecast a total number of approximately 703,000 quasars in S-PLUS brighter than $r=20.5$, with $\sim 94\%$ purity.

We also estimate the photometric redshift probability distribution for the point-like sources classified as quasars. The method devised finds the best fit to the S-PLUS data using a linear combination of quasar eigenspectra derived from a principal component analysis \citep{2004AJ....128.2603Y, 2012MNRAS.423.3251A}  plus a reddening law. We also test the performance of our photo-z code on the spectroscopic sample of point-like sources, by computing the photo-z precision $\sigma_{z}=\mathsf{med}|z_{photo}-z_{spec}|/(1+z_{spec})$. For the magnitude-limited samples $r<20.5$, $r<19.0$ and $r<18.0$ we obtain, respectively, $\sigma_{z}=(0.0655, 0.0545, 0.0220)$.  

\subsection{Determination of Morphological Parameters} 
\label{morphologicalparameters}

S-PLUS will provide a large sample of nearby galaxies for morphological studies and \ac{SED} analyses.  We will use MS data to perform parametric \citep[e.g.,][]{Vika2015} and non-parametric  \citep{Ferrari2015}  multi-band morphological analyses.  Although measurements of the S\'ersic index and effective radius
as a function of wavelength can be used to perform an automated and robust classification of galaxies
\citep{Vika2015},  \citet{Ferrari2015} show that the non-parametric code
\textsc{Morfometryka} is ideal to distinguish between elliptical and spiral classes with a mis-match between
classes smaller than 10\%. Combining parametric and non-parametric approaches we will be able to 
classify all well-resolved MS galaxies.
SED-fitting codes will be used to extract stellar population parameters, attenuations, and stellar masses from the observed SEDs \citep[e.g.,][]{2015A&A...582A..14D, 2015PASP..127...16M, deamorim}. When combined with the accurate photometric redshifts, the morphology measurements, spectral energy distribution modelling, and estimates of the environments will produce a very rich data set for studying galaxy evolution.

\begin{figure*}
\includegraphics[scale = 0.6]{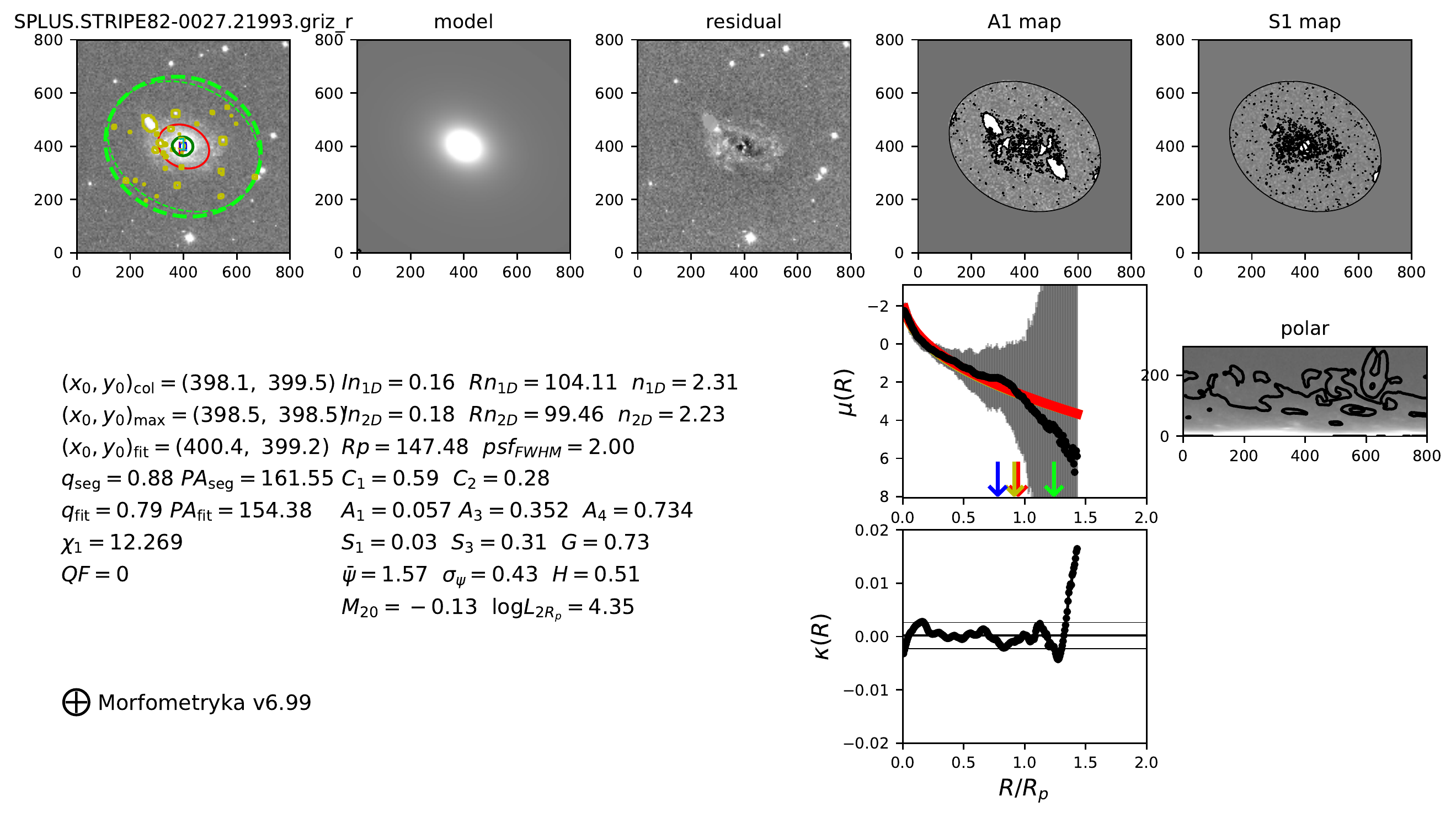}\\
\includegraphics[scale = 0.6]{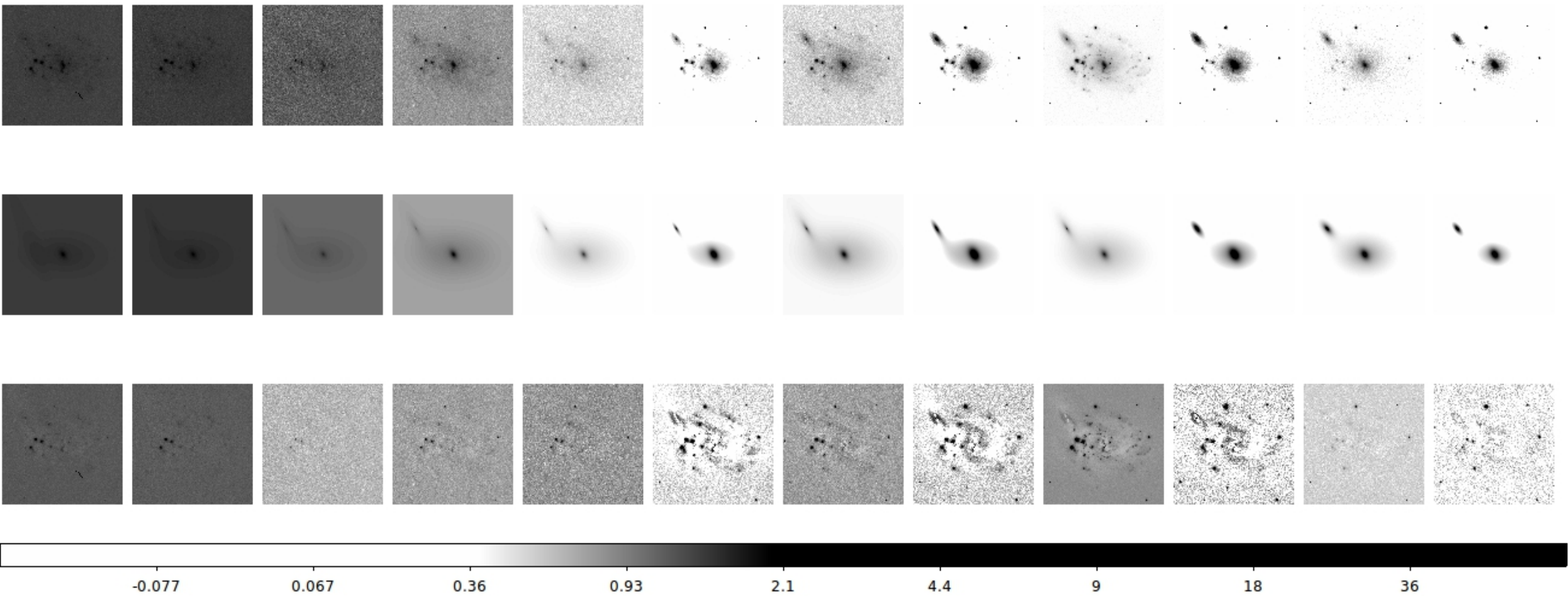}
\caption[\textsc{Morfometryka} (upper panel) and MegaMorph-GALFITM (bottom panel) results]{\textbf{Upper panel:} \textsc{Morfometryka} (top, from left to right): Original image; 2D S\'{e}rsic model image; the residual between image and model; asymmetry map used to compute A1; and smoothness map used to compute S1. \textit{Bottom:} Various measurements (see text for details); Brightness profile (arbitrary units) and model fits; polar map used to compute image gradients and $\sigma_{\psi}$; brightness profile curvature. \textbf{Lower panel:} {MegaMorph-GALFITM: } First row - galaxy images in the 12 S-PLUS bands ($u$, J0378, J0395, J0410, J0430, $g$,  J0515,  $r$, J0660, $i$, J0861 and $z$ respectively); second row - galaxy models as fitted with GALFITM; third row - residuals. The colour bar shows relative intensity measurements with darkest colours indicating largest fluxes.}
\label{fig:MFMTK}
\end{figure*}

In this section, we present preliminary results obtained by applying the codes \textsc{Morfometryka}  \citep{Ferrari2015} and  MegaMorph \citep{Bamford2011} on one image of a bright spiral galaxy in Stripe 82, NGC 0450, and its companion, using data from DR1, to provide an example of what is planned for the entire survey.
\textsc{Morfometryka} does not require any initial  input for the fit, except that the galaxy must be roughly centred on the image stamp and an image of the PSF \citep{Ferrari2015} must be available. The program estimates the sky background iteratively, segments the image, performs basic photometry, and measures morphometric parameters.   
One example of a typical output from  \textsc{Morfometryka} (fully explained in Ferrari et al. 2015) is shown in the top panel of Fig. \ref{fig:MFMTK}. These include single S\'ersic 1D and 2D fit parameters and non-parametric morphometric parameters (concentration, asymmetry, clumpyness, Gini, the second moment of the light distribution, entropy, spirality, and light-profile curvature).  
These parameters can be  combined to assign a morphological class to each galaxy, or they can be used to yield information about the structure of the galaxy. For example, the concentration varies critically among different galaxy classes;  the polar map, used to compute the image-gradient and $\sigma_{\psi}$, has a nearly flat profile for ellipticals, whereas for  spiral galaxies it exhibits  peaks corresponding to the  spiral arms, and  in S0 galaxies it  may have some variation due, for example, to the presence of a bar.
Subsequently, one uses the \textsc{Morfometryka} outputs to create the initial input file to run MegaMorph-GALFITM. In Fig. \ref{fig:MFMTK}, bottom panel, upper-row, we show the images of the galaxy in 12 bands, and in the middle-row,  we show the models obtained fitting all bands simultaneously using MegaMorph-GALFITM. The  galaxy has been fitted with two components, a disk (exponential profile) and a bulge (S\'ersic profile). The residuals are shown in the third row.
Considering these results, exemplified in only one case here, we plan to devise a galaxy morphological classification method, based on the derived parameters and best fitting models. 

\subsection{Example of IFU-like Science with S-PLUS}
\label{IFUscience}

S-PLUS will provide large field-of-view observations, similar to low-resolution integral field spectroscopy, for thousands of nearby galaxies, whose stellar populations are of great interest for galaxy formation and evolution studies. An important goal of S-PLUS is to explore this capability to determine accurate stellar population parameters, such as ages, metallicities, and possibly their radial gradients for extended sources, overcoming known problems such as the age-metallicity degeneracy \citep{1994ApJS...95..107W}, which complicates the differentiation of stellar populations when only optical colours are used. In particular, \citet{2019A&A...622A.181S} have shown that stellar populations derived with the 12-band Javalambre photometric system are very dependent on the choice of models and methods. 

\begin{figure}
\includegraphics[width=\linewidth]{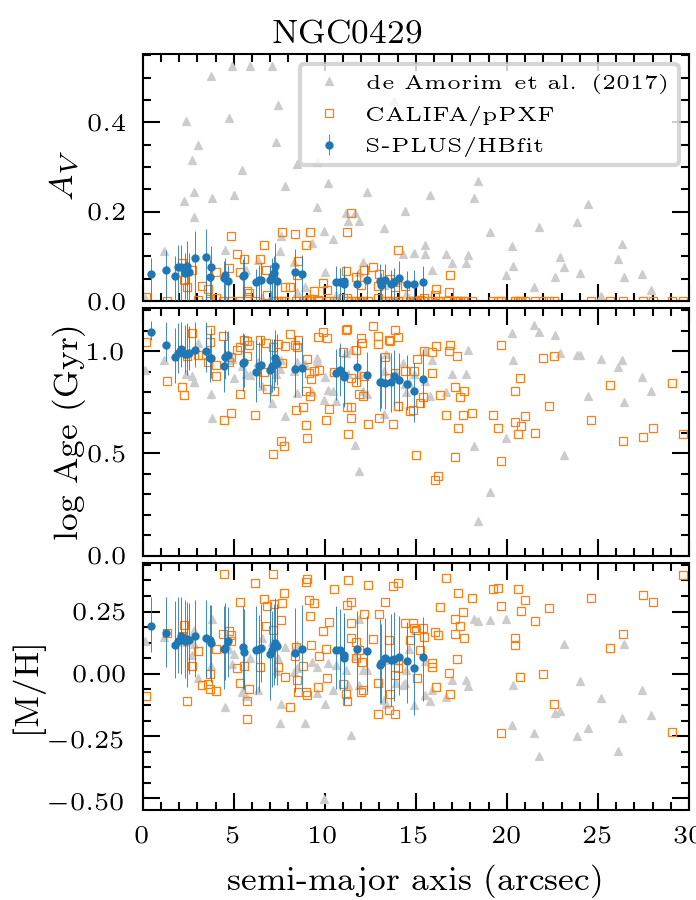}
\caption[]{Comparison of the radial profile of dust attenuation (top), mass-weighted ages (middle) and metallicity (bottom) of NGC 429, one of the galaxies in the STRIPE 82 also observed by CALIFA. Blue circles indicate the results obtained with our new hierarchical Bayesian methods using S-PLUS data, whereas orange squares indicate the results of the CALIFA data cubes using \textsc{pPXF}, both using the same stellar population models from \citep{2010MNRAS.404.1639V}. Gray triangles indicate the results made available by \citet{deamorim} using CALIFA data and a slightly different version of the stellar population models.}
\label{fig:ngc429}
\end{figure}

In this context, we developed a novel method to derive stellar populations for multi-band photometric surveys in general (Barbosa et al., in prep.), which we apply to S-PLUS. The main idea is to use a hierarchical Bayesian method that allows the modelling of all locations inside a galaxy simultaneously, such that a consistent modelling for the whole galaxy is obtained without completely erasing the information of the gradients. To test the new method, we have been using galaxies in the Stripe 82 region also observed by the CALIFA survey \citep{2012A&A...538A...8S}, whose stellar population were made available by \citet{deamorim}. Fig. \ref{fig:ngc429} shows the results of dust attenuation and stellar population gradients for NGC 429 using our method in comparison with those observed by \citet{deamorim}. However, given that we have been using single stellar population models from \citet{2010MNRAS.404.1639V}, which have a larger metallicity coverage than \citet{deamorim}, we have also determined ages and metallicities of the CALIFA galaxies independently, using the \textsc{pPXF} code \citep{2017MNRAS.466..798C}. The good agreement of extinctions, ages and metallicities of our photometric observations with spectroscopic results, in particular when the same single stellar population models are adopted, indicates that we are able to properly constrain the stellar populations using the S-PLUS data, allowing a better census of the metallicities and ages in the local universe. 

\section{Summary}

T80S is a 0.8m robotic telescope with a wide-field camera (\SI{2}{\deg\squared}) that uses 5 broad- and 7 narrow-band filters placed over the main spectral features of stars and galaxies. Its first main goal is to conduct S-PLUS, started in August 2017 and expected to reach completion in five years. The  main characteristics of the telescope and the survey are summarised in the following.

\begin{itemize}

\item S-PLUS is a 12-band optical survey aiming at imaging $\sim$\SI{8000}{\deg\squared} of the sky at high Galactic latitudes and $\sim$\SI{1300}{\deg\squared} over the disk and bulge of our Galaxy. It complements a twin project in the Northern Hemisphere, J-PLUS, being carried out with the T80/JAST, located on Cerro Javalambre, Spain.

\item The combination of a wide field-of-view telescope+camera and a 12-band filter set will allow the study of a large number of scientific topics, from Solar System to Cosmology. 

\item The first public data of  S-PLUS has been released together with this paper. They comprise 170 fields that cover about $\sim$\SI{336}{\deg\squared} of Stripe 82, in 12 bands. The data reaches a depth of r $\sim$ 21 AB mag in the broad-bands and r $\sim$ 20.5 AB in the narrow band filters, for sources detected with a significance larger than S/N$>3$. The bright saturation limit of the data is r $\sim$ 12. The data are available at NOAO data lab. 

\item The typical photo-z precision derived from S-PLUS, especially for galaxies with $r< 20.0$, surpasses that of other overlapping photometric surveys, making it possible to revisit membership analyses of nearby groups and clusters of galaxies. We forecast that, after imaging $\sim$\SI{8000}{\deg\squared} of the sky, a total of $\sim$ 2 million, $\sim$16 million and $\sim$ 32 million galaxies will be measured in the S-PLUS survey with photo-z precisions of 
$\sigma{_z}$$<$ 1.0\%, $\sigma{_z}$$<$ 2.0\% and $\sigma{_z}$$<$ 2.5\%, respectively.
  
\item Some of the main niches of S-PLUS, highlighted in this paper, are: (1) Mapping the nearby universe, (2)  Performing a pixel-by-pixel SED analysis of the sky (i.e., IFU-like science) for resolved nearby galaxies to study stellar populations, gas and dust (3) Finding metal-poor and carbon-enhanced metal-poor stars, and (4) Identifying large numbers of new quasars with precise redshifts. 

\end{itemize}

For all the science examples given in this paper, the tools developed for S-PLUS will ultimately be used for J-PAS, using deeper data and more precise photo-zs, given that J-PAS has 54 narrow-band and 5 broad-band filters.

S-PLUS also provides a rich laboratory for extension efforts. Examples could include teaching and hands-on science projects using S-PLUS data in schools, presentations to community organizations, individual studies such as citizen science efforts, offering educational content via interactive Web sites, or simply engaging the public through social media. S-PLUS, thus, offers a great toolbox to engage young students in STEM (Science, Technology, Engineering, Mathematics) and natural sciences.

Beyond S-PLUS, the plan is to use T80S as a dedicated telescope specifically to do survey-like projects, targeted at a variety of science cases that could be useful for a large number of astronomers from the involved communities.

\section*{Acknowledgments}
The S-PLUS project, including the T80S robotic telescope and the S-PLUS scientific survey, was founded as a partnership between the Funda\c{c}\~{a}o de Amparo \`{a} Pesquisa do Estado de S\~{a}o Paulo (FAPESP), the Observat\'{o}rio Nacional (ON), the Federal University of Sergipe (UFS), and the Federal University of Santa Catarina (UFSC), with important financial and practical contributions from other collaborating institutes in Brazil, Chile (Universidad de La Serena), and Spain (Centro de Estudios de F\'{\i}sica del Cosmos de Arag\'{o}n, CEFCA).  The members of the collaboration are grateful for the support received from the Conselho Nacional de Desenvolvimento Cient\'{\i}fico e Tecnol\'{o}gico (CNPq; grants 312333/2014-5, 306968/2014-2, 142436/2014-3, 459553/2014-3, 400738/2014-7, 302037/2015-2, 312307/2015-2, 300336/2016-0, 304184/2016-0, 304971/2016-2, 401669/2016-5, 308968/2016-6, 309456/2016-9, 421687/2016-9, 150237/2017-0, 311331/2017-3, 304819/2017-4, and 200289/2017-9), FAPESP (grants 2009/54202-8, 2011/51680-6, 2014/07684-5, 2014/11806-9, 2014/13723-3, 2014/18632-6, 2016/17119-9, 2016/12331-0, 2016/21532-9, 2016/21664-2, 2016/23567-4, 2017/01461-2, 2017/23766-0, and 2018/02444-7), the Coordena\c{c}\~{a}o de Aperfei\c{c}oamento de Pessoal de N\'{\i}vel Superior (CAPES; grants 88881.030413/2013-01 and 88881.156185/2017-01), the Funda\c{c}\~{a}o de Amparo \`{a} Pesquisa do Estado do Rio de Janeiro (FAPERJ; grants 202.876/2015, 202.835/2016, and 203.186/2016), the Financiadora de Estudos e Projetos (FINEP; grants 1217/13-01.13.0279.00 and 0859/10-01.10.0663.00), the Direcci\'{o}n de Investigaci\'{o}n y Desarrollo de la Universidad de La Serena (DIDULS/ULS; projects PR16143 and PTE16146 and the Programa de Investigadores Asociados), and the Direcci\'{o}n de Postgrado y Post\'{\i}tulo. TCB, VMP, and DDW acknowledge the support from the Physics Frontier Center for the Evolution of the Elements (JINA-CEE) through the US National Science Foundation (grant PHY 14-30152). JLNC is grateful for financial support received from the Southern Office of Aerospace Research and development (SOARD; grants FA9550-15-1-0167 and FA9550-18-1-0018) of the Air Force Office of the Scientific Research International Office of the United States (AFOSR/IO). YJT and RAD acknowledge support from the Spanish National Research Council (CSIC) I-COOP+ 2016 program (grant COOPB20263), and the Spanish Ministry of Economy, Industry, and Competitiveness (MINECO; grants AYA2013-48623-C2-1-P and AYA2016-81065-C2-1-P). RAOM acknowledges support from the Direcci\'{o}n General de Asuntos del Personal Acad\'{e}mico of the Universidad Nacional Aut\'{o}noma de M\'{e}xico (DGAPA-UNAM) through a postdoctoral fellowship from the Programa de Becas Posdoctorales en la UNAM. 

Funding for the SDSS and SDSS-II has been provided by the Alfred P. Sloan Foundation, the Participating Institutions, the National Science Foundation, the U.S. Department of Energy, the National Aeronautics and Space Administration, the Japanese Monbukagakusho, the Max Planck Society, and the Higher Education Funding Council for England. The SDSS Web Site is http://www.sdss.org/. The SDSS is managed by the Astrophysical Research Consortium for the Participating Institutions. The Participating Institutions are the American Museum of Natural History, Astrophysical Institute Potsdam, University of Basel, University of Cambridge, Case Western Reserve University, University of Chicago, Drexel University, Fermilab, the Institute for Advanced Study, the Japan Participation Group, Johns Hopkins University, the Joint Institute for Nuclear Astrophysics, the Kavli Institute for Particle Astrophysics and Cosmology, the Korean Scientist Group, the Chinese Academy of Sciences (LAMOST), Los Alamos National Laboratory, the Max-Planck-Institute for Astronomy (MPIA), the Max-Planck-Institute for Astrophysics (MPA), New Mexico State University, Ohio State University, University of Pittsburgh, University of Portsmouth, Princeton University, the United States Naval Observatory, and the University of Washington.

We are grateful for the contributions of CTIO staff in helping in the construction, comissioning and maintenance of the telescope and camera and we are particularly thankful to the CTIO director, Steve Heathcote, for his support at every phase, without which this project would not have been completed. We thank C\'{e}sar \'{I}\~{n}iguez for making the 2D measurements of the filter transmissions at CEFCA. We warmly thank David Crist{\'o}bal-Hornillos and his group for helping us to install and run the reduction package \textsc{jype} version 0.9.9 in the S-PLUS computer system in Chile. We warmly thank Mariano Moles, Javier Cenarro, Tamara Civera, Sergio Chueca, Javier Hern\'{a}ndez Fuertes, Antonio Mar\'{i}n Franch, Jesus Varella and Hector Vazquez Ramio - the success of the S-PLUS project relies on the dedication of these and other CEFCA staff members in building OAJ and running J-PLUS and J-PAS. We deeply thank Rene Laporte and INPE, as well as Keith Taylor, for their contributions to the T80S camera.

\nocite{*}
\bibliographystyle{mnras} 
\bibliography{reference}

\appendix
\section{Transformation equations between Filter systems.}

The Southern Hemisphere is covered by several photometric surveys (see Fig. \ref{T80Ssinergies}). Although different surveys may overlap, the combination of the datasets is not straightforward, due to the differences between the filter systems. In this section, we provide the expected colour terms between S-PLUS and other surveys in the Southern Hemisphere (DES \& KiDS) with similar filters.  In addition, we provide simple transformation equations to convert S-PLUS magnitudes to Gaia magnitudes ($Gb$, $G$, $Gr$), using the same methodology presented in \citet{Molino14} for the ALHAMBRA survey.  

\subsection{Gaia}

\begin{figure}
\includegraphics[width=9cm]{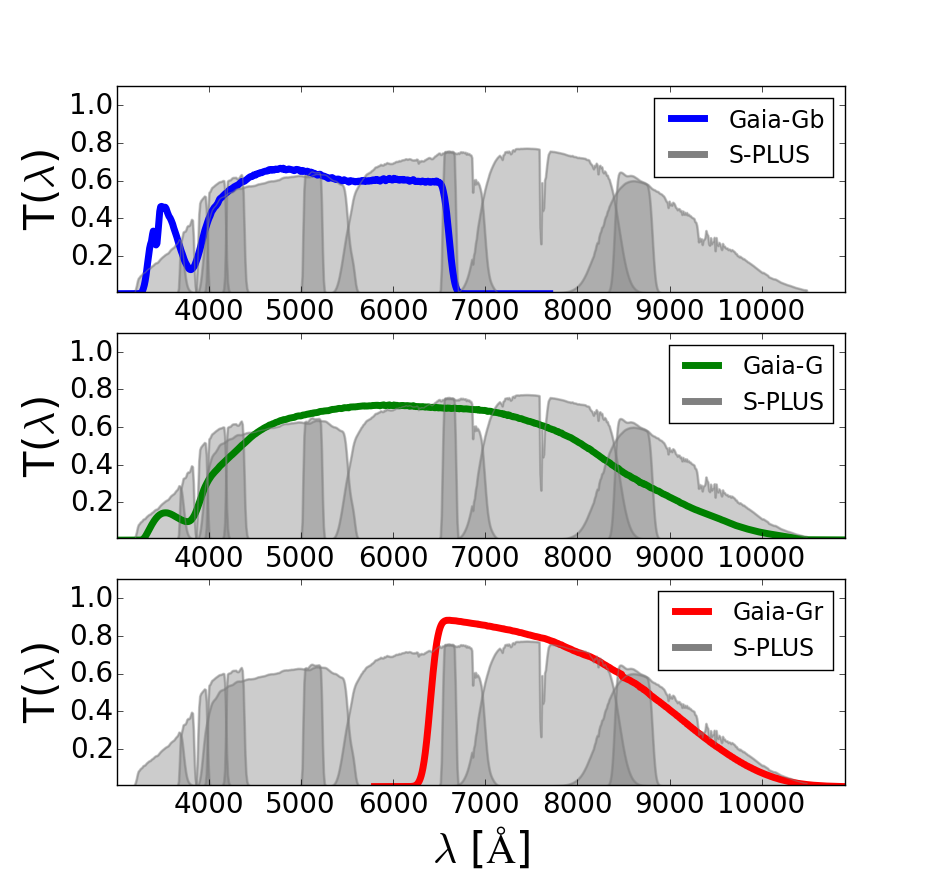}
\caption[Gaia2Splus]{Comparison between the Gaia $Gb$ (blue), $G$ (green), and $Gr$ (red) filters and the S-PLUS filters (light grey).}
\label{gaia2splus}
\end{figure}

Gaia integrates the flux detected by the low-resolution blue and red photometers (BP and RP) to provide photometric estimates in three bands: $G$ (unfiltered light), $Gb$ (blue light) and $Gr$ (red light) as illustrated in Fig. \ref{gaia2splus}. In order to convert S-PLUS magnitudes into Gaia magnitudes, we provide simple transformation equations (\ref{Gb}, \ref{G}, and \ref{Grp}), accurate up to a 3\% level. 

\begin{dmath}
\label{Gb}
Gaia_{\,Gb} = - 0.086 \times m_{u} + 0.087 \times m_{J0378} \\
+ 0.015 \times m_{J0395} - 0.016 \times m_{J0410} \\
- 0.018 \times m_{J0430} + 0.751 \times m_{g} \\
- 0.114 \times m_{J0515} + 0.439 \times m_{r} \\ 
- 0.064 \times m_{J0660} + 0.163
\end{dmath}

\begin{dmath}
\label{G}
Gaia_{\,G} = - 0.033 \times m_{J0395} - 0.029 \times m_{J0410} \\
- 0.004 \times m_{J0430} + 0.349 \times m_{g} \\ 
- 0.053 \times m_{J0515} + 0.314 \times m_{r} \\ 
- 0.004 \times m_{J0660} + 0.286 \times m_{i} \\
- 0.018 \times m_{J0861} + 0.178 \times m_{z} + 0.253
\end{dmath}

\begin{dmath}
\label{Grp}
Gaia_{\,Gr} = + 0.078 \times m_{r} + 0.073 \times m_{J0660} \\
+ 0.544 \times m_{i} + 0.008 \times m_{J0861} \\ 
+ 0.296 \times m_{z}  + 0.020
\end{dmath} 

\subsection{SDSS and KiDS}
\label{KiDS2SPLUS}
Here we present the expected colour terms between SDSS fiters ($ugriz$) used in SDSS and KIDS and the S-PLUS ($ugriz$) filters, using two libraries of templates. For stars, we rely on six stellar models from the Pickles library \citep{Pickles}. For galaxies, we rely on the BPZ templates (Ben\'itez 2000), using a redshift grid  $z$=(0.00, 0.05, 0.20). The different models and the estimated colour terms are shown in Table \ref{KiDS-SPLUS-colours}.

\begin{table*}
\begin{minipage}{140mm}
\begin{center}
\caption{Estimated colour-terms between the SDSS and the S-PLUS ($ugriz$) broadband filters for stars and galaxies. For the former we relied on six stellar models from the Pickles library; for the latter the templates from the BPZ code.} 
\label{KiDS-SPLUS-colours}
\begin{tabular}{c c c c c c}     
\hline\hline   
Model & $u_{SDSS}-u_{S-PLUS}$ & $g_{SDSS}-g_{S-PLUS}$ & $r_{SDSS}-r_{S-PLUS}$ & $i_{SDSS}-i_{S-PLUS}$ & $z_{SDSS}-z_{S-PLUS}$ \\
\hline
Stars &   &      &      &      &      \\
\hline
o5v   &   0.00    &  -0.02   &   -0.01   &   -0.04   &   0.04  \\
b5iii &  -0.08    &  -0.01   &   -0.01   &   -0.03   &   0.03  \\
a5v   &   -0.18   &   0.00   &   0.00    &   -0.02   &   0.00  \\
f5v   &   -0.07   &   0.01   &   -0.00   &   0.00    &   0.01  \\
g5v   &   -0.03   &   0.02   &   0.01    &   0.01    &   -0.01 \\
k0v   &   -0.02   &   0.03   &   0.01    &   0.01    &   -0.01 \\
k7v   &   -0.04   &   0.09   &   0.01    &   0.04    &   -0.04 \\
m5v   &   -0.09   &   0.12   &   0.06    &   0.16    &   -0.12 \\
\hline
Galaxies       &      &      &      &      &     \\ 
\hline
Ell  ($z=0.00$) & -0.01   &   0.04   &   0.02   &   0.03   &   -0.07 \\
Sbc ($z=0.00$)  & -0.07   &   0.02   &   0.02   &   0.04   &   -0.06 \\
Scd ($z=0.00$)  & -0.06   &   0.01   &   0.01   &   0.02   &   -0.03 \\
Im  ($z=0.00$)  & -0.05   &   -0.01  &   0.01   &   0.02   &   -0.04 \\
SB  ($z=0.00$)  & -0.03   &   -0.04  &   -0.04  &   0.01   &   -0.07 \\
\hline
Ell  ($z=0.05$) & -0.05   &   0.02   &   0.02   &   0.03   &   -0.07 \\
Sbc ($z=0.05$)  & -0.06   &   0.02   &   0.01   &   0.04   &   -0.06 \\
Scd ($z=0.05$)  & -0.04   &   0.01   &   0.01   &   0.02   &   -0.03 \\
Im  ($z=0.05$)  & -0.04   &  -0.00   &   0.01   &   0.02   &   -0.04 \\
SB  ($z=0.05$)  & -0.01   &  -0.02   &   0.15   &  -0.03   &   -0.06 \\
\hline
Ell  ($z=0.20$) & -0.14   &   0.11   &   0.03   &   0.04   &   -0.05 \\
Sbc ($z=0.20$)  & -0.05   &   0.05   &   0.02   &   0.03   &   -0.06 \\
Scd ($z=0.20$)  & -0.03   &   0.04   &   0.01   &   0.02   &   -0.03 \\
Im  ($z=0.20$)  & -0.01   &   0.02   &  -0.01   &   0.02   &   -0.04 \\
SB  ($z=0.20$)  & -0.01   &   0.00   &  -0.04   &   0.01   &   -0.02 \\
\hline                    
\end{tabular}
\end{center}
\end{minipage}
\end{table*}   

\subsection{DES}
\label{DES2SPLUS}
Similar to what was done in  \S\ref{KiDS2SPLUS}, we compute the expected colour terms between the DES ($griz$) and the S-PLUS ($griz$) broad-band filters. The estimated colour terms are shown in Table \ref{DES_SPLUS_colours}.

\begin{table*}
\begin{minipage}{140mm}
\begin{center}
\caption{Estimated colour terms between the DES and the S-PLUS ($g$,$r$,$i$,$z$) broad-band filters. The table includes the colour terms for stars and galaxies. For the former, we relied on six stellar models from the Pickles library; for the latter on templates from the BPZ code.}
\label{DES_SPLUS_colours}
\begin{tabular}{c c c c l}    
\hline\hline   
Model & $g_{DES}-g_{S-PLUS}$ & $r_{DES}-r_{S-PLUS}$ & $i_{DES}-i_{S-PLUS}$ & $z_{DES}-z_{S-PLUS}$ \\
\hline
Stars &     &      &      &      \\
\hline
o5v   &   -0.02   &   0.05   &   0.05   &   0.10   \\
b5iii &   -0.01   &   0.03   &   0.03   &   0.06   \\
a5v   &   -0.00   &   0.02   &   0.02   &   0.00   \\
f5v   &   0.01   &   -0.01   &   -0.00   &   0.01  \\
g5v   &   0.02   &   -0.02   &   -0.01   &   -0.03 \\
k0v   &   0.03   &   -0.03   &   -0.01   &   -0.03 \\
k7v   &   0.06   &   -0.07   &   -0.05   &   -0.08 \\
m5v   &   0.06   &   -0.17   &   -0.13   &   -0.25 \\
\hline
Galaxies       &      &      &      &     \\ 
\hline
Ell  ($z=0.00$) & 0.03   &   -0.05   &   -0.03   &   -0.14 \\
Sbc ($z=0.00$)  & 0.02   &   -0.04   &   -0.04   &   -0.14 \\
Scd ($z=0.00$)  & 0.02   &   -0.03   &   -0.02   &   -0.06 \\
Im  ($z=0.00$)  & 0.01   &   -0.03   &   -0.02   &   -0.10 \\
SB  ($z=0.00$)  & 0.01   &   -0.03   &   -0.01   &   -0.15 \\
\hline
Ell  ($z=0.05$) & 0.05   &   -0.06   &   -0.03   &   -0.16 \\
Sbc ($z=0.05$)  & 0.02   &   -0.04   &   -0.04   &   -0.13 \\
Scd ($z=0.05$)  & 0.02   &   -0.03   &   -0.02   &   -0.06 \\
Im  ($z=0.05$)  & 0.01   &   -0.02   &   -0.02   &   -0.10 \\
SB  ($z=0.05$)  & 0.01   &   -0.08   &   0.06    &   -0.14 \\
\hline
Ell  ($z=0.20$) & 0.07   &   -0.06   &   -0.04   &   -0.11 \\
Sbc ($z=0.20$)  & 0.04   &   -0.04   &   -0.03   &   -0.13  \\
Scd ($z=0.20$)  & 0.04   &   -0.03   &   -0.02   &   -0.07 \\
Im  ($z=0.20$)  & 0.02   &   -0.02   &   -0.02   &   -0.09 \\
SB  ($z=0.20$)  & 0.00   &   -0.02   &   -0.03   &   -0.04 \\
\hline                    
\end{tabular}
\end{center}
\end{minipage}
\end{table*}   

\subsection{From narrow to broad S-PLUS filters.}

\begin{table*}
\begin{minipage}{140mm}
\begin{center}
\caption{Estimated colour terms between the S-PLUS ($u$,$g$,$r$,$z$) broad-band and narrow-band filters (J0378, J0515, J0660, and J0861). As in previous tables, we relied on six stellar models from the Pickles library.}       
\label{SPLUS_NB2BBs}
\begin{tabular}{c c c c c l}     
\hline\hline   
Model & $J0378-u_{S-PLUS}$ & $J0515-g_{S-PLUS}$ & $J0660-r_{S-PLUS}-$ & $J0861-z_{S-PLUS}$ \\
\hline
Stars &   &      &      &      \\
\hline
o5v   &    0.06   &   0.20   &   0.13   &   0.00 \\
b5iii &   -0.23   &   0.10   &   0.11   &   0.00 \\
a5v   &   -0.54   &  -0.03   &   0.11   &   -0.00 \\
f5v   &   -0.31   &   -0.12  &  -0.00   &   0.00  \\
g5v   &   -0.26   &   -0.17  &  -0.05   &   0.01 \\
k0v   &   -0.24   &   -0.17  &  -0.09   &   0.02 \\
k7v   &   -0.27   &   -0.11  &  -0.26   &   0.00 \\
m5v   &   -0.26   &   -0.33  &  -0.52   &   0.03 \\
\hline      
\end{tabular}
\end{center}
\end{minipage}
\end{table*}

Finally, in this section we provide the internal colour terms for the overlapping narrow-band (J0378, J0515, J0660, and J0861) and the closest broad-band ($u$,$g$,$r$,$z$) filters in the S-PLUS system. These coefficients are shown in Table \ref{SPLUS_NB2BBs}, using six stellar models from the Pickles library. 

\clearpage
\section{Affiliations}
\label{sec:affiliations}
$^{7}$ Department of Physics, University of Notre Dame, Notre Dame, IN 46556, USA \\
$^{8}$ JINA Center for the Evolution of the Elements (JINA-CEE), USA\\
$^{9}$Departamento de F\'isica Matem\'atica, Instituto de F\'isica, Universidade de S\~{a}o Paulo, SP, Rua do Mat\~{a}o 1371, S\~{a}o Paulo, Brazil\\
$^{10}$Centro de Estudios de F\'isica del Cosmos de Arag\'on (CEFCA), Plaza San Juan, 1, E-44001, Teruel, Spain\\
$^{11}$X-ray Astrophysics Laboratory, NASA Goddard Space Flight Center, Greenbelt, MD 20771, USA\\
$^{12}$Center for Space Science and Technology, University of Maryland, Baltimore County, 1000 Hilltop Circle, Baltimore, MD 21250, USA\\
$^{13}$Observat\'orio do Valongo, Universidade Federal do Rio de Janeiro, Ladeira Pedro Ant\^onio 43,\\
Rio de Janeiro, RJ, 20080-090, Brazil\\	
$^{14}$Departamento de F\'isica Te\'orica e Experimental, Universidade Federal do Rio Grande do Norte, CP 1641, Natal, RN, 59072-970, Brazil\\
$^{15}$Donostia International Physics Center (DIPC),  Manuel Lardizabal Ibilbidea, 4, San Sebasti\'an, Spain \\
$^{16}$Department of Physics \& Astronomy, University of North Carolina at Chapel Hill, Chapel Hill, NC 27599-3255, USA\\	
$^{17}$Instituto de Matem\'atica Estat\'istica e F\'isica, Universidade Federal do Rio Grande, Rio Grande, RS, 96201-900, Brazil\\
$^{18}$Department of Astronomy, University of Florida, 211 Bryant Space Center, Gainesville, FL 32611, USA\\
$^{19}$Instituto de F\'{i}sica y Astronom\'{i}a, Universidad de Valpara\'{i}so, Gran Breta\~{n}a 1111, Valpara\'{i}so, Chile\\
$^{20}$Universidade Estadual de Santa Cruz, DCET, Rodovia Jorge Amado km 16, Ilh\'{e}us 45662-000, Bahia, Brazil\\
$^{21}$Departamento de F\'isica y Astronom\'ia. Universidad de La Serena.  Avenida Juan Cisternas 1200 Norte, La Serena, Chile\\		
$^{22}$Instituto de Astronom\'ia, Universidad Nacional Aut\'onoma de M\'exico, AP 70-264, 04510 CDMX, Mexico  \\
$^{23}$ Instituto de F\'\i sica, Universidade Federal do Rio de Janeiro, C. P. 68528, CEP 21941-972, Rio de Janeiro, RJ, Brazil\\
$^{24}$Planet\'ario, Instituto de Estudos Socioambientais, Universidade Federal de Goi\'as, Goi\^ania, 74055-140, Brazil	\\	
$^{25}$Instituto de F\'isica, Universidade Federal de Goi\'as, Goi\^ania, 74001-970, Brazil \\
$^{26}$ Departamento de F\'{\i}sica, Centro Universit\'ario da FEI; Av. Humberto de Alencar Castelo Branco, 3972, 09850-901, S\~ao Bernardo co Campo-SP, Brazil \\
$^{27}$Campus Duque de Caxias, Universidade Federal do Rio de Janeiro, Rodovia Washington Luiz km 104.5, Duque de Caxias, RJ, 25265-970, Brazil\\
$^{28}$Universidade Federal do ABC (UFABC), Rua Santa Ad\'elia, 166, 09210-170, Santo Andr\'e - SP, Brazil\\
$^{29}$Centro de Astronom\'{i}a (CITEVA), Universidad de Antofagasta, Av. Angamos 601, Antofagasta, Chile\\	
$^{30}$Instituto Milenio de Astrof\'{i}sica, Santiago, Chile\\
$^{31}$Departamento de F\'{i}sica - ICEx - UFMG, Av. Ant\^{o}nio Carlos, 6627, 30270-901 Belo Horizonte, MG, Brazil\\
$^{32}$Univ. Grenoble Alpes, CNRS, IPAG, 38000 Grenoble, France\\
$^{33}$Aix Marseille Universit\'e, CNRS, LAM (Laboratoire d'Astrophysique de Marseille), Marseille, France\\
$^{34}$Department of Physics and Astronomy, University of Pennsylvania, 45 Philadelphia, PA 19104, USA\\
$^{35}$Instituto de Astrof\'{\i}sica de Andaluc\'{\i}a (IAA-CSIC), Glorieta de la Astronom\'{\i}a s/n, E-18008, Granada, Spain\\
$^{36}$Centre for Astrophysics and Cosmology, Science Institute, University of Iceland, Dunhagi 5, 107 Reykjavik, Iceland\\
$^{37}$Departamento de F\'{\i}sica Te\'orica, Universidad Aut\'onoma de Madrid, 28049, Madrid, Spain\\
$^{38}$Instituto de Investigaci\'on Multidisciplinario en Ciencia y Tecnolog\'ia, Universidad de La Serena. Avenida Juan Cisternas \#1015, La Serena, Chile \\
$^{39}$AURA Observatory in Chile, Cisternas 1500, La Serena, Chile\\
$^{40}$Instituto de Ci\^{e}ncias Matem\'aticas e de Computa\c c\~ao, Universidade de S\~ao Paulo, Avenida Trabalhador S\~ao-carlense 400, S\~ao Carlos, SP, Brazil\\
$^{41}$IFAS FYCS Department, University of Florida, PO Box 110310, 3041 McCarty D, Gainesville, FL 32611, USA\\
$^{42}$Steward Observatory, University of Arizona, 933 N. Cherry Ave, Tucson, AZ, USA\\
$^{43}$Universidade do Vale do Para\'{i}ba, Av. Shishima Hifumi, 2911, Cep 12244-000, S\~{a}o Jos\'{e} dos Campos, SP, Brazil\\
$^{44}$Facultad de Cs. Astron\'omicas y Geof\'{\i}sicas, UNLP, Paseo del Bosque s/n, B1900FWA, La Plata, Argentina\\
$^{45}$Instituto de Astrof\'{i}sica de La Plata, UNLP, CONICET, Paseo del Bosque s/n, B1900FWA, La Plata, Argentina\\
$^{46}$Departamento de Ci\^{e}ncia da Computa\c{c}\~{a}o, Instituto de Matem\'{a}tica e Estat\'{i}stica da USP,
Cidade Universit\'{a}ria, 05508-090, S\~ao Paulo, SP, Brazil\\
$^{47}$Centro Brasileiro de Pesquisas F\'isicas, Rua Dr. Xavier Sigaud 150,  Rio de Janeiro, RJ, CEP 22290-180, Brazil\\
$^{48}$Institut de Ci\`encies del Cosmos, Universitat de Barcelona (UB-IEEC), Mart\'\i\ i Franqu\`es 1, 08028 Barcelona, Catalonia, Spain. \\
$^{49}$Gemini Observatory/AURA, Southern Operations Center, Casilla 603 La Serena, Chile\\
$^{50}$Inter-University Centre for Astronomy and Astrophysics, Pune 411007, India\\
$^{51}$Consejo Nacional de Investigaciones Cient\'{i}ficas y T\'ecnicas, Godoy Cruz 2290, C1425FQB, CABA, Argentina\\
$^{52}$Universidade Federal do Paran\'a, Campus Jandaia do Sul, Rua Dr. Jo\~ao Maximiano, 426, Jandaia do Sul-PR, 86900-000, Brazil\\
$^{53}$Institute for Astronomy, Astrophysics, Space Applications and Remote Sensing, National Observatory of Athens, Penteli, 15236, Athens, Greece\\
$^{54}$Instituci\'o Catalana de Recerca i Estudis Avan\c cats, Barcelona, Catalonia, Spain.

\label{lastpage}

\end{document}